\documentclass[9pt, a4paper]{article}
\usepackage[utf8]{inputenc}
\usepackage{graphicx}	% Including figure files
\usepackage{amsmath}	% Advanced maths commands
\usepackage{amssymb}	% Extra maths symbols
\usepackage{latexsym}
\usepackage{hyperref}	% Permits to insert hypertext links
\usepackage{mathtools}
\usepackage{rotating}
\usepackage{epsfig}
\usepackage{tabularx}
\usepackage{natbib}
\usepackage{xcolor}
\usepackage{aas_macros}
\usepackage[
left=1.5cm,
right=1.5cm,
top=2.00cm,
bottom=2.00cm
]{geometry}
\usepackage{comment}
\usepackage{authblk}
\usepackage{titlesec}
\usepackage[labelfont=bf,figurename=Fig.,font=small,labelsep=period]{caption}
\usepackage{abstract}
\usepackage{bm}
\usepackage[T1]{fontenc}
\usepackage[english]{babel}
\usepackage{svg}
\usepackage{tikz}

\titleformat{\section}
  {\normalfont\large\bfseries}{\thesection.}{1em}{}
\titleformat{\subsection}
  {\normalfont\normalsize\bfseries}{\thesubsection.}{1em}{}
\titleformat{\subsubsection}
  {\normalfont\normalsize\bfseries}{\thesubsubsection.}{1em}{}

\usetikzlibrary{shapes.geometric, arrows}

\tikzstyle{startstop} = [rectangle, rounded corners, minimum width=3cm, minimum height=1cm,text centered, draw=black, fill=red!30]
\tikzstyle{io} = [trapezium, trapezium left angle=70, trapezium right angle=110, minimum width=3cm, minimum height=1cm, text centered, draw=black, fill=blue!30]
\tikzstyle{process} = [rectangle, minimum width=3cm, minimum height=1cm, text centered, text width = 2.5cm,  draw=black, fill=orange!30]
\tikzstyle{decision} = [diamond, minimum width=2cm, minimum height=1cm, text centered,text width = 2cm, draw=black, aspect=2, fill=green!30]
\tikzstyle{arrow} = [thick,->,>=stealth]

\title{\huge{\textbf{The Aegis Orbit Determination and Impact Monitoring System and services of the ESA NEOCC web portal}}}

\author[1,2]{M. Fenucci\thanks{Corresponding author: \url{marco.fenucci@ext.esa.int}}}
\author[1,3]{L. Faggioli}
\author[1,3]{F. Gianotto}
\author[4]{D. Bracali Cioci}
\author[5]{J. L. Cano}
\author[1]{L. Conversi}
\author[1,3]{M. Devog\`{e}le}
\author[5]{G. Di Girolamo}
\author[1]{C. Drury}
\author[1,3]{D. F\"{o}hring}
\author[1,6]{L. Gisolfi}
\author[5,7]{R. Kresken}
\author[1,3]{M. Micheli}
\author[1]{R. Moissl}
\author[8,9]{F. Oca\~{n}a}
\author[1,2]{D. Oliviero}
\author[1,10]{A. Porru}
\author[8,11]{P. Ramirez-Moreta}
\author[12,13]{R. Rudawska}
\author[4]{F. Bernardi}
\author[4]{A. Bertolucci}
\author[4]{L. Dimare}
\author[4]{F. Guerra}
\author[4]{V. Baldisserotto}
\author[14]{M. Ceccaroni}
\author[3]{R. Cennamo}
\author[15]{A. Chessa}
\author[16]{A. Del Vigna}
\author[17]{D. Koschny}
\author[2]{A. M. Teodorescu}
\author[18]{E. Perozzi}

\affil[1]{ESA ESRIN / PDO / NEO Coordination Centre, Largo Galileo Galilei, 1, 00044 Frascati (RM), Italy}
\affil[2]{Elecnor Deimos, Via Giuseppe Verdi, 6, 28060 San Pietro Mosezzo (NO), Italy}
\affil[3]{Starion Italia, Via di Grotte Portella, 28, 00044 Frascati (RM), Italy}
\affil[4]{Space Dynamics Services s.r.l., via Mario Giuntini, Navacchio di Cascina, Pisa, Italy}
\affil[5]{ESA ESOC / PDO, Robert-Bosch-Straße 5, 64293 Darmstadt, Germany}
\affil[6]{Universit\`a degli studi di Padova, Dipartimento di Fisica e Astronomia, Vicolo dell'Osservatorio, 3, 35122 Padova, Italy}
\affil[7]{CGI Deutschland B.V. \& Ko. KG, Rheinstrasse 95, 64295 Darmstadt, Germany}
\affil[8]{ESA ESAC / PDO, Bajo del Castillo s/n, 28692 Villafranca del Castillo, Madrid, Spain}
\affil[9]{Deimos Space S.L.U., Ronda de Poniente, 19, 28760 Tres Cantos Madrid, Spain}
\affil[10]{Alia Space System Srl, Via San Giuseppe Calasanzio 15, 00044 Frascati (RM), Italy}
\affil[11]{GMV, Isaac Newton 11, Tres Cantos, 28760 Madrid, Spain}
\affil[12]{ESA ESTEC / PDO, Keplerlaan 1, 2201 AZ Noordwijk, The Netherlands}
\affil[13]{Starion Netherlands, Schuttersveld 2, 2316 ZA, Leiden, The Netherlands}
\affil[14]{School of Aerospace, Transport and Manufacturing, Cranfield University, Cranfield MK43 0AL, UK}
\affil[15]{NHAZCA s.r.l., Via Vittorio Bachelet, 12, 00185, Rome, Italy}
\affil[16]{Dipartimento di Matematica, Università di Pisa, Largo  B. Pontecorvo 5, 56127 Pisa, Italy}
\affil[17]{LRT / TU Munich, Boltzmannstraße 15, 85748 Garching bei München, Germany}
\affil[18]{Agenzia Spaziale Italiana, Via del Politecnico 1, 00133 Rome, Italy}

\date{\today}

\begin{document}
% Activate line numbering
%\linenumbers

%\twocolumn[
%  \begin{@twocolumnfalse}
    \maketitle
    \begin{abstract}
        The NEO Coordination Centre (NEOCC) of the European Space Agency is an operational centre that, among other activities, computes the orbits of near-Earth objects and their probabilities of impact with the Earth. The NEOCC started providing information about near-Earth objects in 2012 on a dedicated web portal, accessible at \url{https://neo.ssa.esa.int/}. Since the beginning of the operational phase, many developments and improvements have been implemented regarding the software, the data provided, and the portal. One of the most important upgrades is that the NEOCC is now independently providing data through a newly developed Orbit Determination and Impact Monitoring system, named Aegis. All the data computed by Aegis is publicly available on the NEOCC web portal, and Aegis is also used to maintain all the major services offered. The most important services comprise an orbital catalogue of all known asteroids, a list of possible future impacts with the Earth (also called Risk List), a list of forthcoming close approaches, a set of graphical toolkits, and an on-demand ephemerides service. Many of the services are also available through dedicated APIs, which can be used to automatically retrieve data. Here we give an overview of the algorithms implemented in the Aegis software, and provide a summary of the services offered by the NEOCC that are supported by Aegis. 
    \end{abstract}
%  \end{@twocolumnfalse}
%]
%\saythanks

%% main text
\section{Introduction}
\label{s:intro}
% General introduction on NEAs, orbit determination and impact monitoring 
Near-Earth objects (NEOs) are small celestial bodies with a perihelion distance $q$ smaller than 1.3 au. They comprise near-Earth asteroids (NEAs) and near-Earth comets (NECs). For a comet to be considered NEC, an orbital period shorter than 200 years is also required. The vast majority of these objects originated in the main asteroid belt, and reached the near-Earth space by the coupled effect of gravitational forces and the Yarkovsky semi-major axis drift \citep{granvik-etal_2017}. The major concern about NEOs is the potential threat they pose to life and infrastructure on our planet. While large impact events causing mass extinctions happen statistically over a timescale of millions of years, impacts by smaller objects are much more frequent \citep{brown-etal_2002}. Asteroids the size of a few tens of metres can still cause severe damage on a local scale. It is therefore important to know about these events in advance in order to plan mitigation actions such as evacuations or even asteroid deflection missions, such as the recent Double Asteroid Redirection Test mission \citep{daly-etal_2023}. 

\begin{figure*}[!ht]
    \centering
    \includegraphics[width=0.48\textwidth]{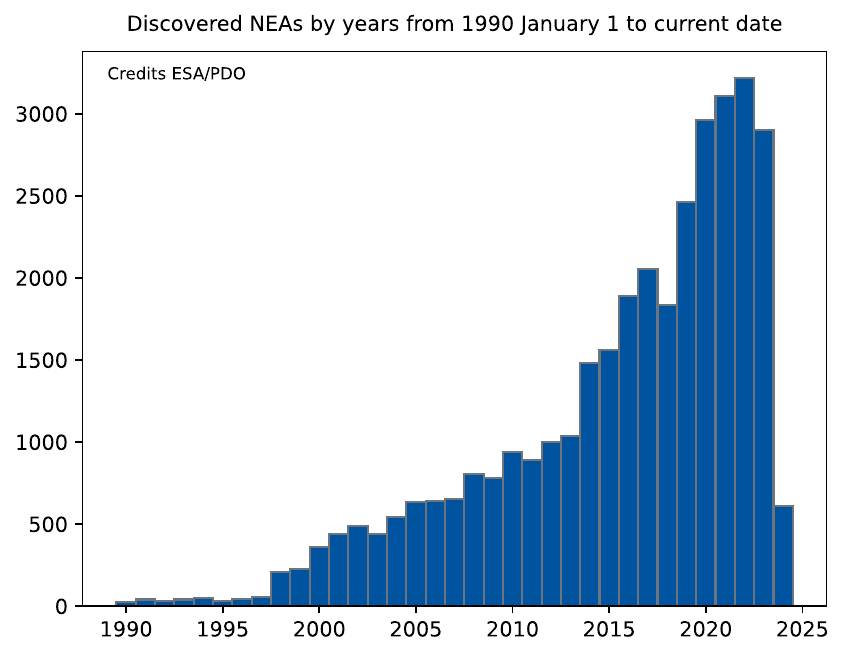}
    \includegraphics[width=0.48\textwidth]{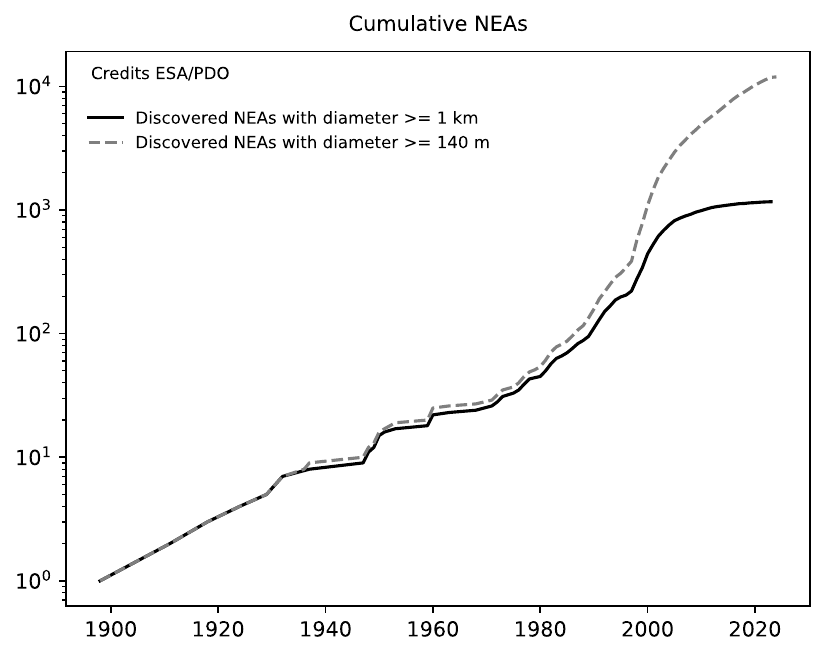}
    \caption{Left panel: distribution of discovered NEOs from 1990 to current date, binned by year. Right panel: cumulative distribution of discovered NEOs larger than 1 km (black thick curve), and of NEOs larger than 140 m (grey dashed curve). In the right panel plot, the diameter is obtained by converting the absolute magnitude $H$ into diameter $D$ with the formula $D = 1329 \textrm{ km}/\sqrt{p_V} \times 10^{-H/5}$ \citep{bowell-etal_1989, pravec-harris_2007}, where the albedo $p_V$ was set to a constant value of 0.1. Plots were produced with data present in the NEOCC database on 2024-01-01.}
    \label{fig:NEAs_Discovery}
\end{figure*}

The first step towards this end is the discovery of new NEOs. While it is estimated that about 87\%-95\% of NEOs larger than 1 km in diameter are known today \citep{nesvorny-etal_2024}, smaller asteroids are largely undiscovered. Population models predict the existence of about $10^5$ NEOs larger than 140 m in diameter \citep{granvik-etal_2018, nesvorny-etal_2023, nesvorny-etal_2024, nesvorny-etal_2024b}, which are still able to cause regional damage, and about $10^6$ NEOs larger than 30 metres in diameter, which may be threatening in the case they impact over a densely populated city. 
Although there are still many unknown NEOs, especially at small sizes, the number of discoveries has been increasing over the past 20 years (see Fig.~\ref{fig:NEAs_Discovery}), mostly on account of the Catalina Sky Survey\footnote{\url{https://catalina.lpl.arizona.edu/}} \citep{fuls-etal_2023}, the Pan-STARRS\footnote{\url{http://www2.ifa.hawaii.edu/research/Pan-STARRS.shtml}} \citep{denneau-etal_2013}, and the ATLAS\footnote{\url{https://atlas.fallingstar.com/}} \citep{tonry-etal_2018} telescopic surveys. The number of known objects is predicted to increase by by roughly an order of magnitude \citep{jones-etal_2018} with the beginning of the operational phase of new generation surveys in the coming years, such as the ESA Flyeye telescope \citep{conversi-etal_2021} and the Vera Rubin Observatory \citep{ivezic-etal_2019}. In addition, two space telescopes specifically dedicated to NEO discovery are foreseen for the next decade: the NEO Surveyor Mission \citep{mainzer-etal_2023} by NASA, and the NEO Mission in the InfraRed \citep[NEOMIR,][]{conversi-etal_2023} by the European Space Agency (ESA). These two missions will contribute to a significant increase of the number of discovered NEOs.  

Once an NEO has been observed or discovered, observations are submitted to the Minor Planet Center\footnote{\url{https://www.minorplanetcenter.net/}} (MPC), the official body recognised by the International Astronomical Union (IAU) for collecting observational data on Solar System small bodies. The subsequent step is the computation of the orbit and the cataloguing of the object. Publicly available catalogues of orbits are maintained by several institutions, such as the MPC, the JPL Solar System Dynamics group by NASA\footnote{\url{https://ssd.jpl.nasa.gov/}} (JPL SSD), NEODyS\footnote{\url{https://newton.spacedys.com/neodys/}} and AstDyS\footnote{\url{https://newton.spacedys.com/astdys/}}, and the Lowell Observatory\footnote{\url{https://asteroid.lowell.edu/}} \citep{moskovitz-etal_2022}. After the orbit is computed, the prediction of possible impacts and the evaluation of the risk are performed by the so-called impact monitoring process \citep[see][for a review]{tommei_2021}.

% ESA Planetary Defence Office
In the effort of supporting NEO related activities, ESA first established the NEO Segment of the Space Situational Awareness (SSA) programme in 2010. Since 2018, it established the current Planetary Defence Office (PDO), which has the mission of: 1) detecting NEOs and determining their dynamical and physical properties; 2) assessing the risk of and predicting possible impacts, warning decision makers and disaster relief forces in case of threats; 3) mitigating the risk through potential reconnaissance and/or deflection missions in the framework of the International Asteroid Warning Network\footnote{\url{https://iawn.net/}} (IAWN) and the Space Mission Planning Advisory Group\footnote{\url{https://www.cosmos.esa.int/web/smpag}} (SMPAG). 

% What is NEOCC and history of NEOCC, I should mention when it has been created
The main operational centre of the ESA PDO is the NEO Coordination Centre\footnote{\url{https://neo.ssa.esa.int/}} (NEOCC), located at the ESA Centre for Earth Observation (ESRIN) near Rome, Italy. The aim of the NEOCC is to coordinate and contribute observing small Solar System bodies, and evaluate the threat coming from NEOs. Since its establishment in May 2013, the NEOCC evolved with several changes and developments. 
The main activities carried out at the NEOCC include the accurate computation of asteroid orbits, which started in November 2018, and the impact monitoring of NEAs, which started in November 2020. 
These tasks are achieved by operating the Aegis Orbit Determination and Impact Monitoring system, a software developed and maintained by SpaceDyS s.r.l.\footnote{\url{https://www.spacedys.com/}} through industrial contracts issued by ESA. The results provided by the Aegis system are completely independent from those obtained by the two other impact monitoring systems, Clomon-2 by NEODyS and Sentry\footnote{\url{https://cneos.jpl.nasa.gov/sentry/}} by the JPL Center for NEO Studies (JPL CNEOS) by NASA. 

The Aegis system automatically updates the local astrometric database by downloading new observations issued by the MPC on a daily basis. In addition, Aegis maintains a catalogue of NEAs comprising orbits with their uncertainties, observations and their orbital fit residuals, close approaches, and ephemerides. A catalogue of orbits of non-NEA asteroids is also maintained. Orbit determination of NECs is not independently performed, as it relies on data published by the JPL SSD. 
More importantly, Aegis is able to compute the impact probabilities of NEAs 100 years into the future, and the results are collected in the so-called NEOCC Risk List. All the data generated by Aegis is publicly available online at the NEOCC web portal. In addition, Aegis is used to produce the input of several tools and services available on the portal, such as the tools of the NEO Toolkit and the ephemerides generation service.
In this paper, we provide an overview of the orbit determination and impact monitoring algorithms implemented in the Aegis software, and summarise all the services supported by Aegis which are available on the NEOCC web portal. We also perform a comparison of the data provided by the NEOCC and by the JPL SSD and CNEOS, and show that the results are consistent.   
\begin{table}[!ht]
    \renewcommand{\arraystretch}{1.1}
    \centering
    \begin{tabular}{ll}
    \hline
    \hline
    Acronym & Definition \\
    \hline
    ADES       & IAU Astrometry Data Exchange \\
    CAFS       & Close Approach Fact Sheet \\
    DLR        & German Aerospace Center \\
    ESA        & European Space Agency \\
    ESRIN      & ESA Centre for Earth Observation \\
    FVT        & Flyby Visualisation Tool \\
    KS         & Kustaanheimo-Stiefel \\
    LOV        & Line of Variations \\
    MPC        & Minor Planet Center \\
    MPEC       & Minor Planet Electronic Circular \\
    MOID       & Minimum Orbit Intersection Distance \\
    NEA        & Near-Earth Asteroid\\
    NEC        & Near-Earth Comet \\
    NEO        & Near-Earth Object \\
    NEOCC      & NEO Coordination Centre \\
    NEOMIR     & NEO Mission in the InfraRed \\
    IAWN       & International Asteroid Warning Network \\
    IP         & Impact Probability \\
    JPL CNEOS  & JPL Center for NEO Studies \\
    JPL SSD    & JPL Solar System Dynamics \\
    JPL SBDB   & JPL Small Body Database \\
    OPT        & Observation Planning Tool \\
    OVT        & Orbit Visualisation Tool \\
    PDO        & Planetary Defence Office \\
    PS         & Palermo Scale \\
    SCDT       & Sky Chart Display Tool \\ 
    SMPAG      & Space Mission Planning Advisory Group \\
    SOVT       & Synodic Orbit Visualisation Tool \\
    SRP        & Solar Radiation Pressure \\
    SSA        & Space Situational Awareness \\
    TP         & Target Plane \\
    VA         & Virtual Asteroid \\
    VI         & Virtual Impactor \\
    \hline
    \end{tabular}
    \caption{List of acronyms used in the paper.}
    \label{tab:acronyms}
\end{table}

The paper is structured as follows. In Sec.~\ref{s:aegis} we briefly introduce how the Aegis software was conceived by ESA.
In Sec.~\ref{s:orb_det} and \ref{s:impact_monitoring} we describe the algorithms for orbit determination and impact monitoring implemented in the Aegis software. In Sec.~\ref{s:devops} we describe the software management approach used at the NEOCC.
The services offered on the NEOCC web portal and supported by Aegis are described in Sec.~\ref{s:portal}. In Sec.~\ref{s:comparison}, we show a comparison of our data with that computed by JPL SSD and CNEOS. Finally, we summarise the content of this paper and list future developments in Sec.~\ref{s:conclusions}.

\section{Genesis of the Aegis software}
\label{s:aegis}
The need of an highly automated approach to monitoring asteroid hazards can be traced back to 1997~XF$_{11}$, when this newly discovered km-size object was predicted to have a very close approach to Earth as early as 2028 by B. G. Marsden, director of the MPC back in 1998 \citep[see IAU Ciruclar 6837][]{marsden_1998}. The subsequent turmoil of discussion both within the scientific community and the media allowed eventually to exclude the impact threat, and called for a systematic approach to NEO impact monitoring \citep{chesley-chodas_2015}.  As a result, in 1999 the NEODyS information system came into operation at the University of Pisa \citep{chesley-milani_1999}. NEODyS focused on providing on-line updated orbit determination and impact monitoring services on a daily basis through the OrbFit \citep{orbfit_2011} and Clomon (later upgraded to Clomon-2) software packages, respectively. Shortly afterwards, the Sentry Earth impact monitoring system was developed at NASA JPL \citep{chamberlin-etal_2001}, thus providing an independent source of data. While being independently developed, the two systems relied on similar methods at the beginning, summarised in \citet{milani-etal_2005}. Both systems had an internal agreement to cross-check their results in case of Palermo Scale \citep[PS,][]{chesley-etal_2002} larger than $-2$, and the United Nations Office for Outer Space Affairs (UNOOSA) took up the challenge of managing governmental interfaces should a threat come true. The Don Quijote mission proposal \citep{milani-etal_2003} laid down the basic design for an asteroid kinetic deflection mission experiment, thus completing at all levels the Planetary Defence mitigation initiatives worldwide \citep{perozzi_2014}.

Within this framework, at the turn of the first decade of the new millennium, ESA started the Space Situational Awareness (SSA) programme, whose aim was to “raise awareness on the population of space objects, the space environment, and the existing threats and risks by timely providing data and services to users, customers, and stakeholders” \citep{drolshagen-etal_2010}. The programme encompassed three segments: SST (Space Surveillance and Tracking), SWE (Space Weather) and NEO (Near-Earth Object). In order to profit of the expertise gained from ten-year NEODyS operations and with the support of the Italian Space Agency, a suite of contracts were awarded by ESA for establishing a centre devoted to NEO monitoring, equipped with state-of-the art orbit determination and impact monitoring software. A road-map was laid out together with the related technology development plan \citep{milani-valsecchi_2011a, milani-valsecchi_2011b}. ESA foresaw to resort initially to the NEODyS services as primary source of data while SpaceDyS s.r.l., a spin-off of the Celestial Mechanics Group of the University of Pisa, would lead the development of evolved versions of OrbFit and Clomon-2. In 2013 the ESA NEO Coordination Centre was established, where the computational and data dissemination activities were complemented by organizing follow-up observations of high-risk objects \citep{dipippo-perozzi_2015}. 

In 2018 the AstOD software was released to ESA, which allowed the NEOCC to independently perform orbit determination of asteroids, and to maintain catalogues of orbits and close approaches with the Earth. Once the AstOD software became operational, ESA also acquired the impact monitoring software in 2020. Since then, the ESA NEOCC started to autonomously provide impact probabilities of NEAs, and was thus no longer dependent on NEODyS data. Moreover, as highlighted in Sec.~\ref{s:orb_det} and Sec.~\ref{s:impact_monitoring}, several improvements and new algorithms were introduced with respect to OrbFit and Clomon-2. In 2022 the NEOCC orbit determination and impact monitoring system was re-named Aegis $-$ the shield used by Zeus and his daughter Athena in Greek Mythology.

\section{Aegis: Orbit Determination and Ephemerides}
\label{s:orb_det}

\subsection{Orbit determination procedure}
\label{ss:OD}
In this section we give an overview and some mathematical details of the algorithms implemented for the orbit determination of asteroids. Most of the algorithms and notations in this section rely on \citet{milani-gronchi_2009}.

\subsubsection{Least-squares method}
The orbital elements of an asteroid are computed by fitting the astrometric observations to the predictions calculated with a dynamical model. From a mathematical point of view, let us consider $m$ observations $r_i$ taken at times $t_i$, $i=1,\dots,m$. The prediction function
\begin{equation}
    r(t) = R(t, \bm{y}(t), \bm{\nu}),
    \label{eq:prediction_function}
\end{equation}
permits to simulate the outcome at the times $t_i$ from a mathematical model $R$. In Eq.~\eqref{eq:prediction_function}, $\bm{y}$ is the state vector of the asteroid, and $\bm{\nu}$ is a vector of kinematic parameters. The state vector $\bm{y}$ is determined through a dynamical model of the form
\begin{equation}
    \begin{cases}
        \dot{\bm{y}} & = \bm{f}(t, \bm{y}, \bm{\mu}), \\
        \bm{y}(t_0) & = \bm{y}_0, 
    \end{cases}
    \label{eq:dynamical_model}
\end{equation}
where $\bm{\mu}$ is a vector of dynamical parameters, $\bm{y}_0$ is the initial condition, and $\bm{f}$ is the vector field determining the motion. With this model, we can compute the observables $r(t_i)$ from Eq.~\eqref{eq:prediction_function}. From the observations and the observables, we can define the vector of residuals $\bm{\xi}=(\xi_1, \dots, \xi_m)$ where
\begin{equation}
    \xi_i = r_i - r(t_i).
    \label{eq:residuals}
\end{equation}
Let us denote with $\bm{x}=(x_1,\dots,x_n)$ a subset of parameters of $\bm{y}_0, \bm{\mu}, \bm{\nu}$, which we want to determine through the fit to the observations. The parameters in the vector $\bm{x}$ are called solve-for parameters. The residuals depend on the solve-for parameters, hence $\bm{\xi} = \bm{\xi}(\bm{x})$. To estimate the parameters in $\bm{x}$, we define the target function
\begin{equation}
    Q(\bm{x}) = \frac{1}{m} \bm{\xi}^T W \bm{\xi},
\end{equation}
where $W$ is an $m \times m$ matrix of weights of the observations. %For example, $W$ can be a diagonal matrix with elements $1/\sigma_i^2$ where $\sigma_i$ is the uncertainty of the $i$-th observation. 

The elements of the matrix $W$ are determined through an astrometric error model based on \citet{veres-etal_2017}.
In the exceptional case of follow-up observations of NEAs with a non-zero impact probability, weights might be manually set by operators. Follow-up observations are routinely performed by NEOCC astronomers and collaborating institutions\footnote{See \url{https://neo.ssa.esa.int/neocc-observing-facilities}}, and they internally send astrometric error values to Aegis operators when needed.
Optical observations also need to be corrected to take into account biases affecting star catalogues. The Aegis software includes the debiasing schemes developed by \citet{farnocchia-etal_2015b} and \citet{eggl-etal_2020}. The scheme that Aegis uses as a default is that of \citet{farnocchia-etal_2015b}, and the transition to \citet{eggl-etal_2020} is foreseen in the next version of the software. Different astrometric measurements are assumed to be uncorrelated, however this may not be generally true \citep{farnocchia-etal_2015b}, especially for observations coming from the same observatory in a short interval of time. To mitigate these correlation effects, we apply a scaling factor of $\sqrt{N/4}$ to the astrometric errors, where $N>4$ is the number of observations from the same observatory within a tracklet of maximum 8 hours.

In addition, the Aegis software is able to take into account time uncertainties by correcting the weights of an optical observation $(\alpha, \delta)$ \citep{farnocchia-etal_2022}. The $2\times2$ weight matrix of the corresponding observation is obtained as 
\begin{equation}
    W = (C_{(\alpha, \delta)}  + C_t)^{-1}, 
\end{equation}
where
\begin{equation}
    C_{(\alpha, \delta)} = 
    \begin{bmatrix}
        \sigma_\alpha^2 & \rho \\
        \rho & \sigma_\delta^2 \\
    \end{bmatrix}, 
    \quad    
    C_{t} = \sigma_{t}^2
    \begin{bmatrix}
        \dot{\alpha}^2 & \dot{\alpha}\dot{\delta} \\
        \dot{\alpha}\dot{\delta} & \dot{\delta}^2 \\
    \end{bmatrix}.
    \label{eq:corr_w}
\end{equation}
In Eq.\eqref{eq:corr_w}, $\sigma_\alpha$ and $\sigma_\delta$ are the a-priori root-mean-square (RMS) error of the observation in $\alpha$ and $\delta$, $\rho$ is their correlation, $\sigma_t$ is the time uncertainty, and $\dot{\alpha}, \dot{\delta}$ represent the apparent motion in the sky. When time uncertainties are not provided by the observers, a default value for all observatories can be set by the software operators. Note that, currently, the OrbFit software is not able to handle time uncertainties. Currently, the operational version of Aegis does not make use of time uncertainties, and they are added ad-hoc by operators only in particular cases, such as imminent impactors or close fly-bys.

\subsubsection{Full differential corrections}
% Differential corrections
The best-fit parameters $\bar{\bm{x}}$ are obtained by minimising the target function $Q$, hence $\bar{\bm{x}}$ satisfies the equation
\begin{equation}
    \frac{\partial Q}{\partial \bm{x}} = 0 = \frac{2}{m}  \bm{\xi}^T W B, \quad B = \frac{\partial \bm{\xi}}{\partial \bm{x}}.
    \label{eq:Qder}
\end{equation}
Equation~\eqref{eq:Qder} is solved with a variant of the Newton method called differential corrections. A step of the differential correction algorithm is obtained through the equation
\begin{equation}
    \bm{x}_{k+1} - \bm{x}_k =\Gamma D, 
    \label{eq:difcor}
\end{equation}
where
\begin{equation}
    D = -B^TW\bm{\xi}, \quad \Gamma =C^{-1}, \quad C =B^T W B. 
\end{equation}
The matrix $C$ is called normal matrix, while $\Gamma$ is the covariance matrix. 
Note that the residuals introduced in Eq.~\eqref{eq:residuals} depend on the units used in the given observation. To define unit-less residuals, we change the notation and we denote with $\bm{\xi}'$ the true residuals of Eq.~\eqref{eq:residuals}, and with $B'$ the partial derivatives of the true residuals. Unit-less normalised residuals are then defined as $\bm{\xi} = P \bm{\xi}'$, where $P$ is such that $P^2 = W$ and it is computed with the Cholewsky decomposition \citep[see e.g.][]{bulirsch-stoer_2002}. Then, the same formula of Eq.~\eqref{eq:difcor} for the differential correction step apply to the normalised residuals, with
\begin{equation}
    B = P B', \quad C = B^T B, \quad D = -B^T \bm{\xi}.
    \label{eq:unitless_residuals}
\end{equation}
Note that at this stage $B, C$ and $D$ in Eq.~\eqref{eq:unitless_residuals} depend on the units chosen for the parameters $\bm{x}$ to determine. To make the differential correction process independent from the units chosen for $\bm{x}$, we proceed in the following way. We first compute the norm $|b_j|, j=1,\dots,n$ of the columns of the matrix $B$. We then define the normalized normal matrix $C^{\textrm{N}}$, which is unit-less, and has elements $C^{\textrm{N}}_{i,j}$ defined by
\begin{equation}
    C^{\textrm{N}}_{i,j} = \frac{C_{i,j}}{|b_i| |b_j|}, \quad i=1,\dots,n, \quad j=1,\dots,n.
\end{equation}
The vector $D$ on the right hand side of Eq.~\eqref{eq:difcor} is also normalized to $D^{\textrm{N}}$, whose elements are defined by $D^{\textrm{N}}_i = D_i/|b_i|, \ i=1,\dots,n$. The correction $\Delta \bm{x}^{\textrm{N}}$ for normalized parameters is then obtained by solving the equation $C^{\textrm{N}} \Delta \bm{x}^{\textrm{N}} = D^{\textrm{N}}$. By defining  the inverse of $C^{\textrm{N}}$ as $\Gamma^{\textrm{N}}$, i.e. a unit-less covariance matrix, the correction is given by
\begin{equation}
    \Delta\bm{x}^{\textrm{N}} = \Gamma^{\textrm{N}} D^{\textrm{N}}.
    \label{eq:normalized_dif_cor}
\end{equation}
Then, the actual correction $\Delta \bm{x}$ to apply to the vector of parameters $\bm{x}$ is obtained by de-normalizing the solution through
\begin{equation}
    (\Delta \bm{x})_i = \frac{(\Delta \bm{x}^{\textrm{N}})_i}{|b_i|}, \quad i=1,\dots,n.
\end{equation}

The differential corrections algorithm in Aegis is implemented by using the normalised quantities as described above, and iterations are stopped when 
\begin{equation}
 \lvert \Delta \bm{x} \rvert_C = \sqrt{(\bm{x}_{k+1} - \bm{x}_k) C (\bm{x}_{k+1} - \bm{x}_k)^T/n} < 10^{-3},
 \label{eq:normC}
\end{equation}
and the procedure is declared as convergent.
% Outlier rejection
The differential corrections algorithm is also endowed with a procedure for the automatic detection and rejection of observational outliers developed in \citet{carpino-etal_2003}, which are excluded from the final fit. 

% Mean epoch
Orbits are fitted at a weighted mean observational epoch $t_0$ determined by
\begin{equation}
    t_0 = \frac{\sum_{i=1}^m \tilde{w}_i t_i}{\sum_{i=1}^m \tilde{w}_i}, \quad \tilde{w_i} = \frac{1}{(\sigma_i^\alpha)^2 \cos^2\delta_i + (\sigma_i^\delta)^2},
    \label{eq:t0}
\end{equation}
where $\delta_i$ is the declination of the $i$-th observation, and $\sigma_i^\delta, \sigma_i^\alpha$ are the standard deviation of the errors in the $i$-th observation in declination (Dec) and right ascension (RA). The values of the standard deviations depend on the observatory, and they are given by the astrometric error model. Note that only optical observations are used for the computation of the mean epoch of Eq.~\eqref{eq:t0}. 

% Initial conditions
Initial conditions for the differential corrections algorithm can be retrieved either from orbital catalogues if the object is already known, or computed with the Gauss or Laplace initial orbit determination methods \citep{milani-etal_2008}, both of which are implemented in the Aegis software.

\subsubsection{Rank deficiency and corrections along the weak direction}
The normalized normal matrix $C^{\textrm{N}}$ has an approximate rank deficiency when some eigenvalues are smaller than a threshold $\varepsilon_1$ \citep{milani-gronchi_2009}. Let us denote with $\lambda_1 < \cdots < \lambda_n$ the eigenvalues of $C^{\textrm{N}}$, and suppose that $\lambda_i < \varepsilon_1$ for $i=1,\dots, h$. We also denote with $V$ the orthogonal matrix whose columns correspond with the eigenvectors $\hat{\mathbf{v}}_1, \dots, \hat{\mathbf{v}}_n$ of $C^{\textrm{N}}$. In this case, differential corrections of Eq.~\eqref{eq:normalized_dif_cor} are performed by using the matrix
\begin{equation}
    \tilde{\Gamma}^{\textrm{N}} = V \tilde{\Lambda} V^T, \quad \tilde{\Lambda} = \textrm{diag}(0,\dots,0,1/\lambda_{h+1}, \dots, 1/\lambda_{n}),
\end{equation}
in place of $\Gamma^{\textrm{N}}$. This algorithm is also called $(n-h)$-parameters corrections.

The solution obtained with the $(n-h)$-parameters corrections is not necessarily the nominal one. To get closer to the nominal solutions, corrections along the weak direction $\hat{\mathbf{v}}_1$ can be performed. In this case, at each iteration, the correction $\Delta \bm{x}^{\textrm{N}}$ is decomposed as $\Delta \bm{x}^{\textrm{N}} = \Delta\bm{x}_\parallel + \Delta \bm{x}_\perp$, where $\Delta\bm{x}_\parallel = (\Delta\bm{x}^{\textrm{N}} \cdot \hat{\mathbf{v}}_1) \hat{\mathbf{v}}_1$ and $\Delta \bm{x}_\perp = \Delta \bm{x}^{\textrm{N}} - \Delta \bm{x}_\parallel$. Then, the normalized correction for the next iteration is defined as
\begin{equation}
    \Delta \bm{x}^{\textrm{N}} = 
    \begin{cases}
    \Delta \bm{x}_\perp & \text{if } \lvert \Delta \bm{x}_\perp \rvert_{C^{\textrm{N}}} > \varepsilon_\perp, \\
    \Delta \bm{x}_\perp + \alpha \Delta \bm{x}_\parallel & \text{if } \lvert \Delta \bm{x}_\perp \rvert_{C^{\textrm{N}}} \leq \varepsilon_\perp \text{and } \lvert \Delta \bm{x}_\parallel \rvert_{C^{\textrm{N}}} > \varepsilon_\parallel.\\
    \end{cases}
    \label{eq:weakdircor}
\end{equation}
In Eq.~\eqref{eq:weakdircor}, $\varepsilon_\parallel$ and $\varepsilon_\perp$ are two tolerances which are set to $10^{-3}$ and to $10^{-5}$ respectively in the Aegis software. In addition, 
\begin{equation}
    \alpha = 
    \begin{cases}
        s/\lvert \Delta \bm{x}_\parallel \rvert_{C^{\textrm{N}}} & \text{if } \lvert \Delta \bm{x}_\parallel \rvert_{C^{\textrm{N}}} < s, \\
        1 & \text{otherwise,}
    \end{cases}
\end{equation}
and $s$ is a parameter that is set to 0.5 as a default. The case left out in Eq.~\eqref{eq:weakdircor} is when both norms are below the two selected tolerances, and the method is therefore declared as convergent. 

\subsubsection{Global algorithm of differential corrections}
\label{ss:glodifcor}
The global algorithm of differential corrections automatically handles all the methods presented above, and is composed of three cycles. By denoting with $k_{\max}$ the maximum number of iterations allowed, and given two tolerances $\varepsilon_1 \leq \varepsilon_2,$ the scheme of the cycles is the following:

\paragraph{Cycle 1.} At each iteration, the normalized normal matrix $C^{\textrm{N}}$ and its eigenvalues $\lambda_1 < \cdots < \lambda_n$ are computed. Then, if $\lambda_1 > \varepsilon_2$, the full corrections of Eq.~\eqref{eq:normalized_dif_cor} are applied. If $\varepsilon_1 < \lambda_1 < \dots < \lambda_h \leq \varepsilon_2$ for some $h \in {1,\dots,n}$, then the corrections of Eq.~\eqref{eq:normalized_dif_cor} are applied with $\Gamma^{\textrm{N}}/2$ in place of $\Gamma^{\textrm{N}}$. Finally, if $\lambda_1 < \dots < \lambda_h \leq \varepsilon_1$, then $(n-h)$-parameters corrections are applied, down to a minimum of $n-h = 4$. Iterations are ended if condition of Eq.~\eqref{eq:normC} is fulfilled at some step, and hence convergence is reached, or when $k \geq k_{\max}$. 
The cycle fully converged if convergence was reached by the full differential corrections of Eq.~\eqref{eq:normalized_dif_cor}, and the global algorithm is therefore terminated. In case this did not happen, cycle 2 is triggered.

\paragraph{Cycle 2.} Differential corrections with increasing number of parameters are attempted, until full differential corrections are performed. Iterations are stopped when Eq.~\eqref{eq:normC} is fulfilled at some step, or when $k \geq k_{\max}$. If the cycle fully converged, the global algorithm stops. If not, there are two possibilities: 1) cycle 1 did not converge, and the global algorithm stops with a failure; 2) the $(n-h)$-parameters corrections converged in cycle 1, hence cycle 3 is triggered in this case.

\paragraph{Cycle 3.} The last cycle consists in attempting the corrections along the weak direction, defined by Eq.~\eqref{eq:weakdircor}.

\vspace{10pt}

The global algorithm of differential corrections implemented in Aegis is a novel approach to automatically handle cases in which rank deficiencies occur, and it represents one of the main advances with respect to the OrbFit software. Note that the choice of the two unit-less tolerances $\varepsilon_1$ and $\varepsilon_2$ can be made because the normalized normal matrix $C^{\textrm{N}}$ does not depend from the units chosen to represent the parameters $\bm{x}$ to be determined.

\subsection{Dynamical model}
% Dynamical model
The force model $\bm{f}$ of Eq.~\eqref{eq:dynamical_model} used for the orbit determination of NEAs includes the gravitational attraction of the Sun, the eight planets from Mercury to Neptune, the Moon, the 16 most massive main-belt asteroids, and Pluto (see Table~\ref{tab:masses_small_bodies} for the values of the masses). The software makes use of the JPL Planetary and Lunar Ephemeris DE441 \citep{park-etal_2021} to retrieve the positions of massive objects. 
%Optionally, the number of massive main-belt asteroids can be increased up to 343, corresponding to the list of asteroids ephemerides provided by the DE441 model. 
%
In addition, the parameterised post-Newtonian relativistic contributions \citep{will_1993}, and the oblateness of the Sun and the Earth are added to the force model. Higher order harmonics of the Earth gravitational field can be optionally added in the case of a deep close encounter with the Earth. 

In some special cases, non-gravitational effects need to be also taken into account to accurately compute the orbit of an NEA, and they can be optionally added to the force field. Non-gravitational forces are accounted for into an acceleration term $\bm{a}$ of the form
\begin{equation}
    \bm{a} = (A_1 \hat{\bm{r}} + A_2 \hat{\bm{t}} )g(r), \quad g(r) = (1\textrm{ au})/r^2.
    \label{eq:nongrav}
\end{equation}
In Eq.~\eqref{eq:nongrav}, the unit vectors $\hat{\bm{r}}$ and $\hat{\bm{t}}$ are the radial and the tangential directions, respectively. 
The term $A_1$ accounts for the solar radiation pressure (SRP), and it is modeled as
\begin{equation}
    A_1 = (1-\Theta)\times \frac{\phi}{c} \times \frac{A}{m},
\end{equation}
where $\phi = 1.361$ kW m$^{-2}$ is the solar radiation energy flux at 1 au, $c$ is the speed of light, $A/m$ is the area-to-mass parameter, and $\Theta$ is the occultation function, which determines whether the object is occulted by a major planet or not \citep{montenbruck-gil_2013}. 
The term $A_2$ accounts for the Yarkovsky effect \citep[see e.g.][]{vokrouhlicky-etal_2015}. %, while $A_3$ for an eventual out-of-plane non-gravitational force \citep{farnocchia-etal_2023, seligman-etal_2023}.
In the orbit determination process, the area-to-mass parameter $A/m$ and $A_2$ are kept among the solved-for parameters, and estimated through the least-square procedure.  
\begin{table}[!ht]
    \renewcommand{\arraystretch}{1.3}
    \centering
    \begin{tabular}{ll}
    \hline
    \hline
    Asteroid & $GM/GM_\odot$ \\
    \hline
 (1) Ceres          & 4.7191422767$\times 10^{-10}$  \\
 (2) Pallas         & 1.0297360324$\times 10^{-10}$  \\
 (3) Juno           & 1.4471670475$\times 10^{-11}$  \\
 (4) Vesta          & 1.3026836726$\times 10^{-10}$  \\
 (7) Hebe           & 8.5890389994$\times 10^{-12}$  \\
 (10) Hygea         & 4.2385986149$\times 10^{-11}$  \\
 (15) Eunomia       & 1.5243642468$\times 10^{-11}$  \\
 (16) Psyche        & 1.1978215785$\times 10^{-11}$  \\
 (31) Euphrosyne    & 8.1331596146$\times 10^{-12}$  \\
 (52) Europa        & 2.0216913527$\times 10^{-11}$  \\
 (65) Cybele        & 7.0687100328$\times 10^{-12}$  \\
 (87) Sylvia        & 1.6337820878$\times 10^{-11}$  \\
 (88) Thisbe        & 8.9653065564$\times 10^{-12}$  \\
 (107) Camilla      & 1.0878696848$\times 10^{-11}$  \\
 (511) Davida       & 2.9345275748$\times 10^{-11}$  \\
 (704) Interamnia   & 2.1327387534$\times 10^{-11}$  \\
 (134340) Pluto     & 7.3504789732$\times 10^{-9}$  \\
    \hline
    \end{tabular}
    \caption{Gravitational parameters $GM$ of the 16 most massive main-belt asteroids included in the dynamical model, and Pluto. Values are reported in the units of the gravitational parameter of the Sun, corresponding to $GM_\odot = 1.3271244004127942 \times 10^{11}$ km$^3$ s$^{-2}$. The gravitational parameters are extracted from the header of the JPL ephemeris DE441.}
    \label{tab:masses_small_bodies}
\end{table}

\subsection{Orbit propagation}
% Orbit propagation
For the numerical propagation of asteroid orbits, the Aegis software uses a 10th order multistep method aimed at solving Eq.~\eqref{eq:dynamical_model}, which is based on a Cowell Backward Difference approach and a predictor-corrector scheme \citep[see e.g.][for details]{hairer-wanner_1993}. Equation~\eqref{eq:dynamical_model} is solved either in Cartesian heliocentric coordinates or by using the Kustaanheimo-Stiefel (KS) regularisation \citep{kustaanheimo-stiefel_1965}. The usage of the KS regularisation makes the propagation less sensitive to the singularities of the two-body problem, and it is suitable for very eccentric orbits or deep close encounters with the Earth.

\begin{figure}[!ht]
    \centering
    \includegraphics[width=0.5\textwidth]{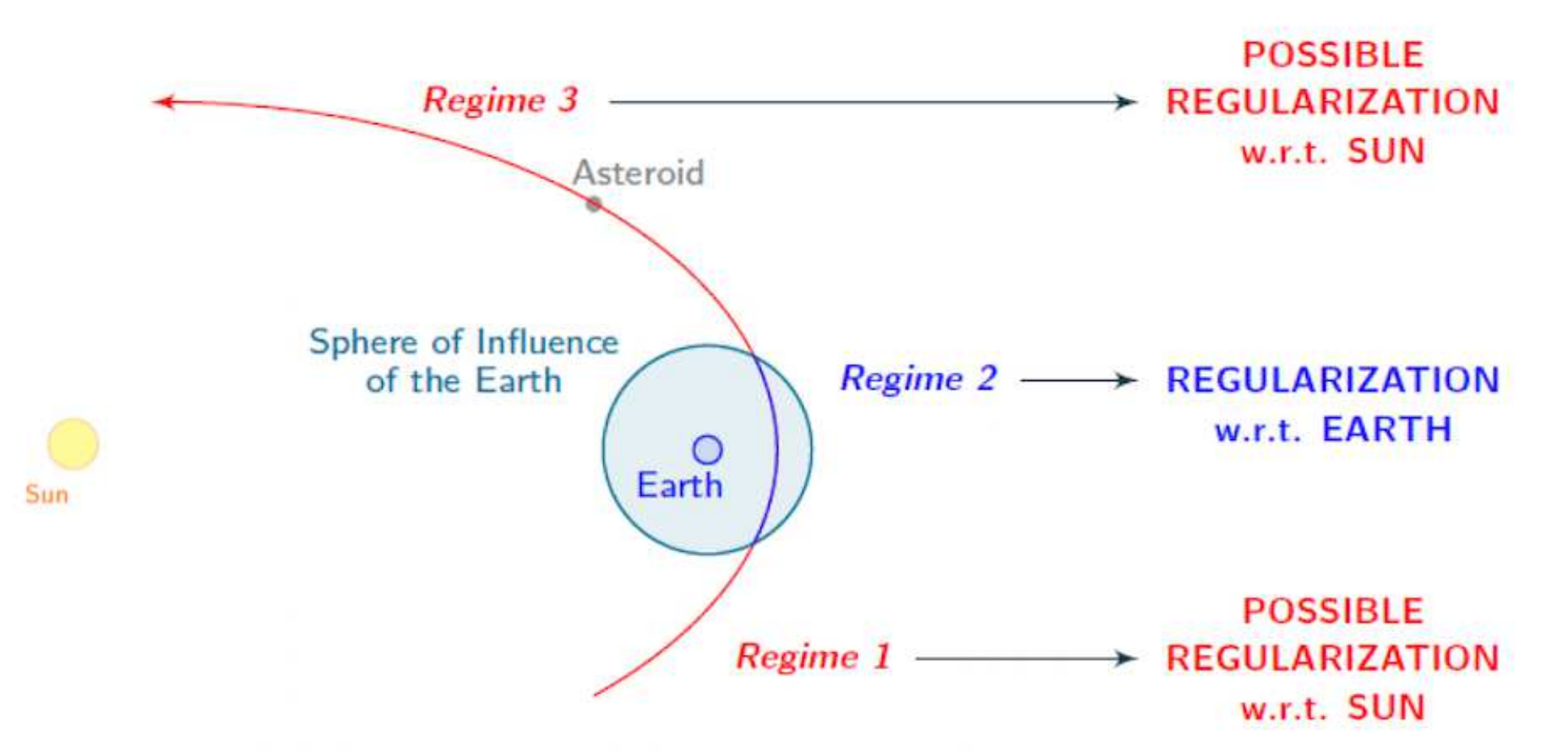}
    \caption{Example of an asteroid having a close approach with the Earth, where the global propagation is split into three different local propagations with different dynamical regimes.}
    \label{fig:propagator}
\end{figure}

The general strategy adopted in Aegis for the numerical propagation of an orbit is to split the entire global propagation into a sequence of local propagations \citep{guerra-etal_2016}. The motion of a Solar System object is obtained by numerically integrating the perturbed Kepler problem with respect to a certain primary body of attraction. Since initial conditions are assumed to be heliocentric, the first primary is always the Sun. When the propagated asteroid is sufficiently close to a third body, that body becomes the new primary. Different cutoff distances for primary switching are used for each planet: 0.05 au for Mercury and Venus, 0.1 au for the Earth, 0.07 au for Mars, and 0.7 au for Jupiter, Saturn, Uranus, and Neptune. Note that the switch of the primary is done for all types of mass-less bodies, e.g. for either NEAs and main-belt asteroids. When switching the primary, the ensemble of the dynamics (i.e. the primary body of attraction, and the perturbations involved), the formulation (Cartesian or KS), and the step-size of the multistep integrator are all subjected to changes. A local propagation is therefore ended when any of the mentioned components change. A scheme of this concept is presented in Fig.~\ref{fig:propagator}.
The KS formulation is used during a close approach with a planet, or in heliocentric regimes when the eccentricity is larger than 0.2, while the Cartesian formulation is used only for heliocentric orbits for eccentricity smaller than 0.2. The step-size is fixed for every local propagation, and the selection is performed following an approach similar to that used in \citet{milani-nobili_1988} when the orbit regime is elliptic, while using a novel approach in a hyperbolic regime during a close encounter which extends that presented in \citet{milani-nobili_1988}. 

Note that the integration centre is either the Sun or a planet of the Solar System, which are accelerating. Therefore, the equations of motion of the mass-less body contain also the acceleration of the body at the integration centre. These are computed by modifying the Fortran interface routines provided for the JPL Ephemerides\footnote{\url{https://ssd.jpl.nasa.gov/ftp/eph/planets/fortran/}}.

The Aegis propagator is another notable difference from the OrbFit software, for which the integration is performed by using the single-step Everhart method with adaptive step-size \citep{everhart_1985}, and without relying on the KS regularisation during close approaches. 
The Aegis propagator permits to save the whole stack of propagation, so that to interpolate on the whole integration time-span after the propagation ended, maintaining the same precision of the integration. This makes the whole integration method more versatile than the single-step Everhart method, and permits it to perform some actions more easily, such as the search of close approaches. 
%This makes the Aegis propagator more stable and reliable in chaotic regimes caused by planetary close encounters.   

\subsection{Ephemerides computation}
\label{ss:ephem}
%Ephemerides computation
Once the nominal orbit $\bm{x}$ and the corresponding covariance matrix $\Gamma_{\bm{x}}$ at time $t_0$ are computed, predictions on the position of the object at a certain time $t^*$ can be made. The nominal prediction $\bm{p}$ can be computed by simply propagating the nominal orbit $\bm{x}$ to the time $t^*$, and then combining the output with the prediction function of Eq.~\eqref{eq:prediction_function}. The covariance matrix $\Gamma_{\bm{x}}$ is mapped to the epoch of the prediction by using a linear approximation, and therefore the corresponding covariance matrix $\Gamma_{\bm{p}}$ is given by
\begin{equation}
    \Gamma_{\bm{p}} = \bigg(\frac{\partial \bm{p}}{\partial \bm{x}}\bigg) \Gamma_{\bm{x}} \bigg(\frac{\partial \bm{p}}{\partial \bm{x}}\bigg)^T.
    \label{eq:covEph}
\end{equation}
Details about this approach can be found in \citet{milani-gronchi_2009}. In a strong non-linear regime, Eq.~\eqref{eq:covEph} may not be accurate for the computation of the uncertainty region of the object. A semi-linear algorithm for the propagation of the uncertainty that works under non-linear dynamical regimes has been developed in \citet{milani_1999}, and it is also implemented in the Aegis software.

\section{Aegis: Impact Monitoring}
\label{s:impact_monitoring}
%The aim of impact monitoring is to understand whether the confidence region of the orbit of an NEA contains some virtual impactors or not. To this purpose, the confidence region is sampled with a finite number of Virtual Asteroids (VAs), all compatible with the observations, and then propagated in the future to search for possible impacts.

\subsection{Target plane}
The asteroid impact hazard analysis is usually performed in the framework of the Target Plane (TP) \citep{valsecchi-etal_2003}. To define the TP at a specific close approach with the Earth at time $\bar{t}$ it is assumed that the geocentric orbit of the asteroid is hyperbolic, with incoming asymptote denoted by $\bm{u}$. 
%, let us consider an asteroid with a close approach with the Earth at time $\bar{t}$, and let $\bm{u}$ be the incoming asymptote of the geocentric hyperbolic orbit of the asteroid. 
%
The TP is defined as the plane containing the centre of the Earth and that is orthogonal to $\bm{u}$. The trace of the close approach $\bm{b}$ is the intersection of the asymptote with the TP, and the value $|\bm{b}|$ corresponds to the unperturbed encounter distance. The norm $|\bm{b}|$ is also called impact parameter. Due to gravitational focusing, the Earth radius is increased by a factor corresponding to $\sqrt{1+v_e^2/|\bm{u}|^2}$, where $v_e$ is the Earth escape velocity, and it is used to compute the actual impact cross section at the given close encounter. 

A reference system $O\xi\eta\zeta$ can be defined by aligning the negative $\zeta$ axis with the heliocentric velocity of the Earth, the positive $\eta$ axis parallel to the asymptotic velocity $\bm{u}$, and $\xi$ such that the reference system is positively oriented. The coordinates $(\xi, \zeta)$ are related to the cross-track and along-track miss distances, respectively. The $\zeta$ coordinates defines therefore how early or late the asteroid is for the minimum orbit intersection distance \citep[MOID,][]{gronchi_2005}, which is approximated by the value of the coordinate $\xi$.

\subsection{Line of Variation method}
\label{ss:LOV}
The complete algorithm that Aegis uses in the operational impact monitoring system is described in detail in \citet{milani-etal_2005}, here we limit ourselves to give a short general description of the main steps to perform. The method relies on the sampling of the Line of Variatons (LOV), which is a 1-dimensional differentiable curve in the orbital elements space identifying the direction with the largest uncertainty. The output of the sampling is a set of ordered Virtual Asteroids (VAs) compatible with the observations. The LOV depends on the choice of the coordinates, and Aegis automatically sets them depending on the curvature of the observational arc of the object. To this purpose, the so-called arc type is determined \citep{milani-etal_2007}, and the selected coordinates are: Cartesian if the curvature is small (arc type less than or equal to 2); Equinoctial if the curvature is moderate (arc type between 2 and 100); Keplerian if the curvature is large (arc type larger than 100). 
The LOV is parameterised with a single variable $\sigma$. 
%
%After the coordinates are chosen, Aegis samples the LOV in the interval $\sigma \in [-5,5]$ with points $\sigma_1, \dots, \sigma_N$ such that the probability of each interval $P([\sigma_i, \sigma_{i+1}])$ is a constant $\text{IP}^*$ \citep{delvigna-etal_2019}. The default value used is $\text{IP}^* = 10^{-7}$, and the maximum step used in $\sigma$ for the sampling is set to 0.01. 
%
After the coordinates are chosen, Aegis samples the LOV in the interval $\sigma \in [-5,5]$ with points $\sigma_1, \dots, \sigma_N$ such that the probability of each interval $(\sigma_i,\sigma_{i+1})$ is a constant $IP^*$ and up to a maximum step-size $\Delta\sigma_{max}$, as described in \citet{delvigna-etal_2019}. The default values used by Aegis are $IP^*=10^{-7}$ and $\Delta\sigma_{\max}=0.01$. The parameter $IP^*$ represents the completeness limit of the LOV method used by our software, which is the impact probability threshold down to which the search of the impact monitoring system is complete. The assumptions for which the search is fully complete could fail in particularly non-linear cases, thus the actual completeness limit achieved is slightly above the theoretical value $IP^*$. An estimate of the loss of completeness close to the limit is given in \citep{delvigna-etal_2019}.

The orbits $\bm{x}_1, \dots, \bm{x}_N$ obtained from the sampling of the LOV, and ordered with respect to the value of $\sigma$, form a set of multiple solutions. Each of them is propagated in the future for 100 years from the current epoch, and all the close approaches with the Earth within 0.2 au are recorded. Each close approach is represented by a trace on the corresponding TP. Due to the fact that the LOV is a smooth curve, also the trace of the LOV on the TP is a smooth curve as well. 
The list of close encounters is split into \textit{showers} and \textit{returns} with the algorithm described in \citet{delvigna-etal_2019}. In this procedure, showers are obtained by clustering the VAs according to the close approach date, while the returns are VAs with consecutive LOV indices within a shower. 

Each return corresponds to a set of TP traces, which is a sampling of the projection of the LOV segment associated to the given return. Each couple of consecutive VAs of a return with TP traces $\bm{y}_i=(\xi_i, \zeta_i)$ and $\bm{y}_{i+1}=(\xi_{i+1}, \zeta_{i+1})$ is analysed to understand the geometry of the LOV trace between them, as explained in \citet{milani-etal_2005}. More precisely, local minima of the distance $r^2(\sigma) = \xi^2(\sigma) + \zeta^2(\sigma)$ are searched for in the interval $[\sigma_i, \sigma_{i+1}]$. Inside a return, only some $\sigma$ intervals between consecutive VAs contain a minimum of $r^2$, and they are identified by a geometric classification of the TP segments. If the minimum approach distance can be small, the minimum distance and the corresponding LOV point $\bm{x}(\sigma^*)$, with $\sigma^* \in (\sigma_j, \sigma_{j+1})$, is determined by using a suitable iterative method, such as the regula falsi of the Newton method with bounded steps. If the corresponding TP trace $\bm{y}(\sigma^*)$ is inside the Earth cross section computed by taking into account the gravitational focusing, then there is a Virtual Impactor (VI) around $\bm{x}(\sigma^*)$, which is a connected set of orbits leading to an impact.
If the point $\bm{y}(\sigma^*)$ is outside the impact cross-section, but the width $w$ of the projection on the TP of the confidence region computed by linearising at $\bm{y}(\sigma^*)$ is large enough, there may be still an intersection between the impact cross-section of the Earth and the confidence ellipse. In case that happens, there is a VI with initial conditions not belonging to the LOV. 

The analysis of the TP traces relies on the principle of the simplest geometery \citep{milani-gronchi_2009}, which assumes that: 1) the LOV projection on the TP in $[\sigma_j, \sigma_{j+1}]$ does not exit the disk of radius 0.2 au centred at the Earth; 2) the LOV geometry is the simplest compatible with with the tangent vectors of the LOV projection at the nodes. This assumption is good when the return contains a large number of points, while problems can arise when a return is composed only by a few points. In this case, the Aegis software applies a densification procedure introduced by \citet{delvigna-etal_2020}, which aims at adding more points to returns with few points, making the principle of simplest geometry reliable again. The densification procedure is another peculiarity of the Aegis software, making it generally more complete than the Clomon-2 system based on OrbFit, in which this procedure is not implemented. 

The impact probability (IP) is generally computed with a 2-dimensional integral on the TP  \citep{milani-etal_2005}. On the other hand, in the case the projection of the LOV on the target plane has consecutive points $\sigma_1, \dots, \sigma_M, M \geq 10$ inside the Earth impact cross-section, the IP is estimated as
\begin{equation}
    \text{IP} = \frac{1}{\sqrt{2\pi}} \sum_{i=1}^M e^{-\frac{\sigma_{M,i}^2}{2}} (\sigma_{i+1} - \sigma_i), \quad \sigma_{M,i} = \frac{\sigma_i + \sigma_{i+1}}{2},
    \label{eq:IP}
\end{equation}
i.e. it is computed by integrating the probability density function defined on the LOV by restriction of the probability density function defined over the whole orbital element space. 
After the impact probability has been estimated, the PS \citep{chesley-etal_2002} and the Torino Scale \citep{binzel_2000} of each VI found are also computed. %The completeness limit of the LOV method used by Aegis, that is the maximum impact probability $\text{IP}_{\min}$ that can escape detection, has been estimated to be $\text{IP}_{\min} \simeq 4.24 \times 10^{-7}$ \citep{delvigna-etal_2019}.

\subsection{Monte Carlo method} 
For specific cases that need to be analyzed in deeper details, Aegis is also able to perform impact monitoring with Monte Carlo algorithms. Two Monte Carlo methods are implemented in the software: the first one based on a sample in the orbital elements space through the covariance matrix, and the second one based on a sample directly in the space of astrometric observations.

The least-squares principle implies that the orbital elements are Gaussian distributed with mean value equal to the nominal orbit $\bar{\bm{x}}$ and covariance given by $\Gamma_{\bar{\bm{x}}}$. Therefore, the confidence region can be sampled with a large number $N$ of VAs by using the covariance matrix $\Gamma_{\bar{\bm{x}}}$ of the orbit \citep[see e.g.][]{fenucci-novakovic_2021}. VAs are then propagated in the future to search for VIs. The cumulative impact probability is then obtained as the number of VIs found divided by the number $N$ of VAs. 

In the non-linear case, which typically occurs when a small number of observations are available, the confidence region can not be approximated with a Gaussian distribution. In this case, the sampling is made directly in the space of the observations, similarly to what described in \citet{milani-etal_2002}. The Aegis software assumes that errors in the astrometric observations are Gaussian distributed, with 1-$\sigma$ uncertainties given either by the error models mentioned in Sec.~\ref{ss:OD}, or by the observer who submitted the errors of the measurements. Uncertainties in time are treated by separating two components: the systematic errors and the random errors. Systematic errors are typically given by synchronisation issues in the telescope system, and they are assumed to be uniformly distributed and constant within the same tracklet. On the other hand, the random error is caused by stochastic effects, and it is assumed to be Gaussian distributed with mean value equal to zero. The actual error is then obtained by summing up these two components. In addition, we assume that errors in time are not correlated with errors in the astrometry. Once the sampling on the observations and in the time is done, the orbit determination is performed for each combination of the sample without using outlier rejection, and the nominal orbit obtained is then propagated in the future to search for possible impacts. The cumulative impact probability is obtained again as the number of VIs found divided by the size $N$ of the sample.

Monte Carlo methods are typically able to find VIs with an impact probability of the order of $\sim$$1/N$, and therefore they are not suitable to be used in an operational scenario, where tens of NEAs need to be analysed on a daily basis. However, they generally rely on less assumptions than the LOV method, and they are therefore useful to verify impact monitoring results when the dynamics is highly non-linear or when probabilities of impact are relatively high. They are also relevant to evaluate the impact risk of objects placed on geocentric orbits, for which the LOV method cannot be applied. 

\subsection{Impact corridor}
\label{ss:IC}
An algorithm for the prediction of the impact corridor of an NEA with a non-zero impact probability with the Earth in the future is implemented in the Aegis software. The algorithm is based on a semi-linear propagation of the confidence region of the orbit. Mathematical details are described in \citet{dimare-etal_2020}. %Despite being an approximation, the semilinear propagation is 5 orders of magnitude less computationally expensive than Monte Carlo approaches. 
The algorithm takes as input the nominal orbit, the corresponding covariance matrix, and the VI representative. In output, it returns the boundary of the impact surfaces determined with 1-$\sigma$, 3-$\sigma$, and 5-$\sigma$ confidence level, at both 100 km altitude and on the ground. The corridor is then approximated by connecting the boundaries at 100 km and on the ground, thus obtaining a tube of possible positions from the atmospheric entry to the ground. 
The propagation to the ground does not take into account the atmospheric drag, thus it represents a rough approximation of the impact area.
The computational load of this algorithm is directly related to the IP of the selected VI, with higher computational cost for lower probabilities. For this reason, the impact corridor is computed only when $\text{IP} \geq 10^{-3}$. 

\begin{figure}[!ht]
    \centering
    \includegraphics[width=0.48\textwidth]{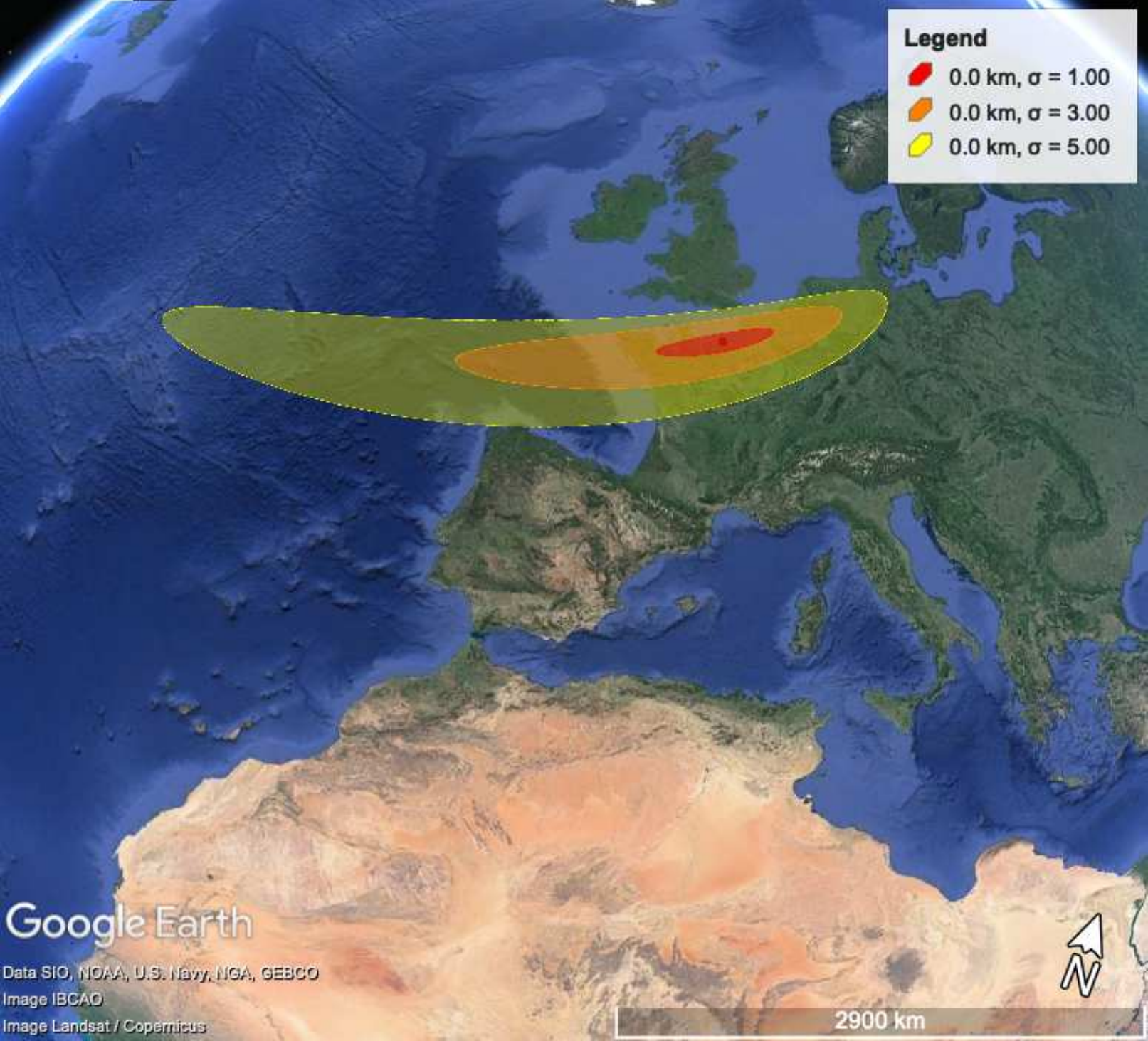}
    \includegraphics[width=0.48\textwidth]{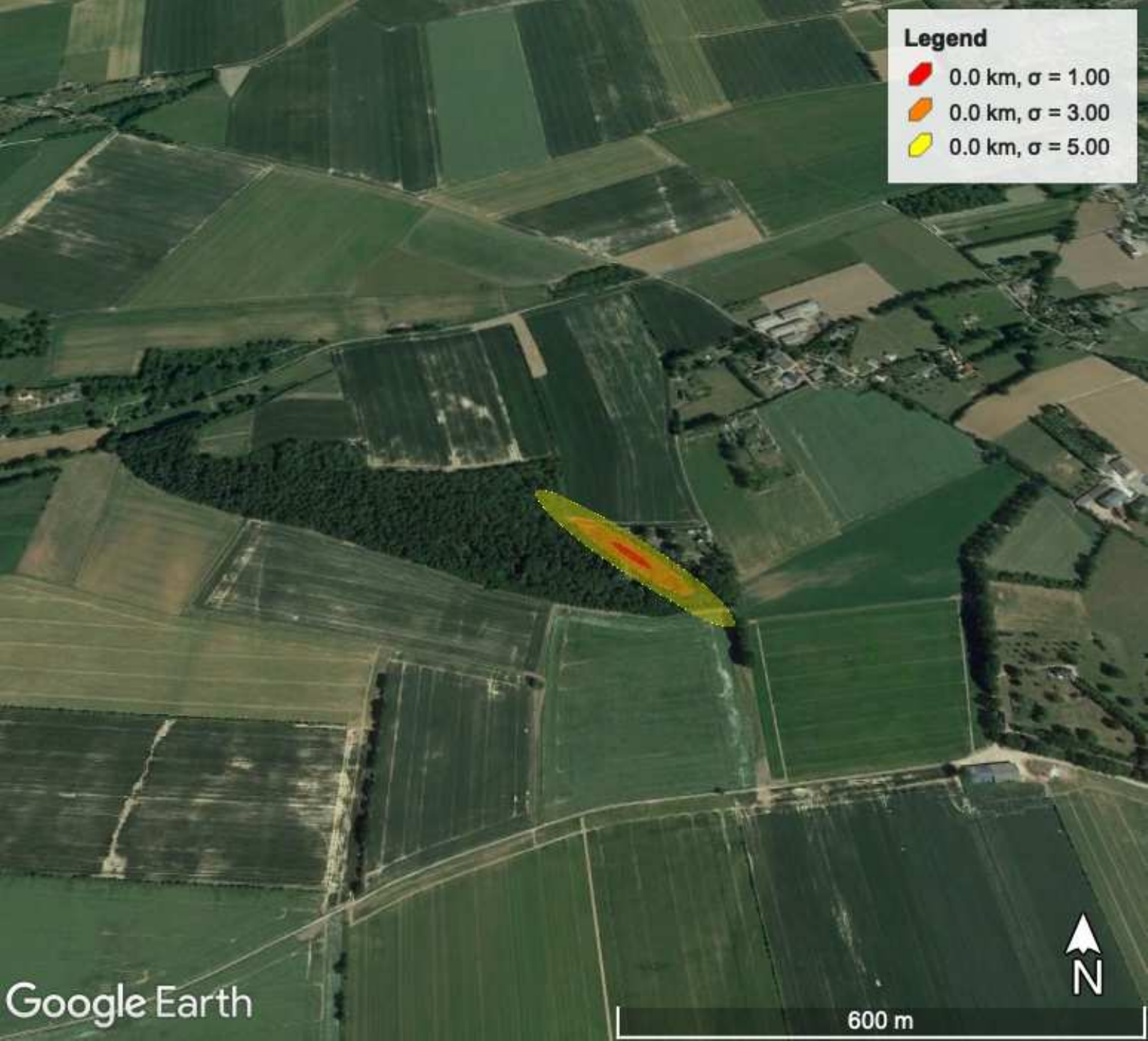}
    \caption{Impact corridor of 2023 CX$_1$ on the ground, a small NEA which impacted Earth 2023-02-13. The left panel shows the results obtained with just the first 7 observations, while the right panel the results obtained with all the observations available. The red, orange, and yellow areas represent the impact area at 1-$\sigma$, 3-$\sigma$, and 5-$\sigma$ level, respectively. The last part of the numerical integration has been performed without taking into account the atmospheric drag.}
    \label{fig:2022WJ1_IC}
\end{figure}

Figure~\ref{fig:2022WJ1_IC} shows the impact corridor computed for asteroid 2023~CX$_1$, a small NEA that impacted the Earth on 13 February 2023, a few hours after its discovery. The top panel shows the impact corridor computed with only the first 7 observations. The impact was already certain, and the impact area at 1-$\sigma$ confidence level covered a large portion of Northern France. The impact corridor was also recomputed after the impact, when more than 300 observations were available, and it is shown in the bottom panel panel of Fig.~\ref{fig:2022WJ1_IC}. With all these observations available, the uncertainty shrunk significantly, and the 5-$\sigma$ impact region was long just about 400 metres.  

\section{Aegis evolution of operational management through DevOps and DataOps}
\label{s:devops}
Aegis has been a key component and actually the driver of the operational evolution of NEOCC, inspiring and leading the adoption of DevOps and DataOps practices within the complex and heterogeneous PDO Segment. 
Being based on OrbFit, which is a monolithic software conceived and developed decades ago, Aegis was subjected to a decay of the maintenance efficiency over the years. Therefore, it required continuous adaptive and evolutionary maintenance not only under the functional point of view, but also under the Information Technology (IT) point of view.
This led to the transition to a new working paradigm at NEOCC, initiated with the implementation of a Space Safety Continuous Integration and Continuous Development (CI/CD) infrastructure, essential to facilitate collaboration among NEOCC operators, industry developers, and ESA software and hardware engineers, enabling them to work seamlessly despite geographical dislocation over different countries. 
This transition has been accompanied by the fine-tuning of our mindset and processes, aimed at efficiently developing, maintaining, and operating our Space Safety software. 
Clear roles and responsibilities have been defined within a DevOps competence quadrant, encompassing functions like: Code Release Management, Automation Expertise, Quality Assurance, Software Development/Testing, and Security Engineering. 
Furthermore, agile software management has enabled us to swiftly respond to operational demands while aligning with modern technical evolution such as hybrid architecture, native cloud integration, and software and hardware containerization. 
This evolution, supported by a well-structured governance approach, remains essential given the complexity of our system components and interfaces.

The CI/CD infrastructure used to maintain and develop the Aegis system is hosted on GitLab, and has been an ESA pioneer activity started in 2017 to evolve the Space Safety software management, using the NEOCC software development, management and evolution demand as a most representative use case. The PDO CI/CD infrastructure was completed at the end of 2019. This allowed NEOCC to enter into the pandemic phase with no Operational breakage, permitting efficient and uninterrupted software management 100\% remotely outside ESA and NEOCC premises, without any impact on the software management process. 

The deployment of the software on a selected environment is efficiently done though an automatic pipeline, however controlled to best coordinate the software management across the sequential stepwise phase from Industry delivery to Operational deployment.
These steps include software build, code quality and security screening, container image creation, and automated regression and integration tests ensuring that all the basic functionalities of the software work correctly. 
Two environments are currently defined for the Aegis system: a pre-operational, and an operational environment. 
The pre-operational environment is used by NEOCC operators to perform additional tests of new software functionalities, while the operational environment hosts the current operational Aegis version which is used to maintain all the services described in Sec.~\ref{s:portal}. Generally, a software release is first deployed on the pre-operational environment, where new functionalities or bug fixes to previous versions are first tested. Once the release passed the testing phase on the pre-operational environment, it is subsequently deployed on the operational environment, and tests are performed once again. 

Additionally, we pioneered a DataOps model to streamline data management, fostering agility and speed in our end-to-end pipeline processes. The implementation of these methodologies, particularly within the PDO Data Hub, has revolutionised our operational approach, laying the groundwork for agile Data Pipelines management and orchestration, which is foreseen to be integrated into the Aegis system. 
Moreover, our focus extends beyond technological advancements to include the enhancement of the human-machine interface, ensuring alignment with the evolving functionality of Aegis and preparing for future advancements in the long way ahead.

\section{Services of the NEOCC web portal}
\label{s:portal}

\subsection{The NEOCC web portal and HTTPS APIs}
\begin{figure*}[!ht]
    \centering
    \includegraphics[width=\textwidth]{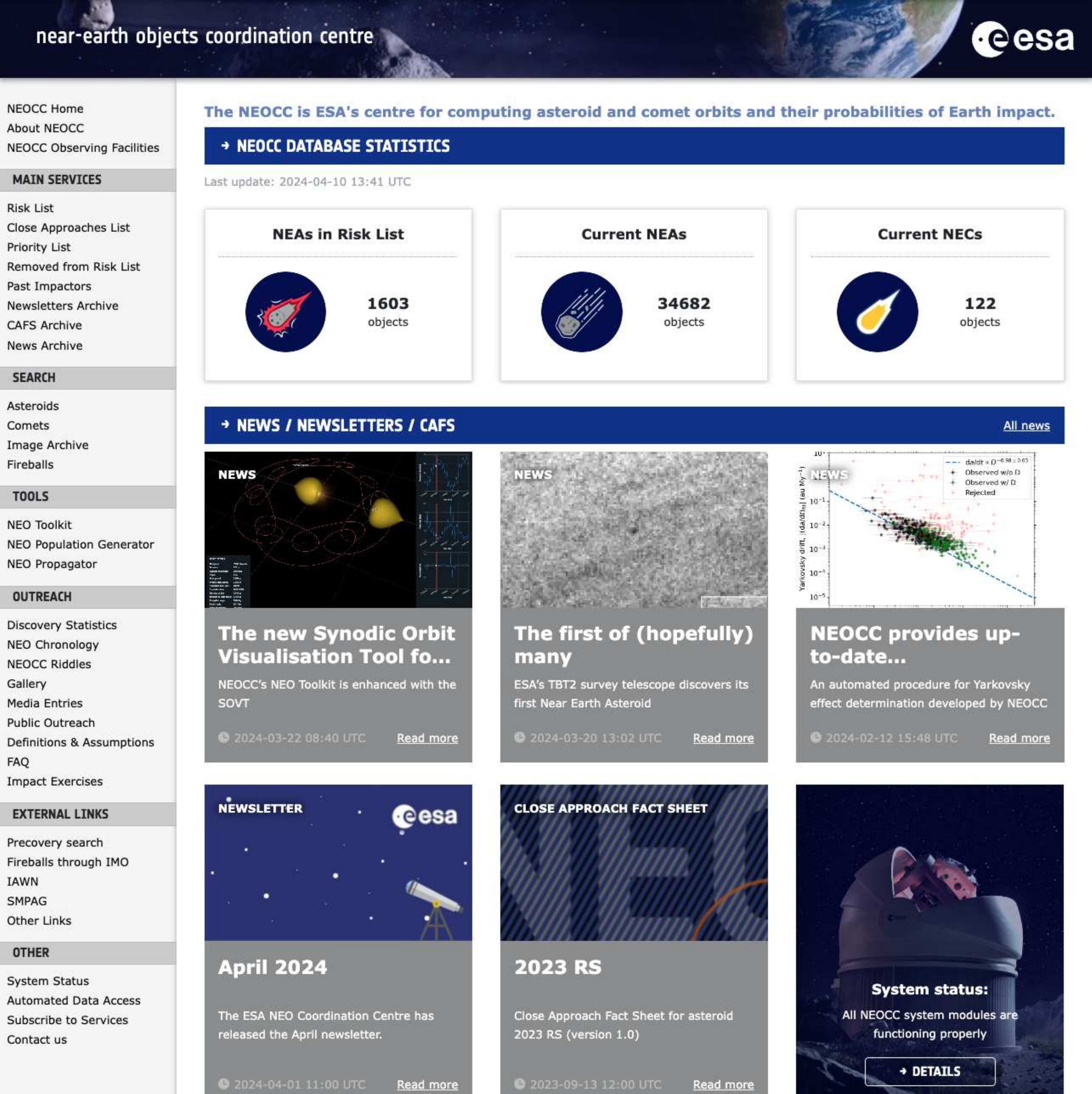}
    \caption{Main page of the NEOCC web portal. Statistics of the orbital database are graphically represented at the top of the web page, while recent news, newsletters, CAFS, and the System Status are shown right below. All the services provided by the NEOCC are accessible through the menu on the left.}
    \label{fig:NEOCC_Portal}
\end{figure*}

 In March 2021, a new version of the web portal adapted to new web standards was released, and all the NEO services that were originally federated by ESA are now independently provided by the NEOCC. The main page of the portal shows some statistics of the orbital database, and the most recent news and newsletters (see Fig.~\ref{fig:NEOCC_Portal}). The portal is currently maintained by Elecnor Deimos\footnote{\url{https://elecnor-deimos.com/}} under ESA contracts.

The NEOCC offers several services, such as an orbital catalogue of all the known asteroids, a list of NEAs with non-zero impact probability with the Earth in the next 100 years, a list of NEAs that passed or will pass close to the Earth, a database of physical properties, and many others. All these services can be accessed on the web portal through an intuitive menu placed on the left side of the web page. 
A help section is provided on many portal pages to explain the given data.
%To help users to better understand the data provided and offer a better experience, a help section on many of the pages of the portal is available.
%
Many of these services are maintained or supported by the Aegis system, and a more detailed description of each of them follows in this section.  

Several services and data available in the web portal are also accessible through dedicated APIs. The automated data access uses the HTTPS protocol and the GET method, meaning that all the requests and all the parameters are transmitted in a URL. All the requests should be prefixed with the NEOCC portal URL \url{https://neo.ssa.esa.int/}. 
These APIs are useful to automatically retrieve data about NEAs, and they can be used by observers for the planning and scheduling of observations, or by researchers for the development or support of NEA-related projects. 
The complete list of the services available through the HTTPS API, the list of the API endpoints, and instructions for users, are available on the Automated Data Access page\footnote{\url{https://neo.ssa.esa.int/computer-access}} of the web portal. Python interfaces to all the APIs are also available to developers through the Astroquery library\footnote{\url{https://astroquery.readthedocs.io}}.

\subsection{Orbit catalogues}

\subsubsection{Catalogues maintenance}
The NEOCC maintains an orbital catalogue of all the asteroids, both NEAs and non-NEAs. Orbits of NEAs are determined with the complete force model described in Sec.~\ref{s:orb_det}. Differently from NEAs, orbits of non-NEAs are computed a simpler force model, which only accounts for the gravitational attraction of the Sun, the eight planets, the Moon, (1) Ceres, (2) Pallas, and (4) Vesta.

% Maintenance of orbital catalogues
The maintenance of asteroid catalogues is completely automated. The Aegis system fetches optical observations from the MPC database, and radar astrometry from the JPL SSD website. 
New observations of known NEAs are downloaded at every daily orbit update issued by the MPC. The orbits of asteroids with new observations are then automatically computed and stored in the internal database, and all the data needed for the NEOCC portal services are updated. 
% Hourly
In addition to the update of known objects, Aegis automatically adds newly discovered NEAs to the database. To this end, the MPC website is queried with every 30 minutes in order to search for new objects announced in recent Minor Planet Electronic Circulars (MPECs). New optical observations are automatically downloaded, and their orbits are computed and stored in the internal database. 
Impact monitoring routines are also automatically triggered when an object is updated in an MPEC or in the daily orbit update. 
A one-dimensional score is computed for every asteroid in order to prioritise impact monitoring runs, and only objects with positive score are processed. The score is similar to that introduced in \citet{milani-etal_2005}, and it is computed as
\begin{equation}
    S = 50 - 25 \cdot \big( 80 \cdot M - \log_{10}(R) \big),
    \label{eq:score}
\end{equation} 
where $R$ is the along-track uncertainty over an integration in the future 100 years, and $M = \min(0, \textrm{MOID} - 3 \sigma_{\textrm{MOID}} )$ where $\sigma_{\textrm{MOID}}$ is the uncertainty of the MOID \citep{gronchi-tommei_2007}. In Eq.~\eqref{eq:score}, both $R$ and $M$ are expressed in au.

% Monthly
Lastly, a general update of the orbital catalogue of all the asteroids, both NEAs and non-NEAs, is performed at every monthly orbital update released by the MPC. During this update, tens of thousands of orbits of asteroids are computed and catalogued. An update on the designations and numbering of the objects is also performed. This process lasts about 24 hours on our machines. 
% Yarkovsky effect update
After processing the monthly update, the list of NEAs for which a positive determination of the Yarkovsky effect can be obtained is also updated by using the procedure described in \citet{fenucci-etal_2023}. 
All the automatic processes are supervised by operators, that are in charge of performing occasional manual maintenance interventions whenever needed.

A catalogue of comets and NECs is also maintained, but their orbits are not independently computed by the NEOCC. The orbital data of these objects are made available to ESA by the JPL through its SSD and then shown in the NEOCC web portal.

\subsubsection{Data in orbital catalogues}
Information about the orbit of an asteroid includes the nominal orbital elements $\bar{\bm{x}}$, the corresponding covariance matrix $\Gamma_{\bar{\bm{x}}}$, and the normal matrix $C$ obtained from the differential correction algorithm (see Sec.~\ref{s:orb_det}). Orbits are available in Equinoctial or Keplerian elements, and are given at the middle observational epoch computed with Eq.~\eqref{eq:t0}, and near the current epoch. Note that the current epoch is updated every 200 days. The observations used in the orbital fits, their weights, and the residuals obtained for the computed orbits, are also made available to users. 
The non-gravitational parameter $A_2$ due to the Yarkovsky effect is also regularly determined by the NEOCC \citep{fenucci-etal_2023}, while the parameter $A_1$ due to SRP is determined for a small number of selected NEAs. Data about these parameters are provided as well, when present. 

Orbital data can be accessed by users either from a web-page dedicated to each asteroid, or through the HTTPS API. 
In addition to orbital data, the page of an individual asteroid provides information about physical properties such as albedo, rotation period, colors, and taxonomy (when available, together with the original reference of the data), observations, close approaches with massive bodies, and possible future impacts with the Earth with the corresponding probability. The total length of the observational arc, the RMS of the weighted residuals, the number of observations used in the fit, and the list of observations of the object with the corresponding residuals, are all available to the user from this page. 

The web portal offers a search functionality that permits queries of the database, where it is possible to provide user-specified constraints for different orbital parameters, physical properties, observational data, and observational conditions. To the best of our knowledge, this wide range of search constraints is not offered by any other publicly available similar services. For instance, it is possible to search for all Atira asteroids (an orbital property) that are currently brighter than magnitude 21 (an observational characteristic), and that are larger than 100 m (a physical property). In addition to orbital and observational properties, users can combine some of the available physical properties and quantities related to the risk assessment of that specific object. The latter functionality makes it easy to select currently observable objects from our risk list. These advanced search functionalities can be extremely useful for observers who need to select targets for a current telescope run, since they allow an easy filtering of detectable objects that also match other constraints relevant for the specific science case at hand.

The HTTPS API permits to retrieve all the data displayed in the portal about a single asteroid. In addition, a catalogue named \texttt{allneo.lst} containing the list of currently known NEAs is available. 
Catalogues of the nominal Keplerian orbital elements of all the known NEAs near the current epoch, and near the middle epoch of the observational arc, are also available through the files named \texttt{neo\_kc.cat} and \texttt{neo\_km.cat}, respectively. Catalogues containing the Equinoctial elements of all NEAs, the uncertainties of the elements and the covariance matrices are contained in the flat files named \texttt{neo.ctc}, for the orbits near the current epoch, and in \texttt{neo.ctm}, for the orbits near the middle of the observational arc.

\subsection{Risk List}
\label{ss:risklist}
Possible impacts of NEAs with the Earth in the next 100 years are searched for by using the LOV method described in Sec.~\ref{ss:LOV}. 
The risk assessment results obtained by the Aegis software are collected together in the NEOCC Risk List\footnote{\url{https://neo.ssa.esa.int/risk-list}}, that is a catalogue of all NEAs for which a non-zero impact probability has been computed. Since orbits of NEAs continuously change due to the availability of new observations, the Risk List is a dynamical set and it is updated every time a change in the observations of an NEA occurs. 

Each entry of the Risk List contains details on the particular Earth approach which poses the highest risk of impact, as expressed by the PS. Data shown in each row include the date of the possible impact, the range for the estimated size of the impacting asteroid, the impact velocity, and the impact probability. Impact history data of a specific NEA can be accessed in tabular and graphical form. In most cases, the size presented in the table is estimated indirectly from the absolute magnitude, and flagged with an asterisk. When a better measurement is available in the literature, it replaces the estimated value. The NEOCC offers also a visualiser of the LOV on the corresponding TP at each close approach with the Earth, that can be reached from the impact list of the asteroid in question. The Risk List and the table of possible future impacts of a specific NEA are also available through the HTTPS API. 

\begin{figure}[!ht]
    \centering
    \includegraphics[width=0.9\textwidth]{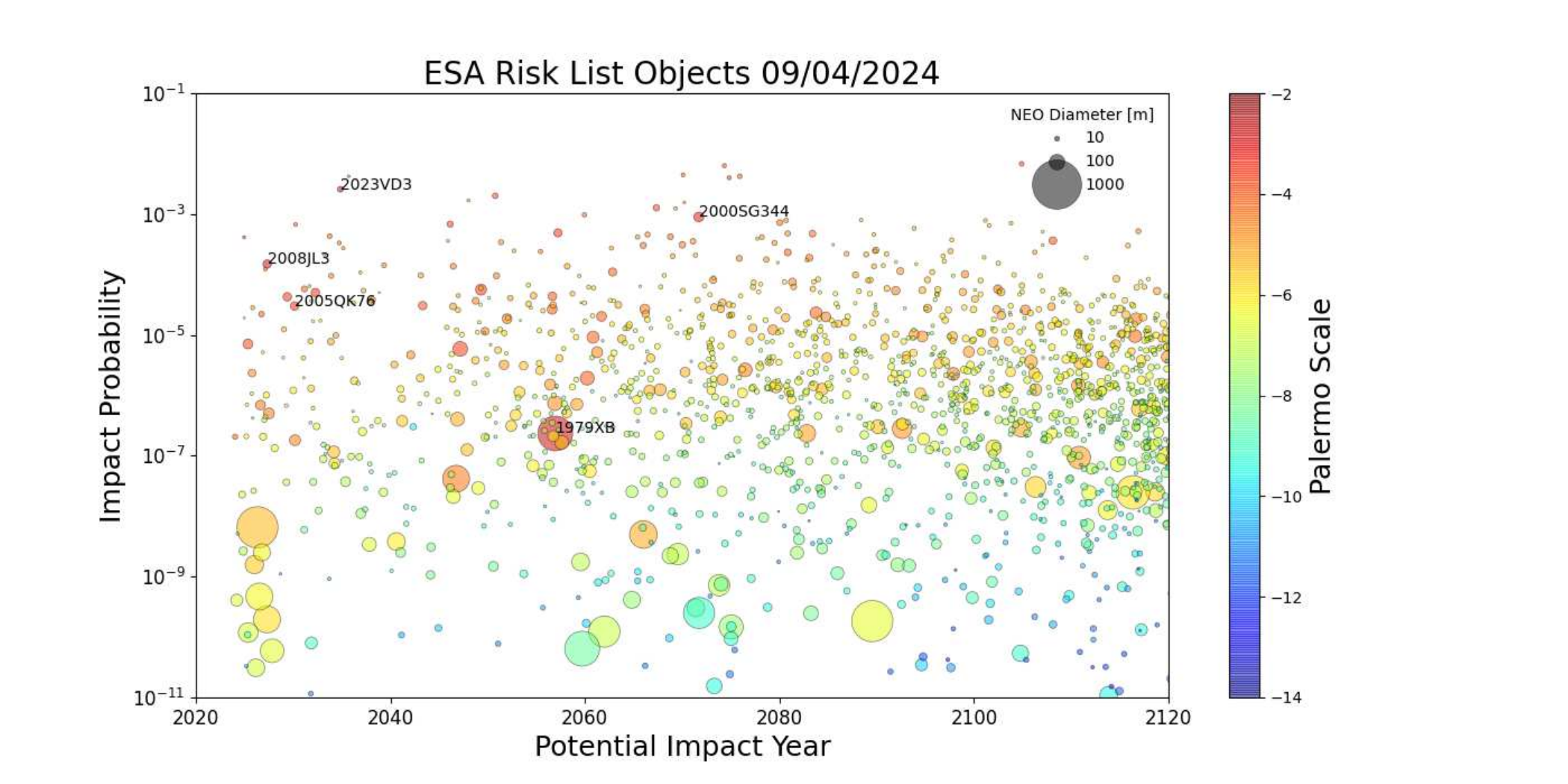}
    \caption{All the objects in the Risk List on 9 April 2024, represented in the plane (Potential Impact Year, Impact Probability). Only the VI with the largest PS is represented. The colour code represents the value of the PS, while the size of the marker is proportional to the diameter of the object.}
    \label{fig:risklist_visual}
\end{figure}

\begin{figure*}[!ht]
    \centering
    \includegraphics[width=\textwidth]{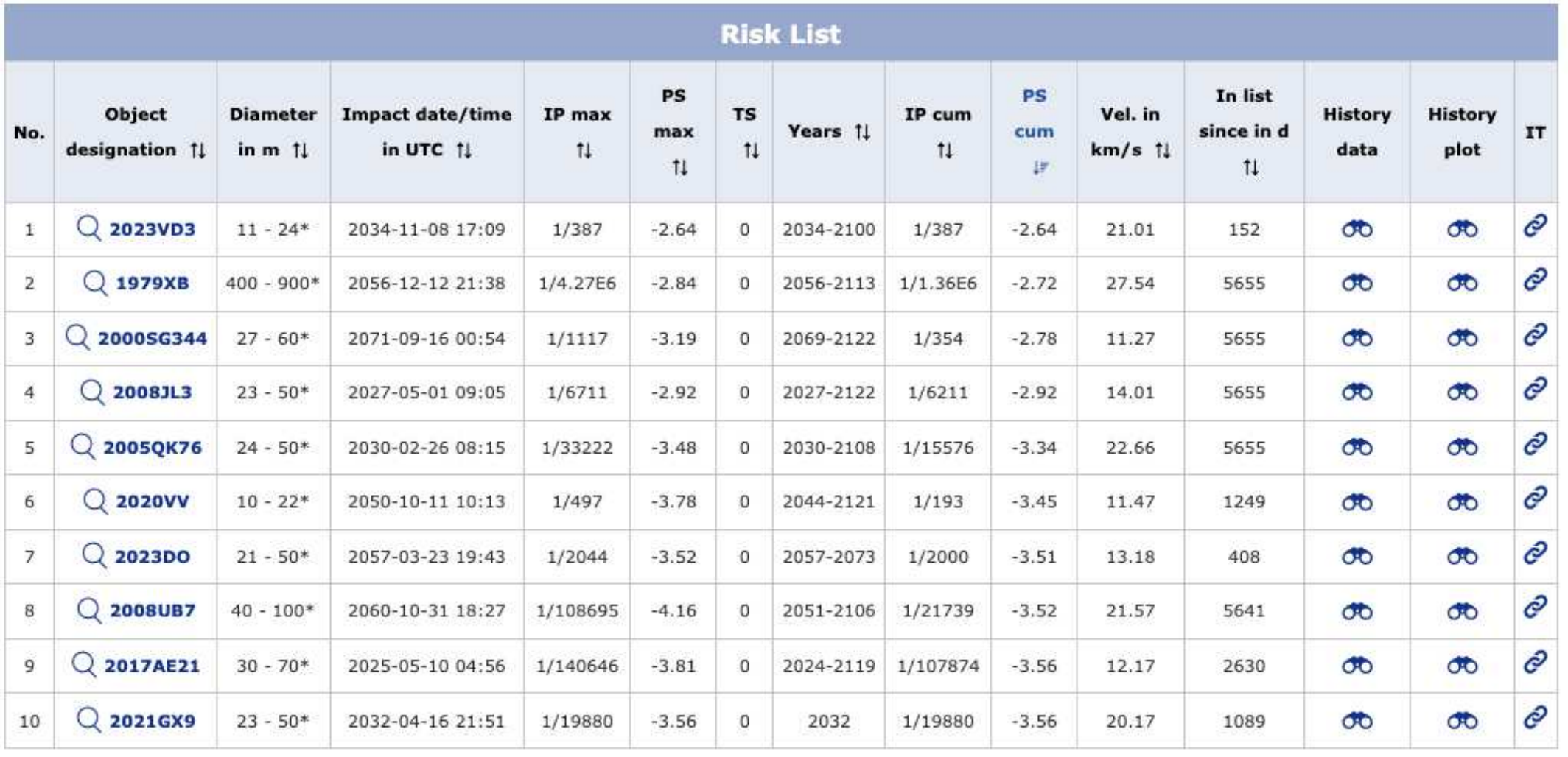}
    \caption{First 10 NEAs in the NEOCC Risk List as of 9 April 2024, ordered by cumulative PS.}
    \label{fig:risklist}
\end{figure*}

 The NEOCC Risk List currently contains 1500 objects, and it is completely independent from similar lists maintained by the other two operational impact monitoring services, the NEODyS Clomon-2 and the JPL Sentry by NASA. 
 Figure~\ref{fig:risklist_visual} shows all the VIs with the largest PS of all the objects of the Risk List on 9 April 2024, in the plane (Potential Impact Year, Impact Probability). Only 5 known NEAs larger than 500 metres are in the Risk List, and a small number of small NEAs have a probability larger than $10^{-3}$ of impacting the Earth in the next century.
 Figure~\ref{fig:risklist} shows the first 10 NEAs in the NEOCC Risk List as of 9 April 2024, ordered by the maximum PS value, as displayed on the NEOCC web portal.

A list containing all the objects that were removed from the Risk List is also maintained and available in the web portal. The removal of an object typically occurs when new observations are collected, and the orbit determination has improved to the level of excluding any future impact threat in the next 100 years. The list includes the date of removal and the highest rated VI of the given NEA, at the time of the removal. Information shown in the page Removed from Risk List are provided since the end of April 2022. 

\subsection{Past Impactors List}
% Past Impactors
The NEOCC maintains a list of NEAs that impacted the Earth that were discovered before the event. To this date, only 10 cases are known: 2008~TC$_3$ \citep{jenniskens-etal_2009}, 2014~AA \citep{farnocchia-etal_2016}, 2018~LA \citep{jenniskens-etal_2021}, 2019~MO, 2022~EB$_5$ \citep{geng-etal_2023}, 2022~WJ$_1$, 2023~CX$_1$ \citep{bischoff-etal_2023}, 
 2024~BX$_{1}$ \citep{spurny-etal_2024}, 2024~RW$_1$, and 2024~UQ. Notably, the impact alert of the last five objects was first triggered by the ESA Meerkat Asteroid Guard \citep{fruhauf-etal_2021, gianotto-etal_2023}, an automated warning service for imminent impactors developed and operated by the ESA NEOCC. This system started the operations at the beginning of 2021, and it continuously scans the MPC NEO Confirmation Page\footnote{\url{https://minorplanetcenter.net/iau/NEO/toconfirm_tabular.html}} in search for unconfirmed NEA objects with a risk of impacting Earth in the following 30 days. 

The list of past impactors includes: 1) an estimate of the diameter of the impacted object; 2) the impact date and time; 3) the impact velocity; 4) the impact energy estimated by Aegis; 5) the impact energy estimated from other sources. Every object also has a dedicated page that can be accessed trough the History column of the table. This page summarises the impact event, giving details of the discovery circumstances and follow-up, the trajectory the space, and the possible meteorite search campaign. The impact corridor computed by Aegis using the method described in Sec.~\ref{ss:IC} is also available on these web pages.

\subsection{Close Approaches List}
Every month dozens of NEAs pass within a distance smaller than 0.05 au from the Earth. A list containing recent and forthcoming close approaches, with details concerning the encounter circumstances, is available on the NEOCC web portal. The maximum brightness reached by the object at the close approach is also shown among the parameters of the table, and it is useful for astronomers to estimate the observability during the encounter. 
The last column of the table provides the Close Approach Frequency Index \citep{cano-etal_2021}, that is an estimate of how often an object of a given size passes within a given distance and velocity from the Earth. An intuitive colour-coded evaluation is provided, and it indicates whether a given close approach can be considered as very frequent (blue), frequent (green), infrequent (yellow), rare (orange), or very rare (red).
For example, the foreseen close approach of Apophis on 13 April 2029 will appear in red since it is predicted to be very rare event.
The lists of past and upcoming close approaches, and the close encounters of a specific NEA, are also available through the HTTPS APIs.

\subsection{NEO Toolkit}
The Aegis software is used to produce all the inputs of the NEO Toolkit\footnote{\url{https://neo.ssa.esa.int/neo-toolkit}} \citep{ramirezmoreta-etal_2023}, a new suite of astronomical tools designed by the NEOCC and developed by Eversis\footnote{\url{https://eversis.com/}} under ESA contracts. The toolkit is composed of five complementary tools, each of them focused on a different goal.

\begin{figure*}[!ht]
    \centering
    \includegraphics[width=\textwidth]{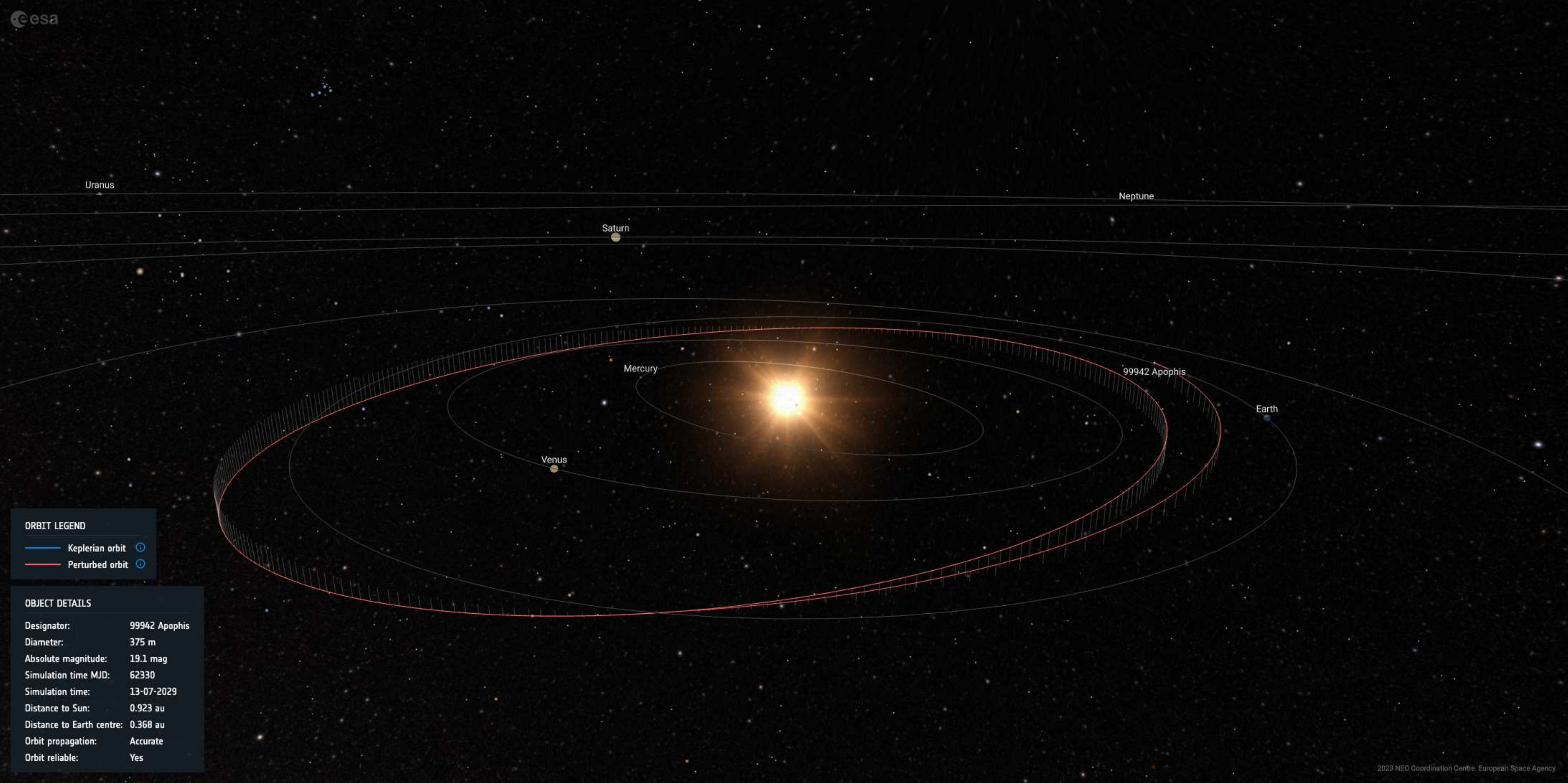}
    \caption{Perturbed orbit of asteroid (99942) Apophis around the epoch 13 July 2029, a few months after the Earth close approach, shown in the OVT.}
    \label{fig:ApophisOrbit}
\end{figure*}

The Orbit Visualisation Tool (OVT) displays the orbit of an asteroid (or a set of asteroids) in a 3D environment of the Solar System. Users can visualise the position and motion of asteroids in relation to the planets and other objects in the Solar System. The tool also enables the visualisation of NEA groups, families, and spectral classes. Figure~\ref{fig:ApophisOrbit} shows the orbit of (99942) Apophis as displayed in the NEO Toolkit OVT.

The Synodic Orbit Visualisation Tool (SOVT) is complementary to the OVT, and displays orbits of asteroids in the synodic reference frame that co-rotates with the Earth. It also shows the visibility of the object according to its size and the limiting visual magnitude of an observer on the Earth \citep{dymock_2007}, so that the observability time intervals can be easily searched for. To the best of our knowledge, this is a unique tool that is not offered by any other popular orbit visualisers. 

The Flyby Visualisation Tool (FVT) produces high-precision graphical representations of NEAs during their close approach to the Earth. The orbits of the Moon and of the geostationary orbit, and the distance to the equatorial plane, can be displayed to better understand the geometry of the close encounter. The uncertainty ellipse of the NEA orbit can be also optionally shown. These simulations can help researchers and developers to better understand the behaviour of NEAs orbits during close encounters. 

The Observation Planning Tool (OPT) allows astronomers to plan their observations by providing precise ephemerides and observational conditions for any observer’s position and time, including target NEA location and motion, its magnitude, orbit uncertainty, Sun elongation, galactic latitude, and background star density (among others).

The Sky Chart Display Tool (SCDT) locates the path in the sky followed by the objects of interest, as observed from any position on the Earth. This tool, in combination with the OPT, is particularly useful for astronomers who wish to observe NEAs, as it allows them to plan their observations and locate the objects in the night sky with high precision.

\subsection{Ephemerides service}
The NEOCC provides users with two different ephemerides services, implementing the ephemerides computation method with linear predictions described in Sec.~\ref{s:orb_det}. The first service offers a pre-computed file containing the geocentric ephemerides of all the NEAs for the coming 30 days, with a time step of 1 day. This file is updated several times per day, and it can help astronomers to find NEAs that are visible in the following weeks to schedule their observation pipelines. This file can be automatically downloaded through the HTTPS APIs.

\begin{figure*}
    \centering
    \includegraphics[width=\textwidth]{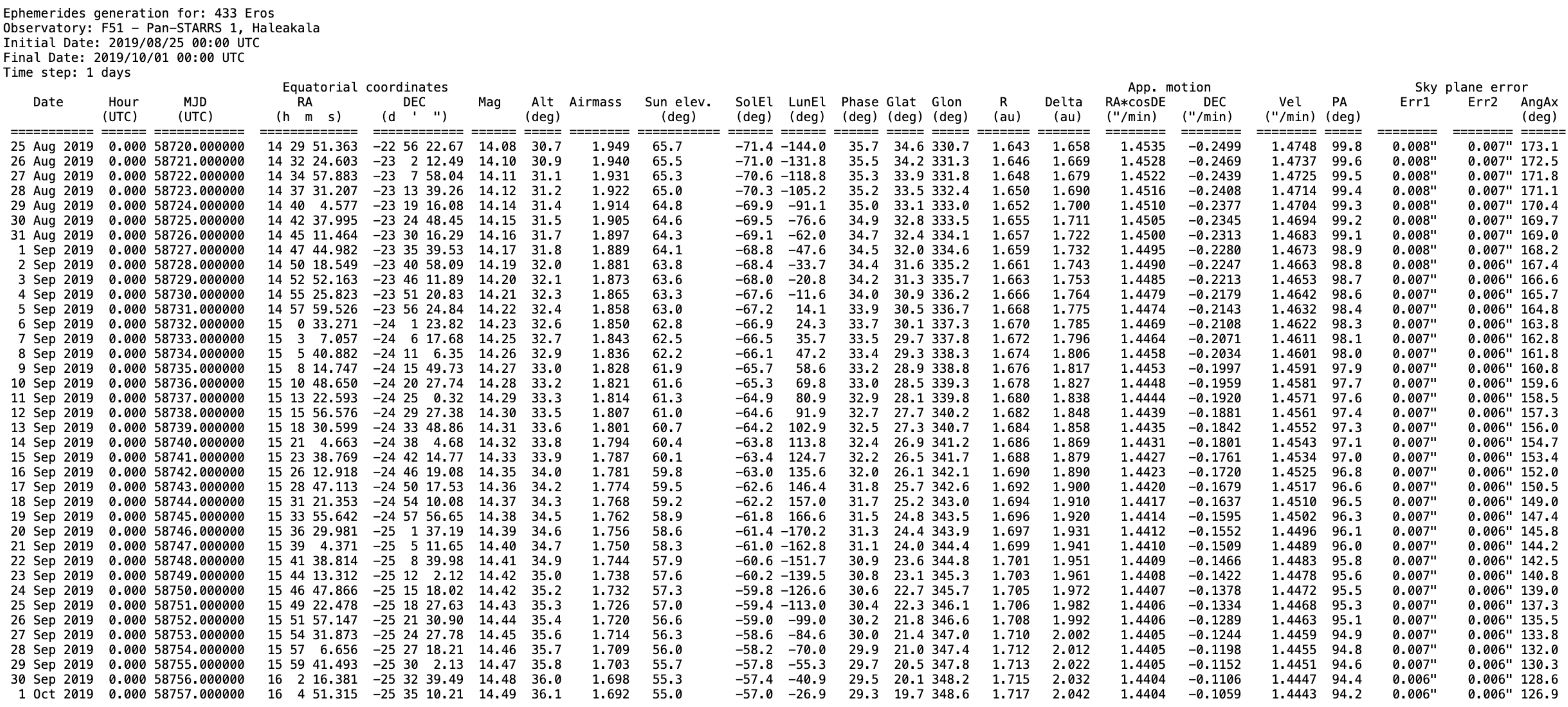}
    \caption{Response of the ephemerides API. In this example, ephemerides of asteroid (433) Eros are requested for the Pan-STARRS 1 telescope (observatory code F51) starting from 2019/08/25 00:00 UTC to 2019/10/01 00:00 UTC, with time-step of 1 day. The table contains information about the equatorial coordinates, the apparent motion of the object in the sky, and about the uncertainty of the prediction. }
    \label{fig:ephem_response}
\end{figure*}

The second service is an ad-hoc ephemerides generator, where the user can specify: 1) the object for which the ephemerides is needed; 2) the observatory code; 3) the initial and the final epochs; 4) the timestep for the output. This service is available on the dedicated page of an asteroid through an intuitive graphical user interface. Users needing more automated access to the ephemerides can send requests by using the HTTPS APIs. Ephemerides are returned to the user in a text table format, containing the equatorial coordinates as a function of time, the visual magnitude, the solar and lunar elongations, the phase angle, the distance from the Sun and the Earth, the apparent motion, and the sky-plane error. The sky plane error is computed from the confidence ellipse projected on the sky through Eq.~\eqref{eq:covEph}, and it is identified by three quantities: 1) the length of the semi-major axis of the ellipse; 2) the length of the semi-minor axis of the ellipse; and 3) the angle between the major axis and the North direction. A typical response from the API is shown in Fig.~\ref{fig:ephem_response}

\subsection{Close Approach Fact Sheets}
\begin{figure}[!ht]
    \centering
    \includegraphics[width=0.8\textwidth]{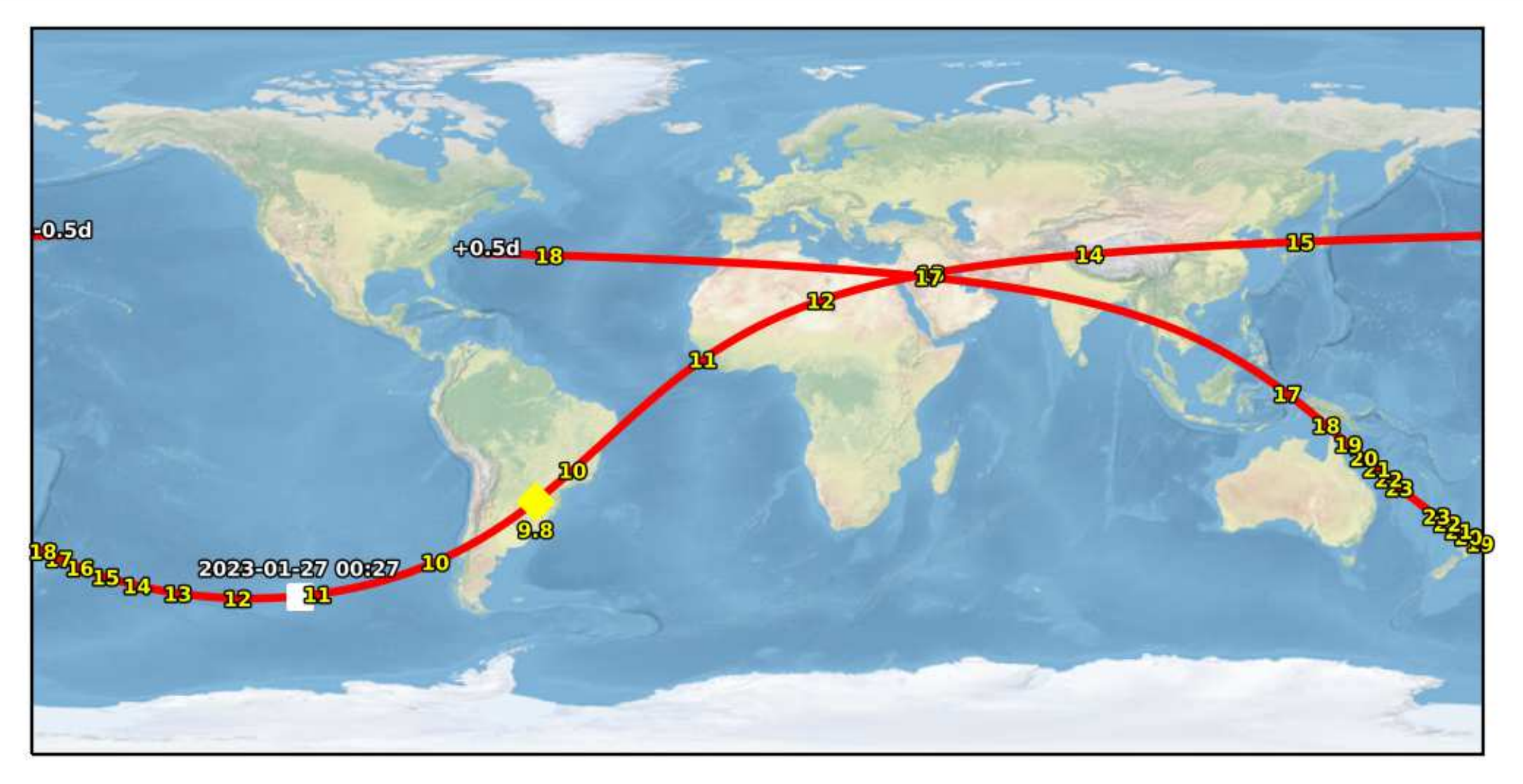}
    \caption{Ground track of asteroid 2023 BU (red curve) during its close approach with Earth on 27 January 2023. The white square corresponds to the epoch of minimum distance, while the yellow square represents the position at the brightness peak. Yellow numbers along the ground track denote the visual magnitude of the NEA as seen from the corresponding point on the Earth. }
    \label{fig:2023BU_ground_track}
\end{figure}

Close Approach Fact Sheets (CAFS) are official notes released by the NEOCC, which contain information about relevant upcoming close approaches of NEAs. 
Typical data available in a CAFS are: the fly-by date and the close approach time, the fly-by distance, the estimated size, and details of the discovery of the NEA. Moreover, orbital elements before and after the close approaches are indicated, and images of the heliocentric orbit and of the fly-by geometry produced with the OVT and FVT are included. To help astronomers to plan their observing schedule, information about the observational conditions of the event, such as the brightness peak and a map of the ground track near the close approach epoch, are also available. Figure~\ref{fig:2023BU_ground_track} shows an example of ground track, published in the CAFS of asteroid 2023 BU close approach happened on 27 January 2023. 
All these data are computed by the Aegis software. 

CAFS are not released with a pre-defined schedule, but they are usually produced as soon as an interesting close approach is found. A total of 13 close approaches have been published in a CAFS so far, and users can automatically receive them by email whenever they are published, upon subscribing to this specific service.

\section{Statistics and comparisons with other services}
\label{s:comparison}

\subsection{Short arc and orbit quality statistics}
\label{ss:Upar}
\begin{figure}
    \centering
    \includegraphics[width=0.48\textwidth]{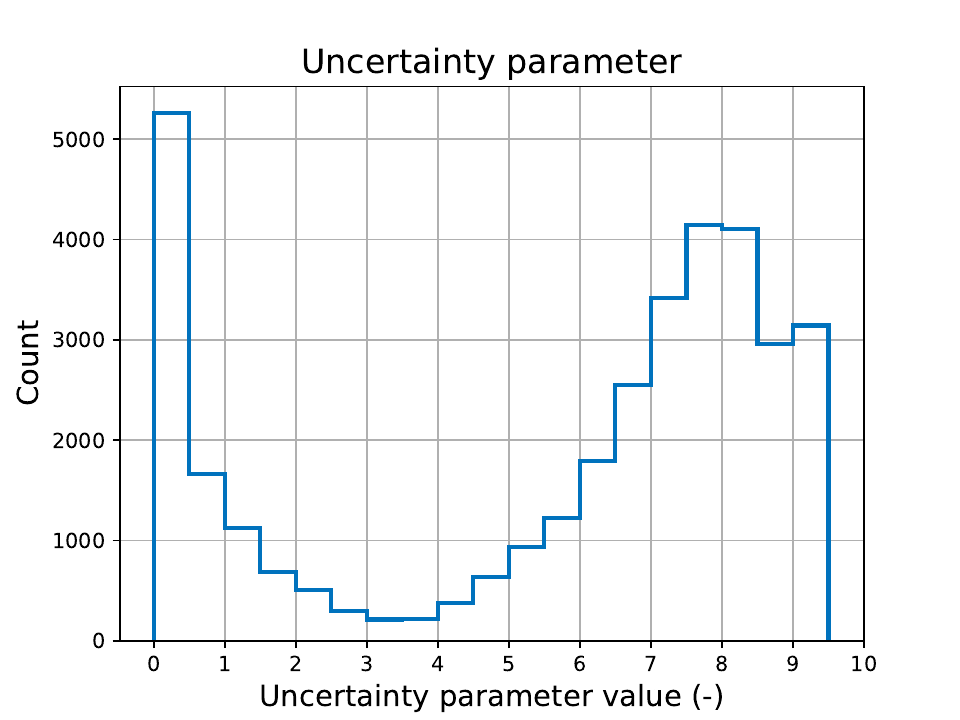}
    \includegraphics[width=0.48\textwidth]{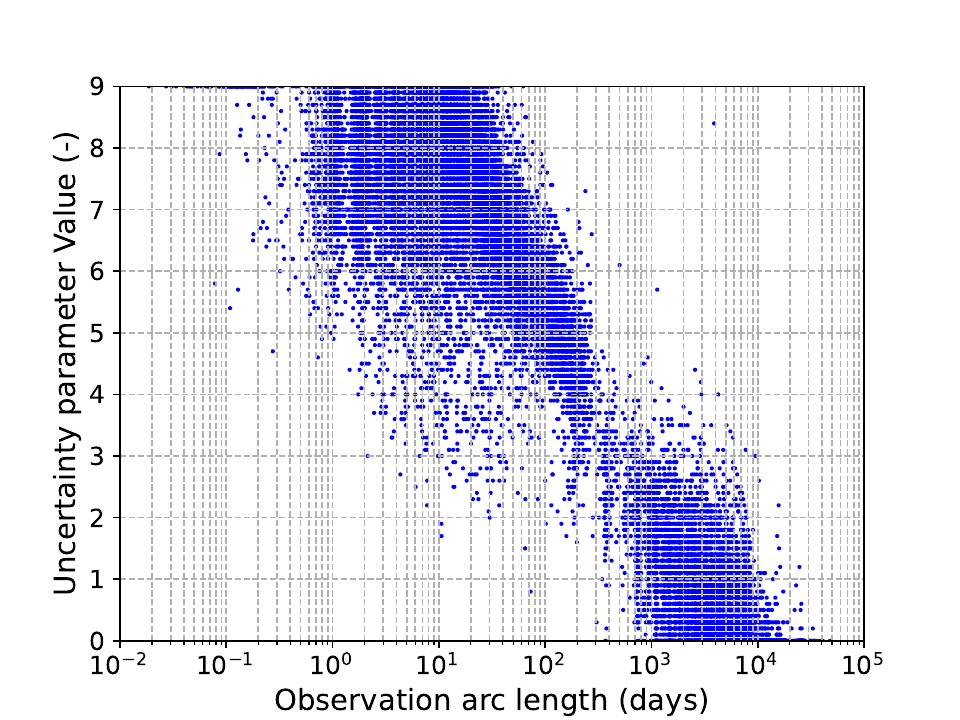}
    \caption{Left panel: distribution of the uncertainty parameter U for all the NEAs. Right panel: distribution unvertainty parameter U vs. arc length of all the NEAs.}
    \label{fig:Upar}
\end{figure}

We report in this section some general statistics on the quality of the orbits of NEAs. To this purpose, we take into account the uncertainty parameter $U$ introduced by \citet{marsden-etal_1978}. This parameter combines the uncertainties of the time of passage through the perihelion and of the orbital period, providing a single number that quantifies the uncertainty of the orbit, and it is also currently used by the MPC for this purpose. This parameter ranges from about 10 for a poorly determined orbit (e.g. a new discovery) to about 0 for a well-characterised one (e.g. a numbered asteroid). The left panel of Fig.~\ref{fig:Upar} shows the distribution of the $U$ parameter for all NEAs in the NEOCC portal. Of about 35~000 NEAs, only about 20 per cent have uncertainty parameter value smaller than 1, while a significant fraction has values larger than 8. This is a consequence of the fact that many newly discovered NEAs may only be observable for a short time, resulting in weakly constrained orbits. This correlation is better seen in the distribution of the uncertainty parameter vs. the arc length \citep{desmars-etal_2011}, shown in the right panel of Fig.~\ref{fig:Upar}. The trend is roughly linear in a semi-logaritmic scale, although some objects are evident as clear outliers. These are likely asteroids that got a particularly dense observational coverage at geometries that favoured the orbit determination process, such as during close fly-bys observed on both sides of their encounter with the Earth. Another reason is that some of these objects were favourable for radar observations, which makes the orbit well constrained even with a short arc. An example is 2024~JZ$_6$, which has an arc of only 22 days and an U parameter of 3.7. In the effort to improve follow-up of newly discovered NEAs, an observatory ranking \citep{bernardi-etal_2021} has been recently developed within the European project NEOROCKS\footnote{\url{https://www.neorocks.eu/}}.

\begin{figure}
    \centering
    \includegraphics[width=0.48\textwidth]{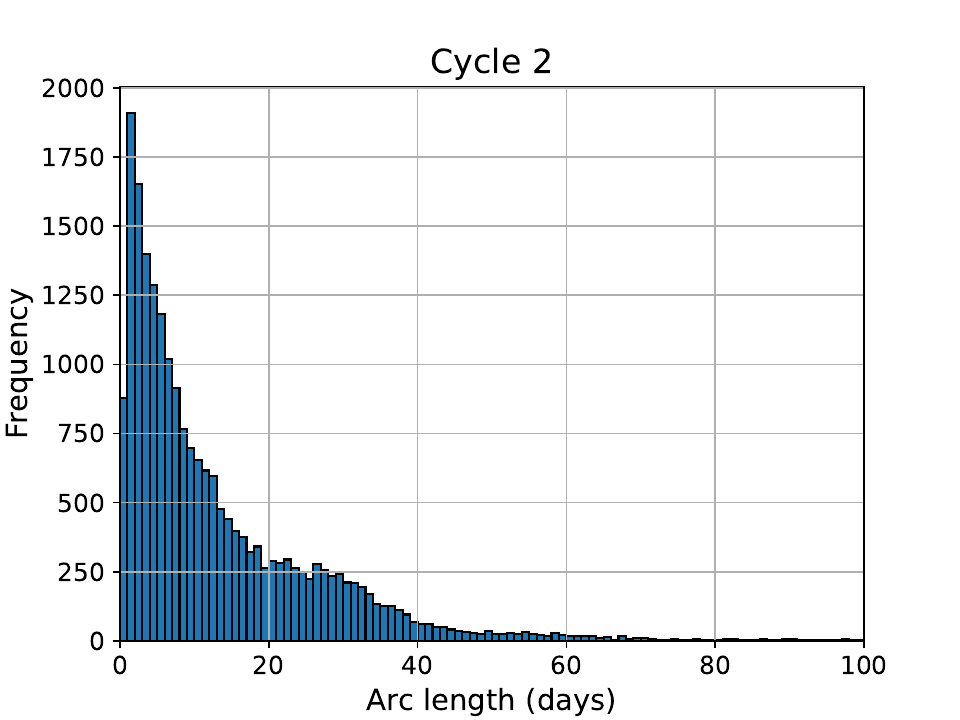}
    \includegraphics[width=0.48\textwidth]{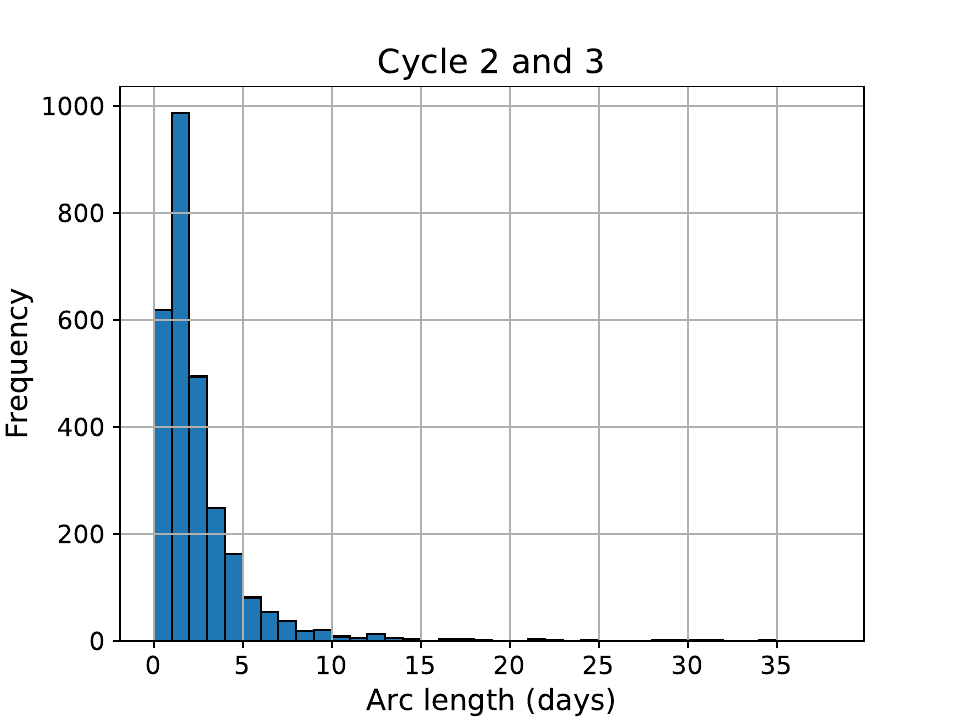}
    \caption{Distributions of the arc length of NEAs which needed the execution of Cycle 2 of the global algorithm of differential corrections (left panel), and which needed the execution of Cycle 2 and 3 (right panel).}
    \label{fig:cycles}
\end{figure}

Short arcs are also known to cause rank deficiencies in the orbit determination process. In Sec.~\ref{ss:glodifcor} we introduced the global algorithm of differential corrections, which comprises 3 different cycles and takes into account possible rank deficiencies. As a test to check when the additional steps of differential corrections are needed, we recomputed the orbits of all NEAs and kept track of the objects which needed to execute Cycle 2 and/or Cycle 3 to get a least-square nominal solution. Of 35~271 NEAs examined, a total of 21~207 executed Cycle 2, while 2~776 executed both Cycle 2 and 3. Figure~\ref{fig:cycles} shows the distributions of the arc length of objects that executed Cycle 2 and Cycle 2 and 3. As can be noted, both distributions are more dense at short arc-length, with a peak for object with $\sim$1 day of observations. Thus, rank deficiencies are more common when data are limited to a short time internal, as expected \citep{milani-gronchi_2009}, and the algorithm introduced in Sec.~\ref{ss:glodifcor} is able to handle these cases.

\begin{figure}
    \centering
    \includegraphics[width=0.48\textwidth]{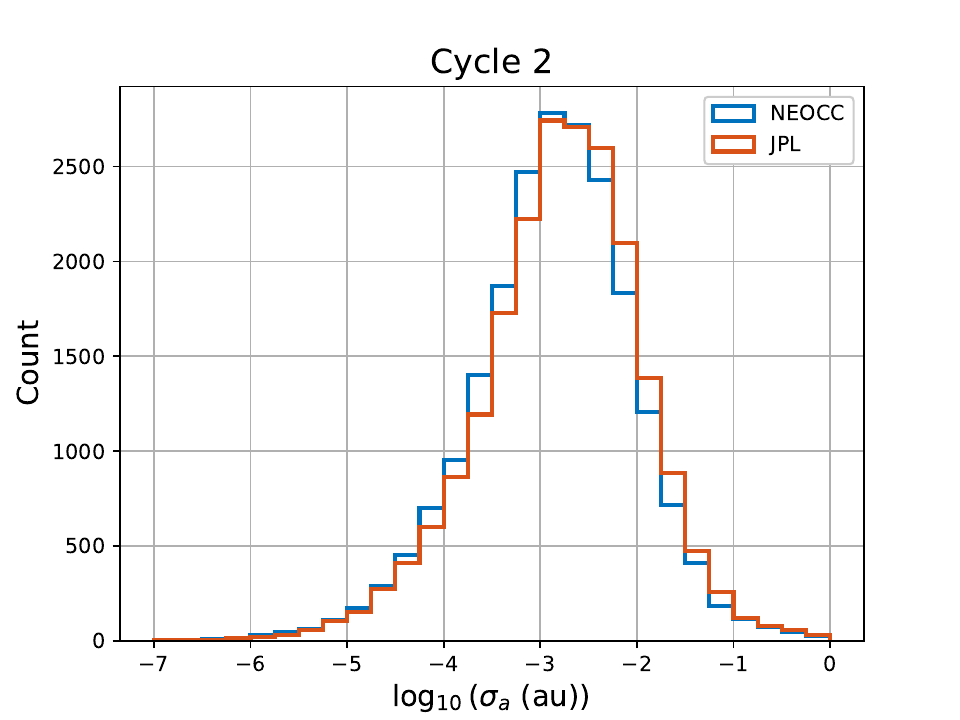}
    \includegraphics[width=0.48\textwidth]{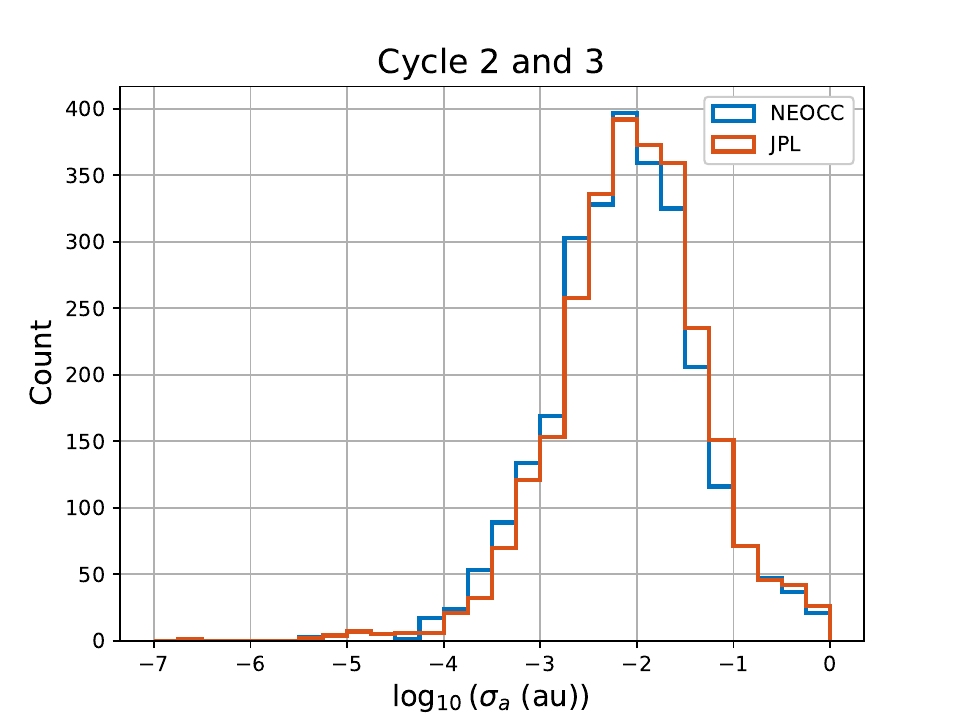}
    \caption{Distributions of uncertainty $\sigma_a$ in the semi-major axis of NEAs which needed the execution of Cycle 2 of the global algorithm of differential corrections (left panel), and which needed the execution of Cycle 2 and 3 (right panel).}
    \label{fig:siga_difcor}
\end{figure}

To further check the quality of the orbit that we are able to compute in these cases, we extracted the uncertainty $\sigma_a$ in the semi-major axis that we computed with Aegis and the corresponding value reported on the JPL SBDB. Note that, for short-arc objects, the JPL SBDB does not propagate the orbital elements near the current epoch (see also next section). Therefore, the values of $\sigma_a$ that we took into account are at the middle-arc epoch, which may differ between NEOCC and JPL. For this reason, we can not make a truly one-by-one comparison between single objects, and we rely on a statistical comparison. Figure~\ref{fig:siga_difcor} shows the distributions of $\sigma_a$ of NEAs which needed both Cycles 2 and/or Cycles 3. The two distributions follow a similar profile, indicating that orbits computed by NEOCC and by JPL are of similar quality, even in the case of short arcs. %In addition, this is an indication that also JPL is able to systematically handle rank deficiencies.

\subsection{Orbit catalogues comparison}
To find possible issues with the orbit determination of certain objects and to validate the results, comparing the orbits obtained independently by different centres is a fundamental task to regularly perform. However, this is not always easy, because the same set of data is needed for a meaningful comparison. For instance, orbital elements at the middle-arc epoch computed from Eq.~\eqref{eq:t0} would be desirable for an accurate comparison with the orbits we compute, because uncertainties do not need to be propagated at a different epoch. However, the middle-arc epoch depends on different factors, such as the observations used in the orbital fit, the error model, the weights used in the differential correction procedure, and on the observations discarded as outliers. All these elements may vary among the different centres of asteroid orbits computation. 
For these reasons, the best that can be done is a comparison near the current epoch, for which the orbital elements are provided by the ESA NEOCC, the JPL Small Body Database (SBDB), the MPC, NEODyS, and the Lowell Observatory. The current epoch is updated every 100 days by the Lowell Observatory database, while every 200 days by the other services, and therefore a comparison with the database provided by Lowell is not always possible. 
Another issue is the set of orbital elements that are publicly available in the databases. While the NEOCC provides orbital elements in both Keplerian and Equinoctial elements, the JPL SBDB provides cometary elements and auxiliary quantities - such as semi-major axis, aphelion distance, mean motion, and orbital period - when defined, the MPC provides only Keplerian elements, and NEODyS only Equinoctial elements in full precision. In addition, the MPC does not provide the orbital elements with full precision yet, and uncertainties are currently not available in the \texttt{MPCORB.DAT} file. All these problems need to be addressed in the future by establishing criteria to present the orbits that are common to all these centres.

\begin{figure*}[!ht]
    \centering
    \includegraphics[trim={0 0cm 0 0}, clip, width=0.48\textwidth]{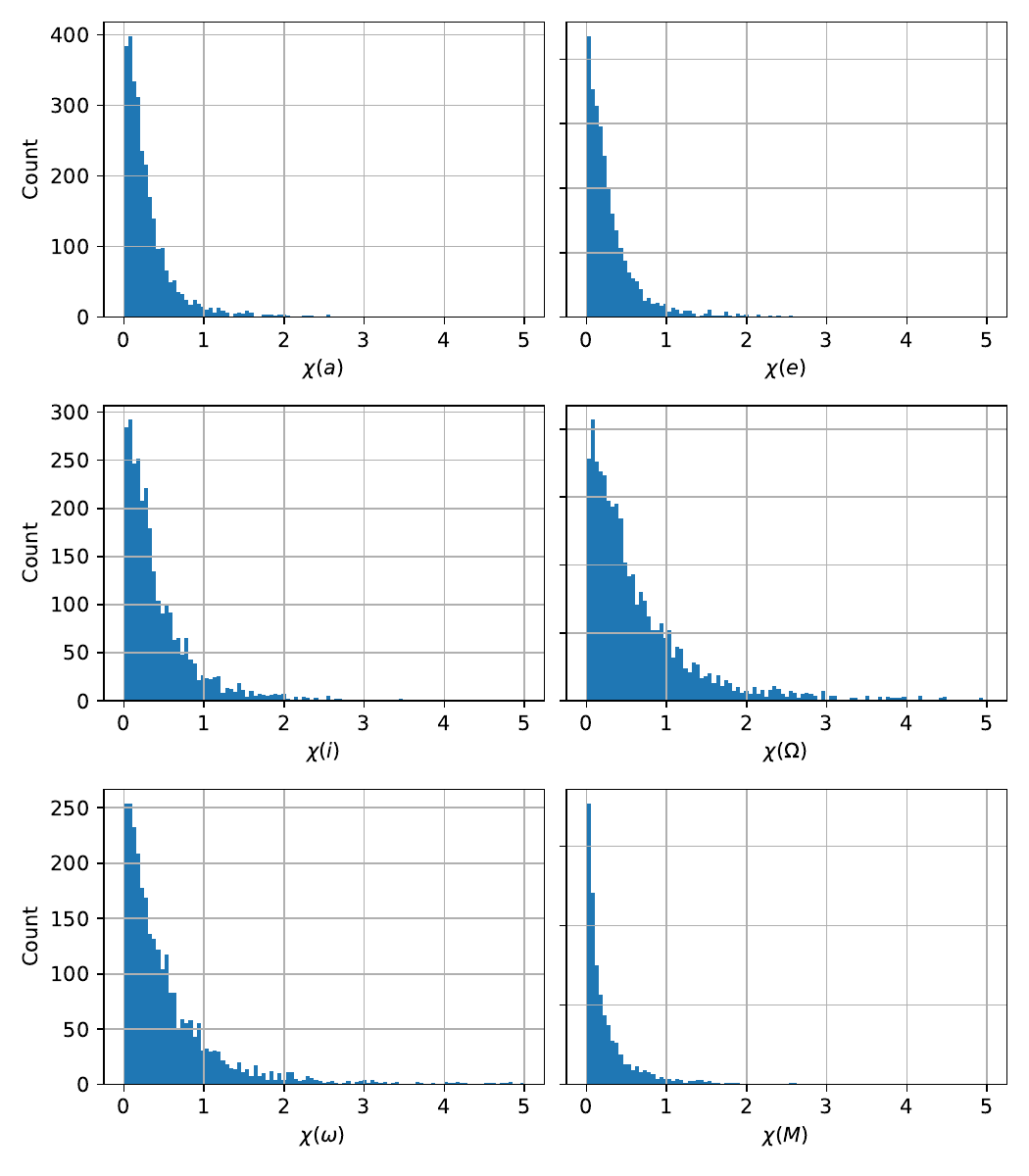}
    \includegraphics[trim={0 0cm 0 0}, clip, width=0.48\textwidth]{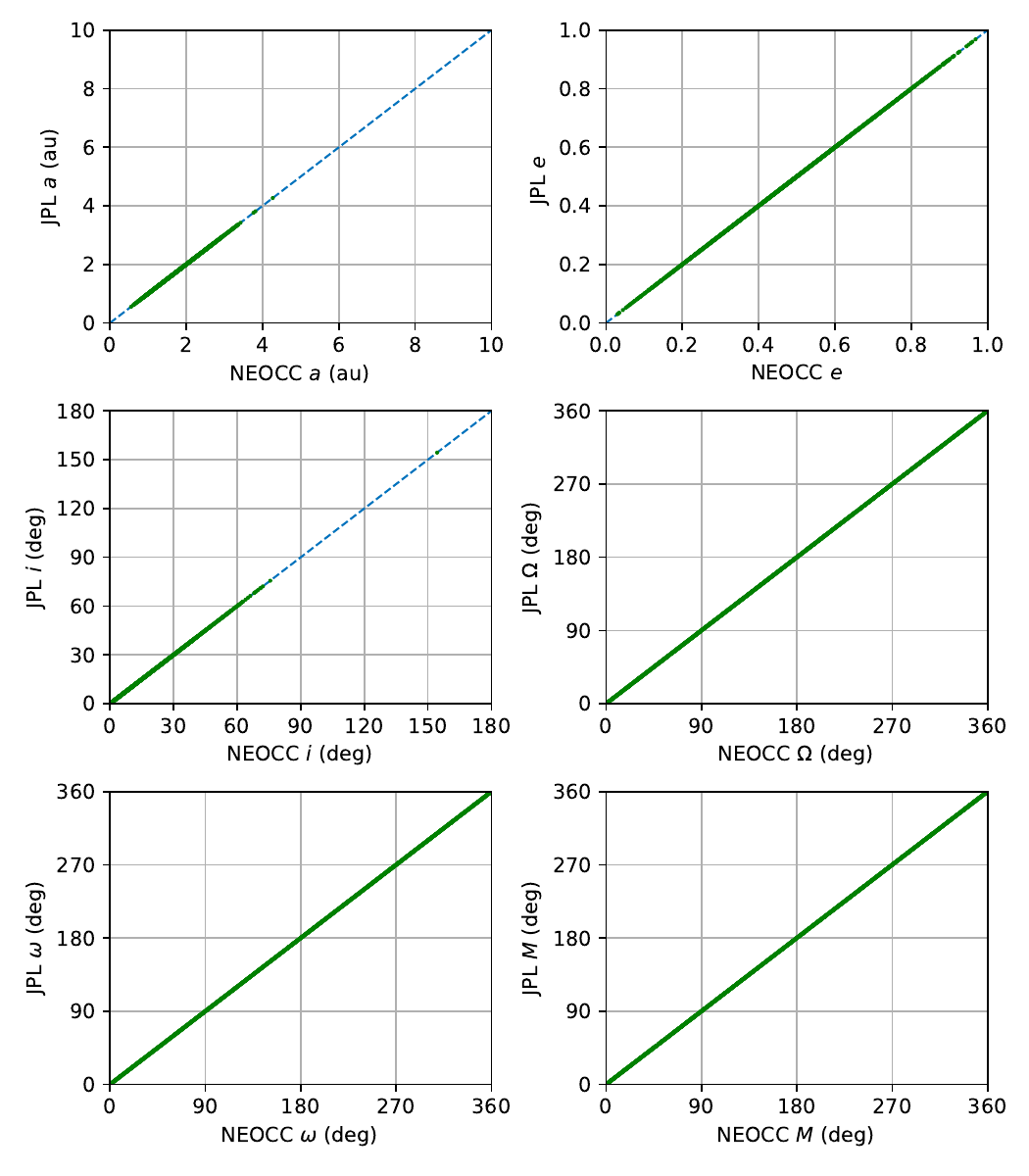}
    \caption{Distributions of $\chi(y)$ for each Keplerian element $y$, obtained from the comparison of the orbits computed by the NEOCC and the JPL (first and second column), and distribution of $y_{\text{NEOCC}}$ vs. $y_{\text{JPL}}$ (third and fourth columns). The results shown here are obtained by taking into account only numbered NEAs. High resolution versions of these figures are provided as supplementary electronic material.}
    \label{fig:Chi_numbered}
\end{figure*}

A comparison with orbits obtained by OrbFit was already performed during a validation phase of Aegis, and they showed an excellent agreement. Due to this and to all the issues mentioned above, here we present a comparison with the JPL SBDB orbits only. 
Keplerian orbital elements for all the NEAs are compared near the current epoch. Orbits of all NEAs were downloaded from the JPL SBDB using the dedicated SBDB API\footnote{\url{https://ssd-api.jpl.nasa.gov/doc/sbdb.html}} on 9 April 2024. Note that the JPL SBDB does not propagate an object to the current epoch when the observational arc is short, and therefore we discarded these object from the comparison. For a comparison between orbits of NEAs with short arc, see Sec.~\ref{ss:Upar}. In addition, other forces such as non-gravitational effects may be included in the model for particular NEAs. However, this set is not uniform between the NEOCC and the JPL SBDB, and we therefore limit ourselves to compare NEAs orbits that have been determined without non-gravitational forces. Note also that a comparison between orbits of NEAs for which non-gravitational parameters are determined has already been performed in \citet{fenucci-etal_2023}.
To evaluate the similarity of the two orbits we compute 
\begin{equation}
    \chi_y = \frac{| y_{\text{NEOCC}} - y_{\text{JPL}}| }{\sqrt{\sigma^2_{y, \text{NEOCC}} + \sigma^2_{y, \text{JPL}} } },
    \label{eq:chi}
\end{equation}
where $y$ is one of the Keplerian elements, $\sigma_y$ is the corresponding uncertainty, and the subscript denotes the institution who computed the orbit. Note that if $\chi_y < 2$, then the hypothesis that the two distributions are the same can not be excluded with 95\% confidence level. To present the results, we divide the list of NEAs in numbered and unnumbered objects. 
The comparison was performed on 5 August 2024, and the epoch at which elements are examined is 60600 MJD. 
Figure~\ref{fig:Chi_numbered} shows the distribution of $\chi_y$ for all the Keplerian elements of the numbered NEAs, as well as the distributions of the orbital elements computed by the NEOCC vs. the orbital elements computed by the JPL SBDB. A total of 2~857 NEAs orbits were compared, while 320 were excluded from the comparison because they included non-gravitational parameters in the model. About 66 per cent of objects have each $\chi_y$ in the interval $(0,1)$, 21 per cent in the interval $(1,2)$, and about 10 per cent have at least a value $\chi_y$ larger than 2.
%
%The vast majority of each $\chi_y$ distribution is contained in the $(0,1)$ interval, a small fraction is contained in $(1,2)$, and only a few cases have a $\chi_y$ value larger than 2.
This indicates a generally good agreement between the orbits of numbered NEAs provided by the two centres. 
For numbered objects with very tight constraints on their orbit (such as radar observations at multiple oppositions), values of $\chi$ larger than 2 may be the consequence of different causes, such as: small differences in the dynamical models; close encounters affecting the propagation to the current epoch; different data weights; number of rejected observations; differences in the propagator. However, without a complete knowledge on the other system, it is not possible to conclude what is the definitive reason for the observed differences in specific cases.

\begin{figure*}[!ht]
    \centering
    \includegraphics[trim={0 0cm 0 0}, clip, width=0.48\textwidth]{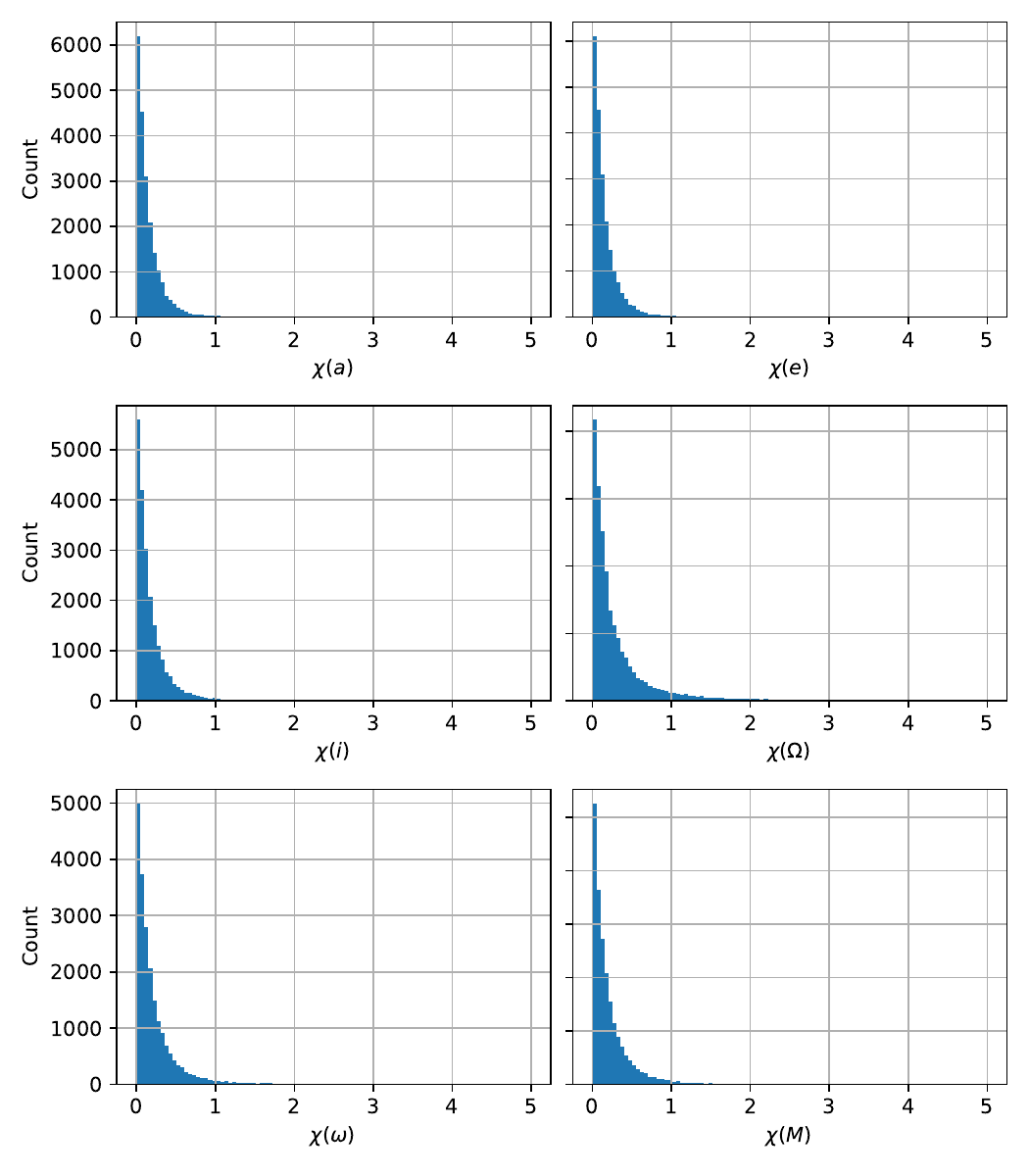}
    \includegraphics[trim={0 0cm 0 0}, clip, width=0.48\textwidth]{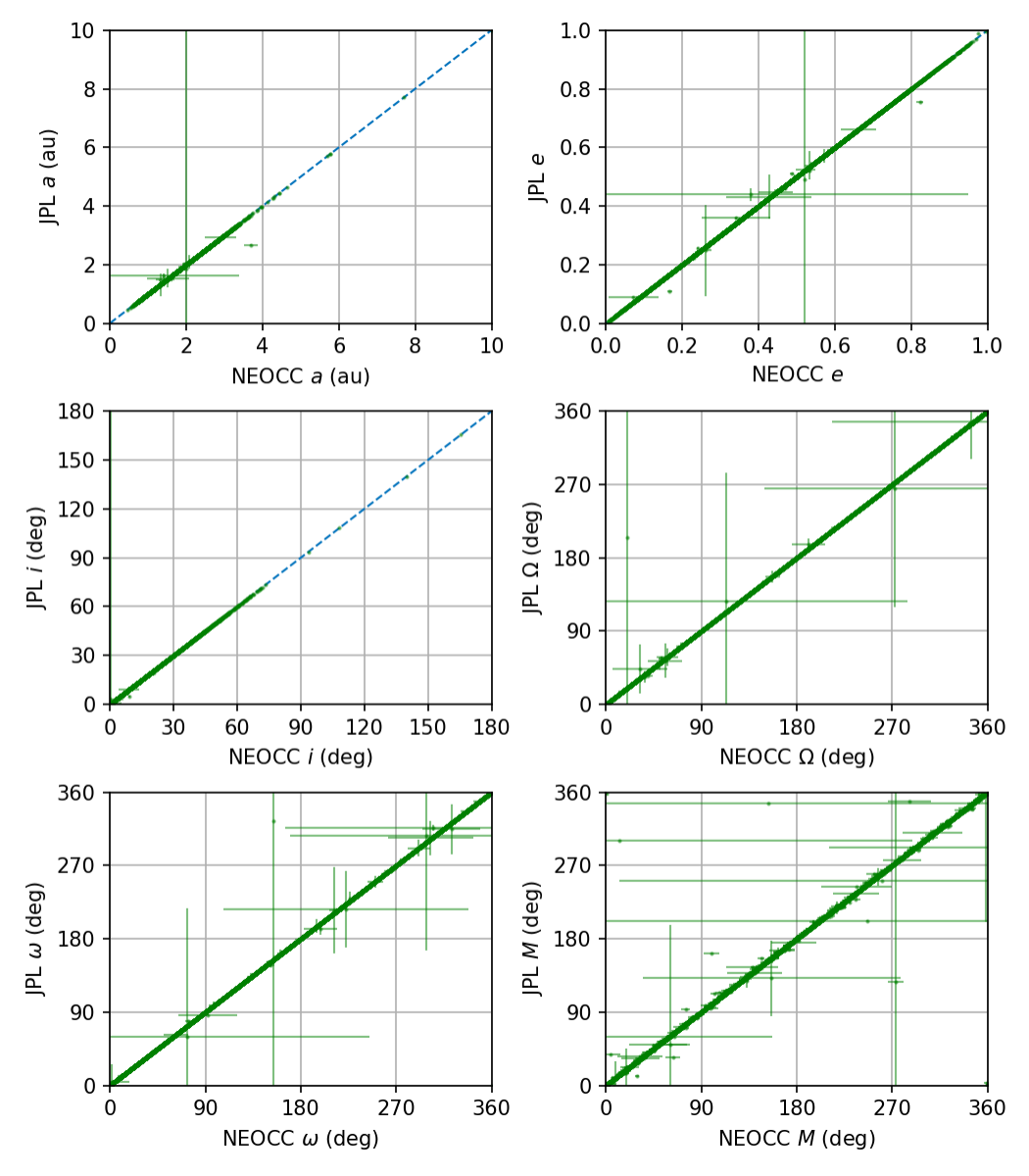}
    \caption{Same as Fig.~\ref{fig:Chi_numbered}, but obtained by taking into account only unnumbered NEAs. High resolution versions of these figures are provided as supplementary electronic material.}
    \label{fig:Chi_unnumbered}
\end{figure*}

Figure~\ref{fig:Chi_unnumbered} shows the results of the comparisons obtained by taking into account only unnumbered NEAs. In this case, 21~285 orbits were compared. Other 10~007 objects were excluded from the comparison because the reference epoch was not the same, while 191 objects because of the presence of non-gravitational perturbations in the orbital fit.
In this case, 88 per cent had all $\chi_y$ in $(0,1)$, 8 per cent had all $\chi_y$ in $(1,2)$, and 4 per cent had at least one element with $\chi_y > 2$.
Note that for unnumbered NEAs the uncertainties may be very large. In turn, these large uncertainties are generally caused by short observational arc, or by propagation near the current epoch that may be affected by close approaches. 
Full tables to reproduce Fig.~\ref{fig:Chi_numbered} and \ref{fig:Chi_unnumbered} are provided as supplementary electronic material.

The results presented above show that the orbits produced by the Aegis system and available on the NEOCC portal are statistically comparable with those provided by the JPL SBDB. In this sense, the two systems are equivalent $-$ although differences in specific single cases may appear due to the reasons explained above $-$, and this adds redundancy to the capabilities of impact monitoring of NEAs. In fact, having two completely identical systems would not be useful, since it would not allow to systematically investigate different impact monitoring techniques, possibly leading to erroneous computations. To keep the consistency in the data and to find possible sources of errors, it is important for all the centres of asteroid orbit computation to perform such comparisons on a regular basis. To this purpose, the MPC is setting up a public service for the comparison of orbits that is meant to be publicly available and kept updated\footnote{\url{https://data.minorplanetcenter.net/orbit-comparison/compare_neos.html}}.

\subsection{Ephemerides comparison}
\begin{figure}
    \centering
    \includegraphics[width=0.95\textwidth]{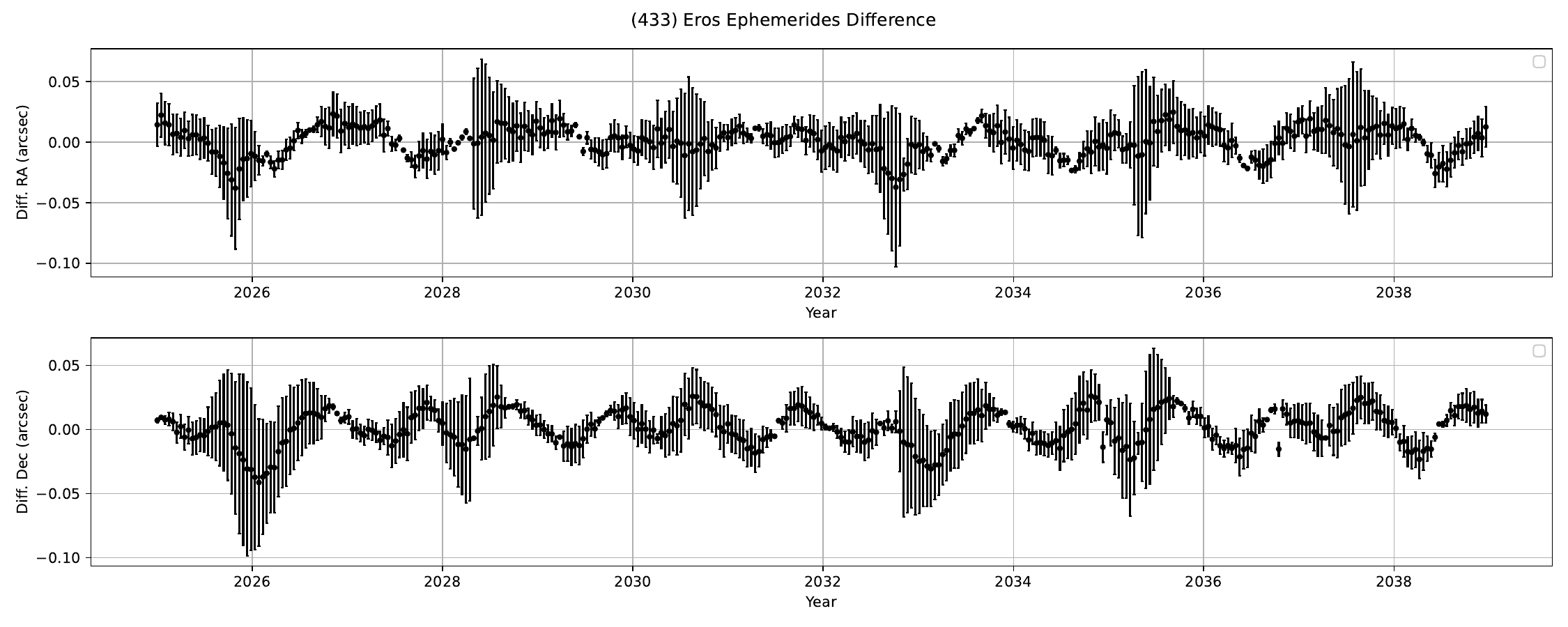}
    \includegraphics[width=0.95\textwidth]{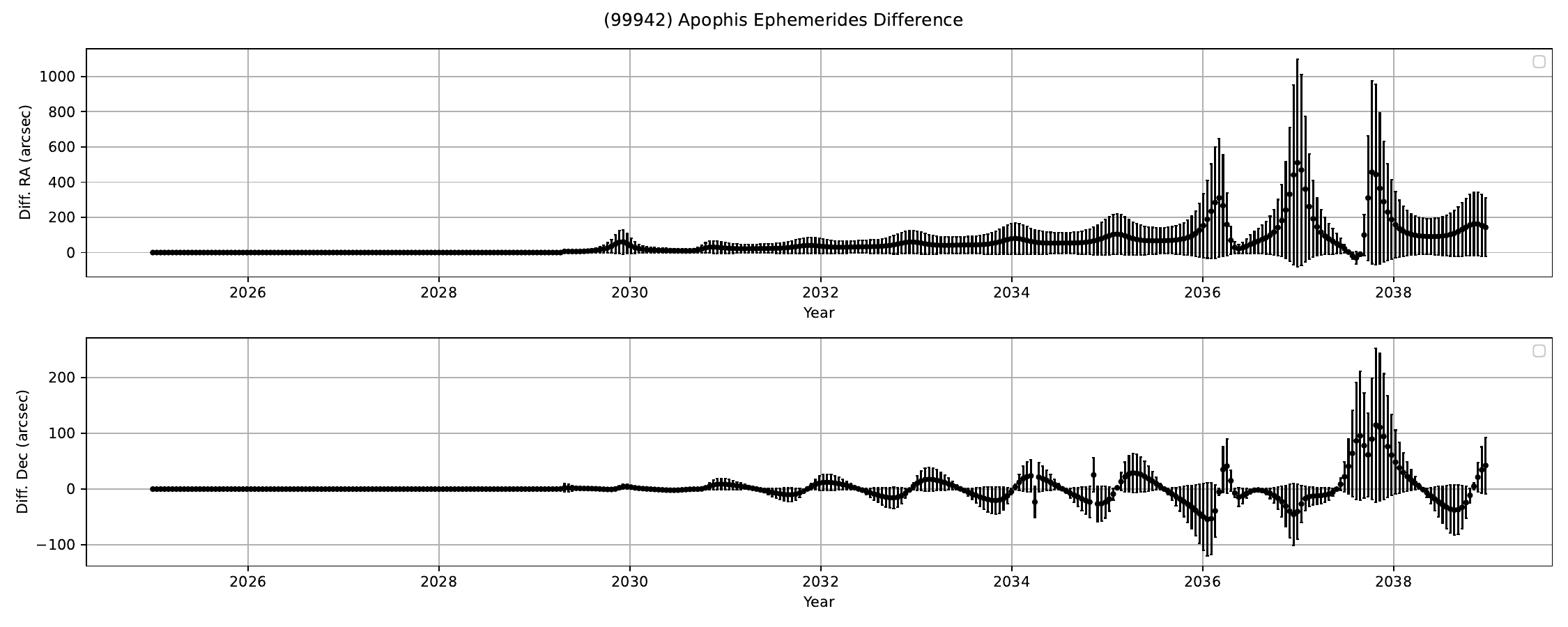}
    \caption{Differences between the celestial coordinates RA and Dec computed by Aegis and JPL Horizons, for asteroid (433) Eros (first and second panels) and (99942) Apophis (third and fourth two panels). Ephemerides as seen from the Geocenter are computed between 1 January 2025 and 1 January 2039, with a time-step of 15 days. Error bars refer to the 3-$\sigma$ uncertainty in the corresponding element computed by Aegis, as obtained as a projection of the uncertainty ellipsoid on RA and Dec, respectively.}
    \label{fig:433_diff}
\end{figure}

Ephemerides are computed from the nominal orbit with the linear propagation described in Sec.~\ref{ss:ephem}, and results from other systems can be also compared. As a representative test, we compared the ephemerides of (433) Eros and (99942) Apophis obtained by Aegis and by JPL Horizons\footnote{\url{https://ssd.jpl.nasa.gov/horizons/app.html\#/}}. We computed the positions in RA and Dec as observed from the Geocenter (IAU MPC code 500) from 1 January 2025 to 1 January 2039, with a cadence of 15 days. Uncertainties are obtained from the uncertainty ellipsoid in the plane of sky as a projection on RA and Dec. The difference between the the results computed by Aegis and JPL Horizons is shown in Fig.~\ref{fig:433_diff}. Error bars refer to the 3-$\sigma$ uncertainties computed by Aegis. For (433) Eros, the two results are in excellent agreement, although the predictions do not agree within 3-$\sigma$ at some specific time. However, note that differences of the order of $\sim$0.01 arcsec are so small that the asteroid would be recovered by any telescope by pointing at either predicted position, so these differences are completely negligible in practical cases.
Results obtained for (99942) Apophis show that differences are very small up until the middle of 2029. Note that (99942) Apophis has an extremely close approach with the Earth on 29 April 2029, which has a minimum distance of about $\sim$38~000 km from Earth centre. After that, differences grow large mainly because of chaotic effects produced by the close approach. However, the two predictions still agree within the 3-$\sigma$ error computed by Aegis throughout the whole integration timespan, indicating therefore that the two predictions are still statistically compatible.   
In conclusion, these two examples show that ephemerides from Aegis and from JPL Horizons are in agreement and of good quality, provided that the initial orbits are statistically compatible and no major chaotic effect is present in the selected ephemeris timespan. On the contrary, chaotic effects or different starting orbits may cause the predictions to significantly differ, and uncertainties should be checked in these cases to understand the reliability of the computed quantities.

\subsection{Risk Assessment comparison}

\subsubsection{Risk List comparison}

Cross-checking the results of different systems is even more crucial for impact monitoring, since mitigation actions need to be discussed and planned well in advance in case large impact probabilities are found. To this purpose, the communication of clear and accurate data is essential. It is therefore needed to ensure the reliability of the impact monitoring systems, and comparing results coming from different and independent implementations is a way to do so. 

The cross checking of impact probabilities results between the ESA NEOCC, the JPL CNEOS, and NEODyS is performed on a daily basis through a dedicated internal mailing list. In case a PS larger than $-2$ is found, all the three centres perform a verification of the results, and take actions in case there are significant differences between them. In these cases, the publication of the data on the web is coordinated by the three institutions,  and it happens only after successful technical verification. 

Here we present a comparison of the impact monitoring results with JPL CNEOS, which are not based on the LOV method \citep{roa-etal_2021} used by Aegis. 
Note that the algorithms implemented in Aegis are more similar to those that were originally implemented in OrbFit for the Clomon-2 system. Because of this, the Aegis software internally underwent a period of about 2 years of validation, where results obtained with Aegis were systematically compared with those obtained by OrbFit as a joined effort by NEOCC operators and Aegis developers. The Aegis software started the Impact Monitoring operations only when the results of the two software reached a satisfactory level of agreement.
The risk files of NEAs in the NEOCC Risk List were retrieved using the HTTPS APIs (see Sec.~\ref{ss:risklist}), while the risk tables of the Sentry risk list were downloaded by using the dedicated Sentry APIs\footnote{\url{https://ssd-api.jpl.nasa.gov/doc/sentry.html}}. The risk files refer to the 9 April 2024, before the MPC daily orbital update was issued. 
To make a simple but still meaningful comparison, we proceed in the following way. We first select a threshold $\text{IP}_{\min}$, then for each NEA we take the VI with the largest IP for both Aegis and Sentry. Let us denote these probabilities with $\text{IP}_{\text{Aegis}}$ and $\text{IP}_{\text{Sentry}}$, and with $d_{\text{Aegis}}$ and $d_{\text{Sentry}}$ the corresponding dates. If both probabilities are smaller than $\text{IP}_{\min}$, the comparison is not performed. If one of them is larger than the threshold, then we proceed in searching close VIs for a comparison. Hence, we select the Sentry VI with the largest IP among those that are closer than 20 days to $d_{\text{Aegis}}$. We denote this probability with $\overline{\text{IP}}_{\text{Sentry}}$.
We also select the Aegis VI with the largest IP among those that are closer than 20 days to $d_{\text{Sentry}}$, and denote its probability with $\overline{\text{IP}}_{\text{Aegis}}$. Then we compute the quantities
\begin{equation}
    r_1 = \log_{10}\bigg(\frac{\text{IP}_{\text{Aegis}}}{\overline{\text{IP}}_{\text{Sentry}}}\bigg), \qquad r_2 = \log_{10}\bigg(\frac{\overline{\text{IP}}_{\text{Aegis}}}{\text{IP}_{\text{Sentry}}}\bigg).
    \label{eq:IPratios}
\end{equation}
Note that when a close VI is not found, the value of $r_1$ ($r_2$) is artificially set to 5 ($-5$).

\begin{figure}[!ht]
    \centering
    \includegraphics[width=0.48\textwidth]{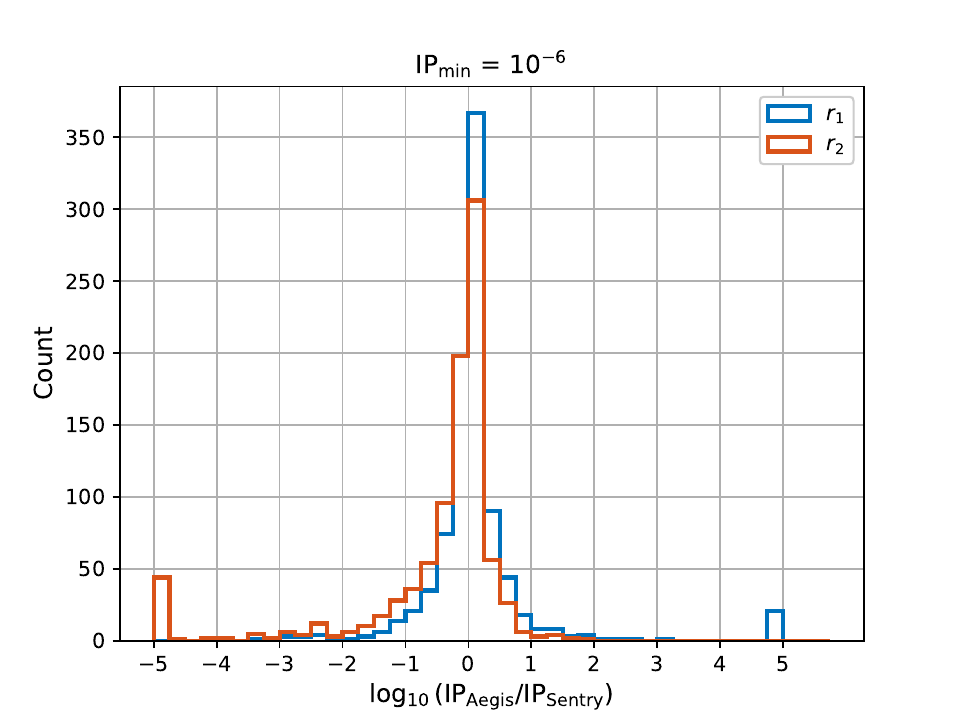}
    \includegraphics[width=0.48\textwidth]{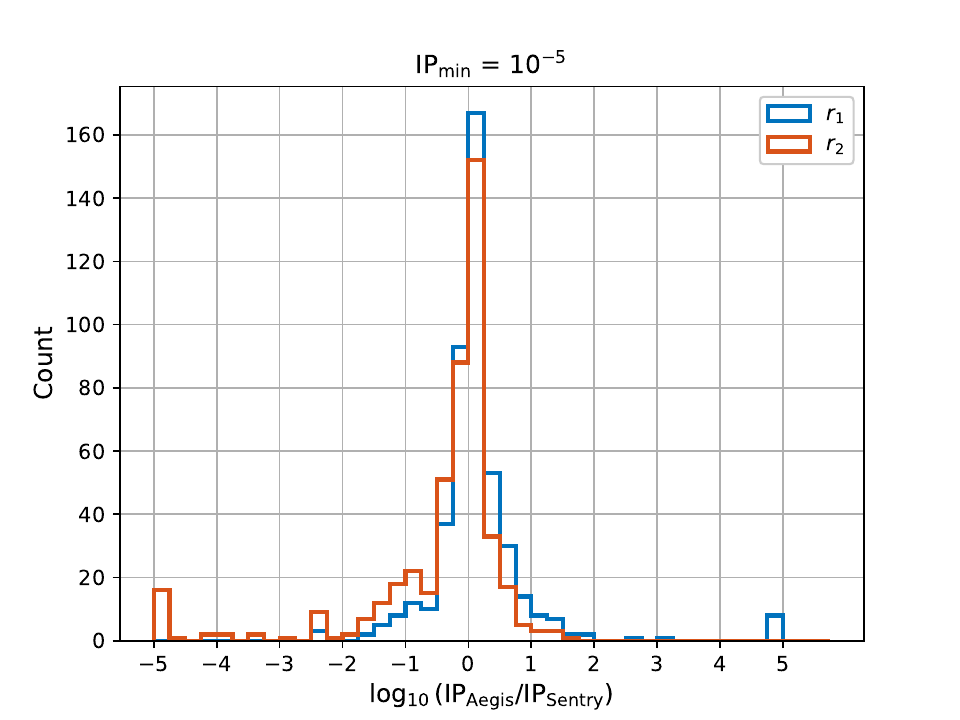}
    \includegraphics[width=0.48\textwidth]{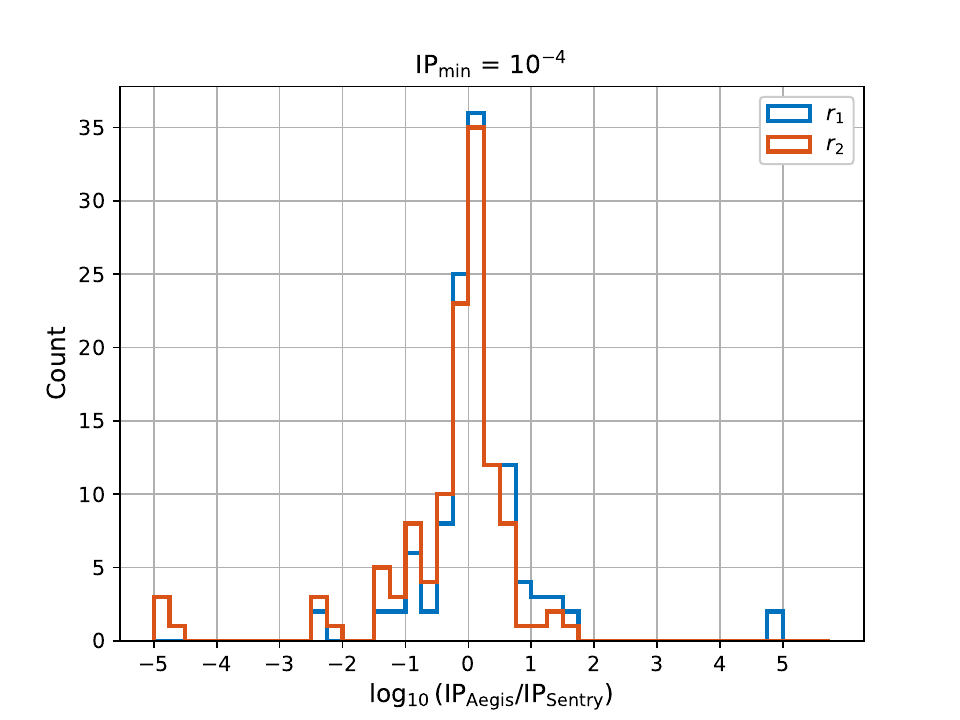}
    \includegraphics[width=0.48\textwidth]{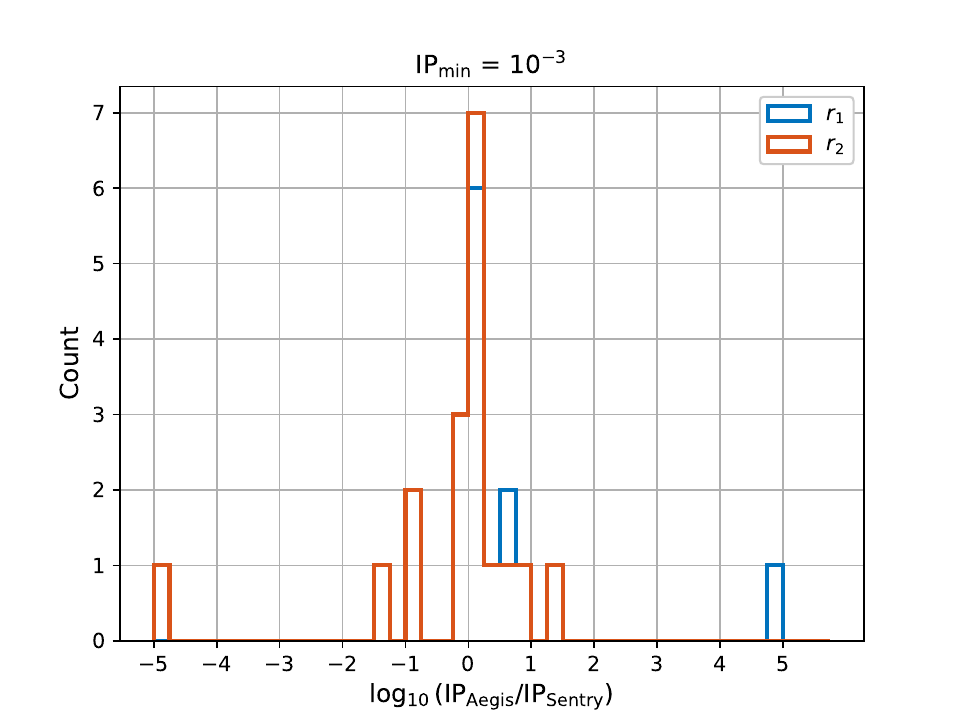}
    \caption{Distributions of the IP ratios $r_1$ (blue histogram) and $r_2$ (red histogram), obtained with equi-spaced bins with width equals to 0.25.}
    \label{fig:IP_ratios}
\end{figure}

We ran the procedure described above with threshold values $\text{IP}_{\min} = 10^{-6}, 10^{-5}, 10^{-4}, 10^{-3}$. The comparison was performed on 1 August 2024, before processing the MPC daily update. 
Figure~\ref{fig:IP_ratios} shows the distributions of the two quantities $r_1$ and $r_2$. 
Table~\ref{tab:percdiff} reports the percentages of $r_1, r_2$ between $-0.5$ and $0.5$, and between $-1$ and $1$.
Depending on the value of $\textrm{IP}_{\max}$, the cases bounded between $-0.5$ and 0.5 range from about 75 per cent down to about 55 per cent. On the other hand, cases that differ by less than an order of magnitude, i.e. values of $r_1, r_2$ between $-1$ and 1, are about 85 per cent.
These values are an indication that the results obtained by the two impact monitoring systems are in good agreement for the vast majority of the cases, since the maximum IPs found differ by a factor of just a few. 
Only a limited number of cases differ by more than one order of magnitude, and we note that both distributions have a more dense tail at negative values. This means that, while VIs at the same date are found by both systems, the impact probability estimated by Sentry seems to be systematically larger than those estimated by Aegis. A deeper analysis on specific single cases should be performed to understand the causes of these discrepancies, but it is out of the scope of the present paper.
In addition, we found some Aegis VIs that are not found by Sentry and vice-versa, and they are seen in the extremal bins of the distributions of Fig.~\ref{fig:IP_ratios} by artificially setting the value of $r_1$ ($r_2$) to 5 (-5). The complete lists used to produce the figures are provided as supplementary electronic material. 
The majority of the large differences in the IP ratios appear for VIs far in the future, and it is reasonable to understand them as an effect of the chaotic nature of the dynamics of NEAs, especially when the uncertainties in the nominal orbit are not very small. 
Another reason for IP differences is in how astrometric and time uncertainties are treated. While the statistical models for astrometric uncertainties used by ESA NEOCC and JPL CNEOS are basically the same, astrometric errors from individual observers who submitted this information to the MPC may be treated differently by the two centres. Time uncertainties may also play a significant role for NEAs that are observed during a close approach, and differences in their treatment can produce differences in the IP of VIs.% This was the case, for instance, of 2023~VD$_3$ \citep{micheli-etal_2024}: the IP is significantly larger when time uncertainties are not taken into account, as for the results obtained by the operational Aegis system, while it lowers down when they are correctly modelled.    
Some of the large differences may also arise from the fact that Aegis and Sentry performed the orbit determination with different dynamical models. For instance, non-gravitational effects may be included in one system but not in the other, and this may completely change the outcome of the impact monitoring \citep{chesley-etal_2014, spoto-etal_2014, delvigna-etal_2019b}. 

\begin{table}[ht]
    \centering
    \begin{tabular}{ccccc}
      \hline
      \hline
      $\textrm{IP}_{\min}$   & $P(r_1 \in (-0.5,0.5)$) & $P(r_1 \in (-1,1)$) & $P(r_2 \in (-0.5,0.5)$) & $P(r_2 \in (-1,1)$) \\
      \hline
       $10^{-6}$  & 78 & 91  & 70  & 83 \\
       $10^{-5}$  & 75 & 89  & 70  & 82  \\
       $10^{-4}$  & 67 & 86  & 66  & 83  \\
       $10^{-3}$  & 55 & 83  & 61  & 83 \\
      \hline
    \end{tabular}
    \caption{Percentages of $r_1$, $r_2$ in $(-0.5, 0.5)$ and in $(-1,1)$, for different values of $\textrm{IP}_{\max}$.}
    \label{tab:percdiff}
\end{table}

For specific example cases, we can comment on the results obtained for $\text{IP}_{\min} = 10^{-3}$. The only differences larger than one order of magnitude appear for asteroids 2016~WN$_{55}$, 2023~VD$_3$ and 2024~BY$_{15}$. These are all problematic cases, and they deserve a more detailed explanation.
Asteroid 2016~WN$_{55}$ has 35 observations covering an arc of about 4 days, all taken from the Wide-field Infrared Survey Explorer \citep[WISE, MPC code C51][]{wright-etal_2010} infrared space telescope. Sentry reports a maximum IP of 0.001 for a VI on May 2030, while Aegis does not have a VI at all in that year. A Monte Carlo run that we performed with the orbit computed by Aegis also did not find any VI in 2030, suggesting that the differences are given by the large uncertainties in the orbits determined by the two centres. The poor quality of the orbit is to be found in the observational dataset. In addition, only 1 over 35 observations has a reported visual magnitude, which makes the computation of the absolute magnitude $H$ (and of the PS as a consequence) problematic. These issues produce differences between the orbits computed by the two centres, thus leading to different impact probabilities. However, it is not possible to determine which one is better, and their reliability is questionable in both cases. 

Asteroid 2023~VD$_3$ has a short observational arc of about 2 days, and the observations were all taken during its close approach with Earth in 2023. Both Aegis and Sentry are able to find a VI on 8 November 2024, however the IPs are significantly different: 0.00258 for Aegis and 0.00014 for Sentry. Cases like this $-$ short observational arc at close approach $-$ are difficult to treat, because of the small amount of data and their quality. In fact, NEAs may have a large sky-plane velocity at Earth close approaches, which makes the astrometry more difficult. In addition, time errors in astrometric measurements may lead to erroneous determination of astrometric uncertainties. While in recent years IAWN organised and coordinated two observational campaigns to estimate time errors of participating observatories \citep{farnocchia-etal_2022, farnocchia-etal_2023b}, there is not a systematic way to handle them yet. Since the operational version of Aegis does not make use of time uncertainties, this could be a reason for the differences seen in the impact monitoring results. This topic, however, deserves more deep studies in the future, and the transition to the new International Astronomical Union Astrometry Data Exchange Standard (ADES) may help in addressing the issue. 

The object 2024~BY$_{15}$ has an orbit very similar to that of Earth, which causes many subsequent close approaches, making the long-term dynamics very complex. Note that Earth encounters with low relative velocity may cause low performances of the LOV method, since it is based on the target plane theory. In addition, non-gravitational forces $A_1, A_2$ and $A_3$ are also needed to properly fit the available data, and their value is compatible with those expected from artificial satellites\footnote{See JPL Solution \#16.}. This object has been the subject of an internal discussion between NEOCC, JPL and MPC, and it was decided to hold the results until further clarification on the nature of the object.

Finally, three objects for which Sentry found an impact probability larger than $10^{-3}$ and for which Aegis did not find any VI are 2006~RH$_{120}$, 2010~RF$_{12}$, and 2020~CD$_3$. The first two are small objects whose dynamics is significantly affected by non-gravitational forces, and the JPL SBDB provides estimates for all the three components $A_1, A_2$ and $A_3$. These two asteroids were recently classified into the new \textit{dark comets} category \citep{seligman-etal_2023, taylor-etal_2024}. Currently, Aegis is not able to fit the out-of-plane component $A_3$, and the orbit of these two asteroids is still computed without non-gravitational forces. As a result, these VIs are not found because of the different future dynamics. On the other hand, 2020~CD$_3$ is a meter-sized NEA with an orbit similar to that of Earth, and the nominal solution also enters gravitational capture in 2044, 2061, 2082, and 2103. Sentry found three VIs with IP larger than $10^{-3}$ in 2082, 2083, and 2084. On the other hand, the LOV method fails because of the really small relative velocities, therefore we relied on a Monte Carlo run to estimate the impact probabilities. With 100~000 orbital clones obtained by sampling the confidence region with the covariance matrix, Aegis found several VIs after 2100 with a cumulative IP of $5.7 \times 10^{-4}$, hence we were not able to find the three Sentry VIs. Note also that the orbit of 2020~CD$_3$ reported on the JPL SBDB has a determination of $A_1$, while the orbit on the NEOCC web portal does not have any non-gravitational effect. This may be the reason that explains the differences in the two results, because non-gravitational components may lead to different dynamical paths. Finally, it is worth noting that these significant differences described above all arise for NEAs smaller than about 20 meters in size, which are less critical from a planetary defence point of view.

As a final note, the fact that results obtained by the two centres may differ because of different assumptions made on the dynamical model, on the accuracy of the astrometric measurements, or on timing errors, adds redundancy to the whole system of impact monitoring.
While the results presented here indicate a general good agreement between the two impact monitoring systems on the vast majority of the cases, a deeper analysis to understand the limitations of the two impact monitoring methods needs to be performed in the future.

\subsubsection{A practical Planetary Defence scenario: 2023 DZ$_2$}
Here we briefly summarise how the three impact monitoring systems (Aegis, Sentry, and Clomon-2) operated in a practical Planetary Defence scenario which happened in recent months: 2023~DZ$_2$. This Tunguska-size NEA was discovered on 27 February 2023 \citep{popescu-etal_2023} from the La Palma observatory (MPC code 950), about a month before a close approach with Earth. The discovery was announced in MPEC 2023-F12\footnote{\url{https://www.minorplanetcenter.net/mpec/K23/K23F12.html}}, which was released on 16 March 2023 and contained 31 observations. Upon discovery, all the three impact monitoring systems found a VI on 27 March 2027, but they computed three different IP values: $7.2 \times 10^{-6}$ for Aegis, $1.3 \times 10^{-4}$ for Sentry, and $1.6 \times 10^{-5}$ for Clomon-2. The corresponding PS values were determined at $-3.49$, $-2.07$, and $-3.00$, respectively, and the results were posted on the web, although the PS computed by Sentry was almost at the cross-check validation limit of $-2$. The differences between Aegis and Clomon-2, which are based on a similar implementation of the LOV method, were explained by the different coordinates used to sample the LOV: Aegis used Equinoctial coordinates, while Clomon-2 used Cartesian coordinates. With a subsequent run performed using Cartesian coordinates, the IP computed by Aegis increased to $8.1 \times 10^{-6}$, which is close to that determined by Clomon-2. A Monte Carlo run over 300~000 clones found a probability of impact of $(5 \pm 1) \times 10^{-5}$ for the 2027 VI, which is slightly larger than what computed by Aegis and Clomon-2. This difference may be due to the settings used to discretize the LOV, which affect how the impact probability is approximated. On the other hand, the differences between the Aegis/Clomon-2 and Sentry results are likely due to the large uncertainties in the orbit computed at this stage, which had a 1-$\sigma$ uncertainty in mean anomaly of more than 1 deg.

On 17 March, 25 additional follow-up observations were announced in MPEC 2023-F20\footnote{\url{https://www.minorplanetcenter.net/mpec/K23/K23F20.html}}, including those performed by NEOCC astronomers. With this data, the uncertainties in the orbital elements dropped by more than two orders of magnitude, and the impact monitoring systems computed an IP of about 1 in 1~000, a PS of about $-1.1$, and a TS of 1. The three results were all consistent, and they were posted on the web without the need of a validation. In the meantime, a world-wide follow-up campaign coordinated by IAWN was also triggered \citep{reddy-etal_2024}. 

\begin{figure}
    \centering
    \includegraphics[width=0.48\linewidth]{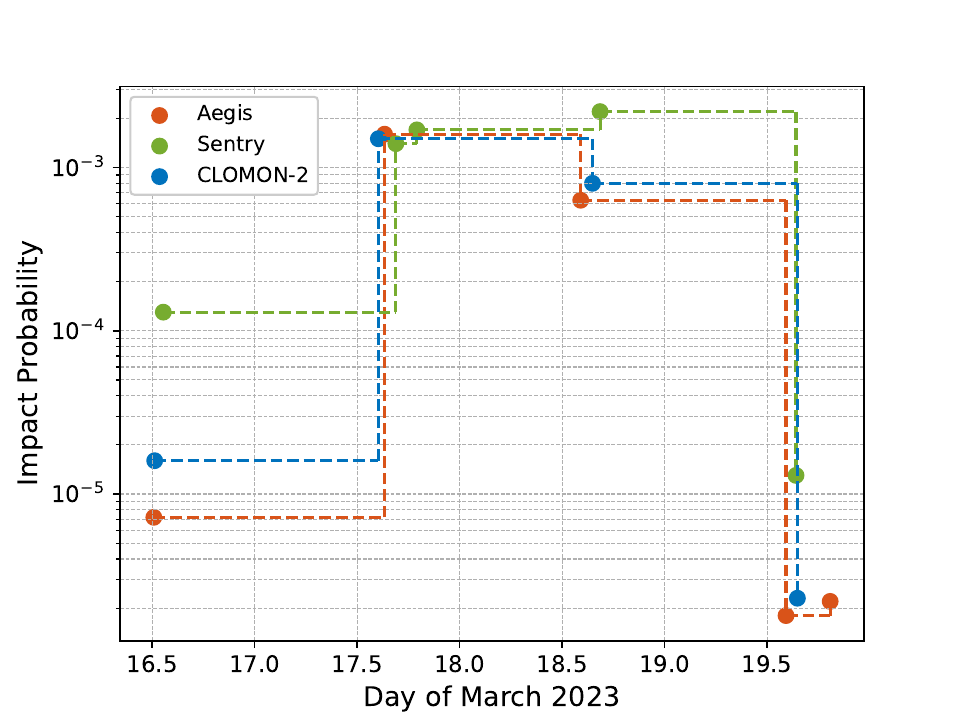}
    \includegraphics[width=0.48\linewidth]{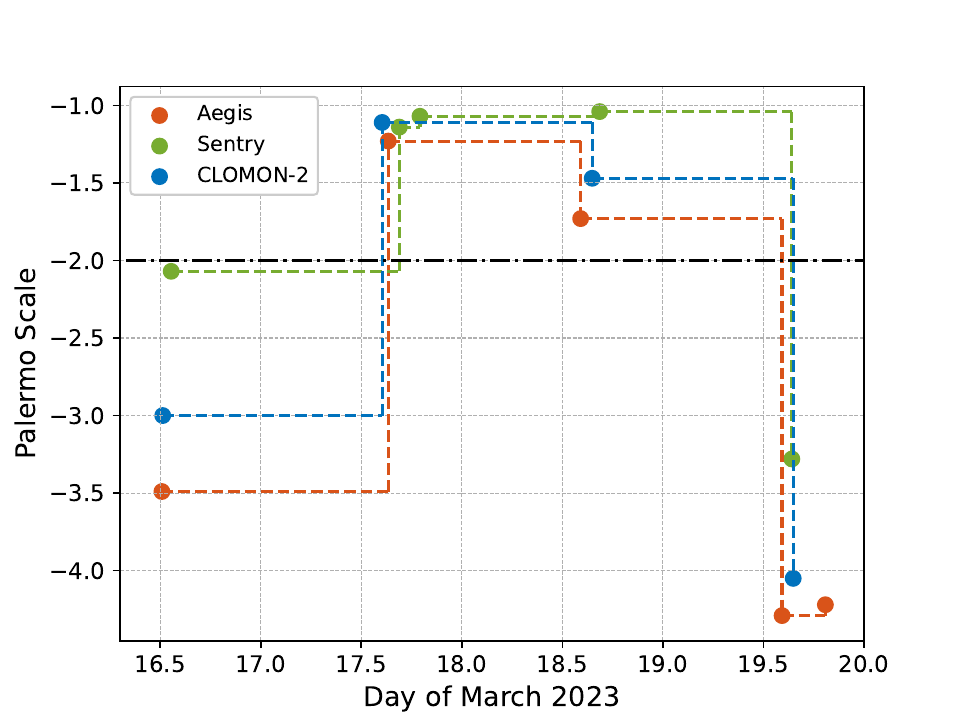}
    \caption{Evolution of the IP (left panel) and PS (right panel) of 2023~DZ$_2$ between 16 and 19 March 2024, as computed by Aegis, Sentry, and Clomon-2. The values refer to the VI of 27 March 2027.}
    \label{fig:2023DZ2_IPPS}
\end{figure}

More observations were announces on 18 March in MPEC 2023-F30\footnote{\url{https://www.minorplanetcenter.net/mpec/K23/K23F30.html}}, which also included precoveries that extended the arc by roughly a month. The IPs computed by Aegis and Clomon-2 slightly dropped, while that of Sentry slightly increased. In any case, the computed PS was still larger than $-2$, and these numbers were considered good enough to be posted on the web. With new follow-up observations announced on 19 March, the IP significantly dropped in all the three systems, while by the 21 March 2023~DZ$_2$ was removed from all the different risk lists. Figure~\ref{fig:2023DZ2_IPPS} shows the evolution of the IP and the PS of the 2027 VI, computed between 16 and 19 March 2023 by the three impact monitoring systems. Note also that results may be updated more than once per day, typically because of manual changes in the astrometric errors of specific observations.

This practical example shows that differences can arise between the three systems, especially when the observational arc is short and uncertainties in the orbit are large. Despite this fact, the availability of independent systems is needed in order to maintain redundancy, and to increase the awareness of follow-up observations to extend the observational arc and remove (or confirm) VIs with high PS, even if results may differ in specific cases.

\section{Summary and future developments}
\label{s:conclusions}
In this paper we introduced and gave an overview of the algorithms implemented in the ESA Aegis Orbit Determination and Impact Monitoring system: an automated software for the computation of asteroid orbits and their probabilities of impact with the Earth. This system is operated by the ESA NEOCC, and all the results obtained with the software are publicly available on the NEOCC web portal. Aegis maintains and supports many services available on the NEOCC web portal, including the orbital catalogue of asteroids, the Risk List, the Close Approaches List, all the tools of the NEO Toolkit and the ephemerides service. We performed a comparison between the orbit of NEAs computed by the NEOCC and those computed by the JPL SSD group, and concluded that the results are generally statistically comparable. We also compared the impact probability computed by the ESA NEOCC and the JPL Sentry impact monitoring systems, and concluded that they agree on the vast majority of the cases. We suggested what could cause large differences in the results, but a deeper analysis is out of the scope of this paper.

Several developments and upgrades are already planned or foreseen in the future. We plan to expand the dynamical model to include other smaller effects, such as the gravitational perturbation of the most massive 343 asteroids and 30 Kuiper Belt Objects \citep{park-etal_2021}, and add the non-gravitational component orthogonal to the plane of motion in Eq.~\eqref{eq:nongrav}. Another substantial upgrade regards orbit determination of comets. Currently, the Aegis system is not able to process data and compute orbits of comets, and the information available on the NEOCC web portal is replicated from the JPL SBDB. One of the plans of the NEOCC for the near future is to be able to independently process data about comets. Suitable methods for orbit determination of such objects will be implemented, and an upgrade to the automatic processing of cometary MPECs is going to be integrated in the current pipelines. 
% Artsat
The NEOCC is also planning to maintain a catalogue of known artificial objects around the Earth, and the Aegis software will be adapted to compute their orbits.
Moreover, several new services and upgrades to the NEOCC web portal are foreseen, such as: a visualiser for the stationary of the MOID between an NEA and the Earth, an observation prediction tool for astronomers, the possibility to choose the planet for the reference system of the SOVT and a fly-by visualiser for other planets. The HTTPS APIs will potentially be modernised to make them more efficient and user-friendly, and to offer additional data with automated access. 

Several important changes are going to be needed to make the Aegis system ready for the beginning of the operations of the Vera Rubin Observatory, that is going to dramatically increase the amount of data to process. Among them, we plan to adopt the new ADES format for Solar System objects as default input data in the Aegis system, which has several advantages with respect to the current 80-columns format. Along with this, the retrieval of new observational data by parsing the MPECs from MPC web pages is foreseen to be discontinued, giving preference to  data-fetching from the new MPC PostGres database, thus changing the paradigm for the maintenance of the orbital catalogue. 
This will allow more frequent orbit updates, providing users more up-to-date information. 
From the hardware point of view, we are planning to migrate from the current system to a set of micro-services that will be able to run on a highly scalable infrastructure, such as the ESA Cloud computing facility. This would enable the availability of more computational resources and a more efficient distribution of jobs on a large number of nodes, thus increasing the overall speed of the system.  

\section*{Acknowledgements}
% Dedication to Andrea Milani
This paper is dedicated to the memory of Professor Andrea Milani, whose invaluable guidance shaped the development of the Aegis software, and for his contribution to the advancement of NEOCC. We also wish to express our sincere gratitude to Giovanni Valsecchi, for his significant contributions to the development and enhancement of the Aegis software. We also thank the two anonymous referees, whose comments helped improving the quality of the manuscript.
The work by D. Bracali Cioci, F. Bernardi, A. Bertolucci, L. Dimare, F. Guerra, V. Baldisserotto, A. Chessa, and A. Del Vigna was conducted under European Space Agency contracts: No. AO/1-9075/17/DE/MRP "P3-COM-VI", No. 4000116371/16/D/MRP "P2-NEO-VII" and its CCN, No. 4000123583/18/D/MRP "P3-NEO-XIII" and its CCN, No. 4000133990/21/D/KS "S1-PD-02".

\section*{Data availability}
The datasets generated during and/or analysed during the current study are available from the corresponding author on reasonable request. Scripts used to generate Fig.~\ref{fig:Chi_numbered}, \ref{fig:Chi_unnumbered} and \ref{fig:IP_ratios} use publicly available data from the ESA NEOCC and the JPL SBDB, and are available from the corresponding author on reasonable request.

\section*{Conflict of interest}
The authors declare no conflict of interest.

\bibliographystyle{apalike85}
\bibliography{holyBib}{}

\begin{thebibliography}{}

\bibitem[{Bernardi} et~al., 2021]{bernardi-etal_2021}
{Bernardi}, F., et~al. (2021).
\newblock {New NEODyS Tools for the EU funded NEOROCKS Project: Observations
  support and Priority Lists}.
\newblock In {\em 7th IAA Planetary Defense Conference}, page~21.

\bibitem[{Binzel}, 2000]{binzel_2000}
{Binzel}, R.~P. (2000).
\newblock {The Torino Impact Hazard Scale}.
\newblock {\em Planetary and Space Science}, 48(4):297--303.

\bibitem[{Bischoff} et~al., 2023]{bischoff-etal_2023}
{Bischoff}, A., et~al. (2023).
\newblock {Saint-Pierre-le-Viger (L5-6) from asteroid 2023 CX$_{1}$ recovered
  in the Normandy, France{\textemdash}220 years after the historic fall of
  L'Aigle (L6 breccia) in the neighborhood}.
\newblock {\em \maps}, 58(10):1385--1398.

\bibitem[{Bowell} et~al., 1989]{bowell-etal_1989}
{Bowell}, E., et~al. (1989).
\newblock {Application of photometric models to asteroids.}
\newblock In {Binzel}, R.~P., et~al., editors, {\em Asteroids II}, pages
  524--556.

\bibitem[{Brown} et~al., 2002]{brown-etal_2002}
{Brown}, P., et~al. (2002).
\newblock {The flux of small near-Earth objects colliding with the Earth}.
\newblock {\em \nat}, 420(6913):294--296.

\bibitem[{Bulirsch} and {Stoer}, 2002]{bulirsch-stoer_2002}
{Bulirsch}, R. and {Stoer}, J. (2002).
\newblock {\em Introduction to Numerical Analysis}.
\newblock Texts in Applied Mathematics 12. Springer New York, 3rd ed edition.

\bibitem[{Cano} et~al., 2021]{cano-etal_2021}
{Cano}, J.~L., et~al. (2021).
\newblock {Evaluation of an NEO Close Approach Frequency Index for Public/media
  Release Purposes}.
\newblock In {\em 7th IAA Planetary Defense Conference}, page~86.

\bibitem[{Carpino} et~al., 2003]{carpino-etal_2003}
{Carpino}, M., et~al. (2003).
\newblock {Error statistics of asteroid optical astrometric observations}.
\newblock {\em \icarus}, 166(2):248--270.

\bibitem[{Chamberlin} et~al., 2001]{chamberlin-etal_2001}
{Chamberlin}, A.~B., et~al. (2001).
\newblock {Sentry: An Automated Close Approach Monitoring System for Near-Earth
  Objects}.
\newblock In {\em AAS/Division for Planetary Sciences Meeting Abstracts \#33},
  volume~33 of {\em AAS/Division for Planetary Sciences Meeting Abstracts},
  page 41.08.

\bibitem[{Chesley} and {Chodas}, 2015]{chesley-chodas_2015}
{Chesley}, S. and {Chodas}, P. (2015).
\newblock {Impact Risk Estimation and Assessment Scales}.
\newblock In {\em Handbook of Cosmic Hazards and Planetary Defense}, pages
  651--662.

\bibitem[{Chesley} et~al., 2002]{chesley-etal_2002}
{Chesley}, S.~R., et~al. (2002).
\newblock {Quantifying the Risk Posed by Potential Earth Impacts}.
\newblock {\em Icarus}, 159(2):423--432.

\bibitem[{Chesley} et~al., 2014]{chesley-etal_2014}
{Chesley}, S.~R., et~al. (2014).
\newblock {Orbit and bulk density of the OSIRIS-REx target Asteroid (101955)
  Bennu}.
\newblock {\em \icarus}, 235:5--22.

\bibitem[{Chesley} and {Milani}, 1999]{chesley-milani_1999}
{Chesley}, S.~R. and {Milani}, A. (1999).
\newblock {NEODyS: an online information system for near-Earth objects}.
\newblock In {\em AAS/Division for Planetary Sciences Meeting Abstracts \#31},
  volume~31 of {\em AAS/Division for Planetary Sciences Meeting Abstracts},
  page 28.06.

\bibitem[{Conversi} et~al., 2021]{conversi-etal_2021}
{Conversi}, L., et~al. (2021).
\newblock {Esa's NEO Coordination Centre Observational Network}.
\newblock In {\em 7th IAA Planetary Defense Conference}, page~22.

\bibitem[{Conversi} et~al., 2023]{conversi-etal_2023}
{Conversi}, L., et~al. (2023).
\newblock {NEOMIR: an NEO early-warning, space-based mission}.
\newblock In {\em 2nd NEO and Debris Detection Conference}, page~86.

\bibitem[{Daly} et~al., 2023]{daly-etal_2023}
{Daly}, R.~T., et~al. (2023).
\newblock {Successful kinetic impact into an asteroid for planetary defence}.
\newblock {\em \nat}, 616(7957):443--447.

\bibitem[{Del Vigna} et~al., 2020]{delvigna-etal_2020}
{Del Vigna}, A., et~al. (2020).
\newblock {Improving impact monitoring through Line Of Variations
  densification}.
\newblock {\em \icarus}, 351:113966.

\bibitem[{Del Vigna} et~al., 2019a]{delvigna-etal_2019}
{Del Vigna}, A., et~al. (2019a).
\newblock {Completeness of Impact Monitoring}.
\newblock {\em \icarus}, 321:647--660.

\bibitem[{Del Vigna} et~al., 2019b]{delvigna-etal_2019b}
{Del Vigna}, A., et~al. (2019b).
\newblock {Yarkovsky effect detection and updated impact hazard assessment for
  near-Earth asteroid (410777) 2009 FD}.
\newblock {\em \aap}, 627:L11.

\bibitem[{Denneau} et~al., 2013]{denneau-etal_2013}
{Denneau}, L., et~al. (2013).
\newblock {The Pan-STARRS Moving Object Processing System}.
\newblock {\em \pasp}, 125(926):357.

\bibitem[{Desmars} et~al., 2011]{desmars-etal_2011}
{Desmars}, J., et~al. (2011).
\newblock {Statistical analysis on the uncertainty of asteroid ephemerides}.
\newblock In {Alecian}, G., et~al., editors, {\em SF2A-2011: Proceedings of the
  Annual meeting of the French Society of Astronomy and Astrophysics}, pages
  639--642.

\bibitem[{Di Pippo} and {Perozzi}, 2015]{dipippo-perozzi_2015}
{Di Pippo}, S. and {Perozzi}, E. (2015).
\newblock {European Operational Initiative on NEO Near Earth Object (NEO)
  Hazard Monitoring}.
\newblock In {\em Handbook of Cosmic Hazards and Planetary Defense}, pages
  615--635.

\bibitem[{Dimare} et~al., 2020]{dimare-etal_2020}
{Dimare}, L., et~al. (2020).
\newblock {Use of the semilinear method to predict the impact corridor on
  ground}.
\newblock {\em Celestial Mechanics and Dynamical Astronomy}, 132(3):20.

\bibitem[{Drolshagen} et~al., 2010]{drolshagen-etal_2010}
{Drolshagen}, G., et~al. (2010).
\newblock {The near-Earth objects segment of the european Space Situational
  Awareness program}.
\newblock {\em Cosmic Research}, 48(5):399--402.

\bibitem[{Dymock}, 2007]{dymock_2007}
{Dymock}, R. (2007).
\newblock {The H and G magnitude system for asteroids}.
\newblock {\em Journal of the British Astronomical Association}, 117:342--343.

\bibitem[{Eggl} et~al., 2020]{eggl-etal_2020}
{Eggl}, S., et~al. (2020).
\newblock {Star catalog position and proper motion corrections in asteroid
  astrometry II: The Gaia era}.
\newblock {\em \icarus}, 339:113596.

\bibitem[{Everhart}, 1985]{everhart_1985}
{Everhart}, E. (1985).
\newblock {\em {An efficient integrator that uses Gauss-Radau spacings}},
  volume 115, page 185.
\newblock Cambridge University Press.

\bibitem[{Farnocchia} et~al., 2016]{farnocchia-etal_2016}
{Farnocchia}, D., et~al. (2016).
\newblock {The trajectory and atmospheric impact of asteroid 2014 AA}.
\newblock {\em \icarus}, 274:327--333.

\bibitem[{Farnocchia} et~al., 2015]{farnocchia-etal_2015b}
{Farnocchia}, D., et~al. (2015).
\newblock {Star catalog position and proper motion corrections in asteroid
  astrometry}.
\newblock {\em \icarus}, 245:94--111.

\bibitem[{Farnocchia} et~al., 2023]{farnocchia-etal_2023b}
{Farnocchia}, D., et~al. (2023).
\newblock {The Second International Asteroid Warning Network Timing Campaign:
  2005 LW3}.
\newblock {\em \psj}, 4(11):203.

\bibitem[{Farnocchia} et~al., 2022]{farnocchia-etal_2022}
{Farnocchia}, D., et~al. (2022).
\newblock {International Asteroid Warning Network Timing Campaign: 2019 XS}.
\newblock {\em Planetary Science Journal}, 3(7):156.

\bibitem[{Fenucci} et~al., 2024]{fenucci-etal_2023}
{Fenucci}, M., et~al. (2024).
\newblock {An automated procedure for the detection of the Yarkovsky effect and
  results from the ESA NEO Coordination Centre}.
\newblock {\em \aap}, 682:A29.

\bibitem[{Fenucci} and {Novakovi{\'c}}, 2021]{fenucci-novakovic_2021}
{Fenucci}, M. and {Novakovi{\'c}}, B. (2021).
\newblock {The Role of the Yarkovsky Effect in the Long-term Dynamics of
  Asteroid (469219) Kamo'oalewa}.
\newblock {\em The Astronomical Journal}, 162(6):227.

\bibitem[{Fr{\"u}hauf} et~al., 2021]{fruhauf-etal_2021}
{Fr{\"u}hauf}, M., et~al. (2021).
\newblock {Meerkat Asteroid Guard imminent impactor warning service of the
  European Space Agency}.
\newblock In {\em 7th IAA Planetary Defense Conference}, page~97.

\bibitem[{Fuls} et~al., 2023]{fuls-etal_2023}
{Fuls}, C., et~al. (2023).
\newblock {Bridging Discoveries and Collaboration: Catalina Sky Survey's NEO
  Survey, Follow-up, and Community Projects}.
\newblock In {\em AAS/Division for Planetary Sciences Meeting Abstracts},
  volume~55 of {\em AAS/Division for Planetary Sciences Meeting Abstracts},
  page 405.08.

\bibitem[{Geng} et~al., 2023]{geng-etal_2023}
{Geng}, S., et~al. (2023).
\newblock {Near-Earth object 2022 EB5: From atmospheric entry to physical
  properties and orbit}.
\newblock {\em \aap}, 670:A27.

\bibitem[{Gianotto} et~al., 2023]{gianotto-etal_2023}
{Gianotto}, F., et~al. (2023).
\newblock {Meerkat Asteroid Guard {\textendash} ESA's imminent impactor warning
  service}.
\newblock In {\em 2nd NEO and Debris Detection Conference}, page~49.

\bibitem[{Granvik} et~al., 2018]{granvik-etal_2018}
{Granvik}, M., et~al. (2018).
\newblock {Debiased orbit and absolute-magnitude distributions for near-Earth
  objects}.
\newblock {\em Icarus}, 312:181--207.

\bibitem[{Granvik} et~al., 2017]{granvik-etal_2017}
{Granvik}, M., et~al. (2017).
\newblock {Escape of asteroids from the main belt}.
\newblock {\em \aap}, 598:A52.

\bibitem[{Gronchi}, 2005]{gronchi_2005}
{Gronchi}, G.~F. (2005).
\newblock {An Algebraic Method to Compute the Critical Points of the Distance
  Function Between Two Keplerian Orbits}.
\newblock {\em Celestial Mechanics and Dynamical Astronomy}, 93(1-4):295--329.

\bibitem[Gronchi and Tommei, 2007]{gronchi-tommei_2007}
Gronchi, G.~F. and Tommei, G. (2007).
\newblock On the uncertainty of the minimal distance between two confocal
  {K}eplerian orbits.
\newblock {\em Discrete and Continuous Dynamical Systems - B}, 7(4):755--778.

\bibitem[{Guerra} et~al., 2016]{guerra-etal_2016}
{Guerra}, F., et~al. (2016).
\newblock {An automatic and robust method for the integration of orbits with
  close approaches}.
\newblock In {\em Stardust Final Conference on Asteroids and Space Debris}.

\bibitem[Hairer et~al., 1993]{hairer-wanner_1993}
Hairer, E., et~al. (1993).
\newblock {\em Solving Ordinary Differential Equations II: Stiff and
  Differential-Algebraic Problems}.
\newblock Solving Ordinary Differential Equations II: Stiff and
  Differential-algebraic Problems. Springer.

\bibitem[{Ivezi{\'c}} et~al., 2019]{ivezic-etal_2019}
{Ivezi{\'c}}, {\v{Z}}., et~al. (2019).
\newblock {LSST: From Science Drivers to Reference Design and Anticipated Data
  Products}.
\newblock {\em \apj}, 873(2):111.

\bibitem[{Jenniskens} et~al., 2021]{jenniskens-etal_2021}
{Jenniskens}, P., et~al. (2021).
\newblock {The impact and recovery of asteroid 2018 LA}.
\newblock {\em \maps}, 56(4):844--893.

\bibitem[{Jenniskens} et~al., 2009]{jenniskens-etal_2009}
{Jenniskens}, P., et~al. (2009).
\newblock {The impact and recovery of asteroid 2008 TC$_{3}$}.
\newblock {\em \nat}, 458(7237):485--488.

\bibitem[{Jones} et~al., 2018]{jones-etal_2018}
{Jones}, R.~L., et~al. (2018).
\newblock {The Large Synoptic Survey Telescope as a Near-Earth Object discovery
  machine}.
\newblock {\em \icarus}, 303:181--202.

\bibitem[Kustaanheimo and Stiefel, 1965]{kustaanheimo-stiefel_1965}
Kustaanheimo, P. and Stiefel, E. (1965).
\newblock Perturbation theory of {K}epler motion based on spinor
  regularization.
\newblock {\em Journal für die reine und angewandte Mathematik}, 218:204--219.

\bibitem[{Mainzer} et~al., 2023]{mainzer-etal_2023}
{Mainzer}, A.~K., et~al. (2023).
\newblock {The Near-Earth Object Surveyor Mission}.
\newblock {\em \psj}, 4(12):224.

\bibitem[{Marsden}, 1998]{marsden_1998}
{Marsden}, B.~G. (1998).
\newblock {1997 XF11}.
\newblock {\em IAU Circular}, 6837:2.

\bibitem[{Marsden} et~al., 1978]{marsden-etal_1978}
{Marsden}, B.~G., et~al. (1978).
\newblock {New osculating orbits for 110 comets and analysis of original orbits
  for 200 comets.}
\newblock {\em \aj}, 83:64--71.

\bibitem[{Milani}, 1999]{milani_1999}
{Milani}, A. (1999).
\newblock {The Asteroid Identification Problem. I. Recovery of Lost Asteroids}.
\newblock {\em \icarus}, 137(2):269--292.

\bibitem[{Milani} et~al., 2002]{milani-etal_2002}
{Milani}, A., et~al. (2002).
\newblock {Asteroid Close Approaches: Analysis and Potential Impact Detection}.
\newblock In {\em Asteroids III}, pages 55--69.

\bibitem[{Milani} et~al., 2005]{milani-etal_2005}
{Milani}, A., et~al. (2005).
\newblock {Nonlinear impact monitoring: line of variation searches for
  impactors}.
\newblock {\em \icarus}, 173(2):362--384.

\bibitem[{Milani} and {Gronchi}, 2009]{milani-gronchi_2009}
{Milani}, A. and {Gronchi}, G.~F. (2009).
\newblock {\em Theory of Orbit Determination}.
\newblock Cambridge University Press.

\bibitem[{Milani} et~al., 2008]{milani-etal_2008}
{Milani}, A., et~al. (2008).
\newblock {Topocentric orbit determination: Algorithms for the next generation
  surveys}.
\newblock {\em \icarus}, 195(1):474--492.

\bibitem[{Milani} et~al., 2007]{milani-etal_2007}
{Milani}, A., et~al. (2007).
\newblock {New Definition of Discovery for Solar System Objects}.
\newblock {\em Earth Moon and Planets}, 100(1-2):83--116.

\bibitem[{Milani} and {Nobili}, 1988]{milani-nobili_1988}
{Milani}, A. and {Nobili}, A.~M. (1988).
\newblock {Integration error over very long time spans.}
\newblock {\em Celestial Mechanics}, 43(1-4):1--34.

\bibitem[Milani and Valsecchi, 2011a]{milani-valsecchi_2011b}
Milani, A. and Valsecchi, G. (2011a).
\newblock Neo segment technology predevelopment plan – part i.
\newblock In {\em WP4110 Report. ESA/ESTEC Contract n. 22929/09/ML/GLC}.

\bibitem[Milani and Valsecchi, 2011b]{milani-valsecchi_2011a}
Milani, A. and Valsecchi, G. (2011b).
\newblock Roadmap for segment technology development and demonstration - part
  i.
\newblock In {\em WP4210 Report. ESA/ESTEC Contract n. 22929/09/ML/GLC}.

\bibitem[Milani et~al., 2003]{milani-etal_2003}
Milani, A., et~al. (2003).
\newblock Near earth object space mission preparation: Don quijote mission
  executive summary.
\newblock In {\em Report of ESA GSP contract 26252/02/F/IZ}.

\bibitem[Montenbruck and Gill, 2013]{montenbruck-gil_2013}
Montenbruck, O. and Gill, E. (2013).
\newblock {\em Satellite Orbits: Models, Methods and Applications}.
\newblock Springer Berlin Heidelberg.

\bibitem[{Moskovitz} et~al., 2022]{moskovitz-etal_2022}
{Moskovitz}, N.~A., et~al. (2022).
\newblock {The astorb database at Lowell Observatory}.
\newblock {\em Astronomy and Computing}, 41:100661.

\bibitem[{Nesvorn{\'y}} et~al., 2023]{nesvorny-etal_2023}
{Nesvorn{\'y}}, D., et~al. (2023).
\newblock {NEOMOD: A New Orbital Distribution Model for Near-Earth Objects}.
\newblock {\em \aj}, 166(2):55.

\bibitem[{Nesvorn{\'y}} et~al., 2024a]{nesvorny-etal_2024}
{Nesvorn{\'y}}, D., et~al. (2024a).
\newblock {NEOMOD 2: An updated model of Near-Earth Objects from a decade of
  Catalina Sky Survey observations}.
\newblock {\em \icarus}, 411:115922.

\bibitem[{Nesvorn{\'y}} et~al., 2024b]{nesvorny-etal_2024b}
{Nesvorn{\'y}}, D., et~al. (2024b).
\newblock {NEOMOD 3: The debiased size distribution of Near Earth Objects}.
\newblock {\em \icarus}, 417:116110.

\bibitem[{Orbfit Consortium}, 2011]{orbfit_2011}
{Orbfit Consortium} (2011).
\newblock {OrbFit: Software to Determine Orbits of Asteroids}.
\newblock Astrophysics Source Code Library, record ascl:1106.015.

\bibitem[{Park} et~al., 2021]{park-etal_2021}
{Park}, R.~S., et~al. (2021).
\newblock {The JPL Planetary and Lunar Ephemerides DE440 and DE441}.
\newblock {\em \aj}, 161(3):105.

\bibitem[Perozzi, 2014]{perozzi_2014}
Perozzi, E. (2014).
\newblock {\em The Near Earth Asteroid Hazard and Mitigation}, pages 87--97.
\newblock Springer International Publishing, Cham.

\bibitem[{Popescu} et~al., 2023]{popescu-etal_2023}
{Popescu}, M.~M., et~al. (2023).
\newblock {Discovery and physical characterization as the first response to a
  potential asteroid collision: The case of 2023 DZ$_{2}$}.
\newblock {\em \aap}, 676:A126.

\bibitem[{Pravec} and {Harris}, 2007]{pravec-harris_2007}
{Pravec}, P. and {Harris}, A.~W. (2007).
\newblock {Binary asteroid population. 1. Angular momentum content}.
\newblock {\em \icarus}, 190(1):250--259.

\bibitem[{Ram{\'\i}rez Moreta} et~al., 2023]{ramirezmoreta-etal_2023}
{Ram{\'\i}rez Moreta}, P., et~al. (2023).
\newblock {The NEO Toolkit: a new set of astronomical tools for the NEO
  community}.
\newblock In {\em 2nd NEO and Debris Detection Conference}, page~46.

\bibitem[{Reddy} et~al., 2024]{reddy-etal_2024}
{Reddy}, V., et~al. (2024).
\newblock {2023 DZ2 Planetary Defense Campaign}.
\newblock {\em \psj}, 5(6):141.

\bibitem[{Roa} et~al., 2021]{roa-etal_2021}
{Roa}, J., et~al. (2021).
\newblock {A Novel Approach to Asteroid Impact Monitoring}.
\newblock {\em \aj}, 162(6):277.

\bibitem[{Seligman} et~al., 2023]{seligman-etal_2023}
{Seligman}, D.~Z., et~al. (2023).
\newblock {Dark Comets? Unexpectedly Large Nongravitational Accelerations on a
  Sample of Small Asteroids}.
\newblock {\em \psj}, 4(2):35.

\bibitem[{Spoto} et~al., 2014]{spoto-etal_2014}
{Spoto}, F., et~al. (2014).
\newblock {Nongravitational perturbations and virtual impactors: the case of
  asteroid (410777) 2009 FD}.
\newblock {\em \aap}, 572:A100.

\bibitem[{Spurny} et~al., 2024]{spurny-etal_2024}
{Spurny}, P., et~al. (2024).
\newblock {Atmospheric entry and fragmentation of small asteroid 2024 BX1:
  Bolide trajectory, orbit, dynamics, light curve, and spectrum}.
\newblock {\em arXiv e-prints}, page arXiv:2403.00634.

\bibitem[{Taylor} et~al., 2024]{taylor-etal_2024}
{Taylor}, A.~G., et~al. (2024).
\newblock {The dynamical origins of the dark comets and a proposed evolutionary
  track}.
\newblock {\em \icarus}, 420:116207.

\bibitem[{Tommei}, 2021]{tommei_2021}
{Tommei}, G. (2021).
\newblock {On the Impact Monitoring of Near-Earth Objects: Mathematical Tools,
  Algorithms, and Challenges for the Future}.
\newblock {\em Universe}, 7(4):103.

\bibitem[{Tonry} et~al., 2018]{tonry-etal_2018}
{Tonry}, J.~L., et~al. (2018).
\newblock {ATLAS: A High-cadence All-sky Survey System}.
\newblock {\em \pasp}, 130(988):064505.

\bibitem[{Valsecchi} et~al., 2003]{valsecchi-etal_2003}
{Valsecchi}, G.~B., et~al. (2003).
\newblock {Resonant returns to close approaches: Analytical theory}.
\newblock {\em \aap}, 408:1179--1196.

\bibitem[{Vere{\v{s}}} et~al., 2017]{veres-etal_2017}
{Vere{\v{s}}}, P., et~al. (2017).
\newblock {Statistical analysis of astrometric errors for the most productive
  asteroid surveys}.
\newblock {\em \icarus}, 296:139--149.

\bibitem[{Vokrouhlický} et~al., 2015]{vokrouhlicky-etal_2015}
{Vokrouhlický}, D., et~al. (2015).
\newblock {\em The Yarkovsky and YORP Effects}, pages 509--532.
\newblock University of Arizona Press.

\bibitem[{Will}, 1993]{will_1993}
{Will}, C.~M. (1993).
\newblock {\em {Theory and Experiment in Gravitational Physics}}.

\bibitem[{Wright} et~al., 2010]{wright-etal_2010}
{Wright}, E.~L., et~al. (2010).
\newblock {The Wide-field Infrared Survey Explorer (WISE): Mission Description
  and Initial On-orbit Performance}.
\newblock {\em \aj}, 140(6):1868--1881.

\end{thebibliography}

\end{document}